\documentclass[onecolumn,notitlepage,10pt,showpacs,prf,floatfix,longbibliography]{revtex4-1}

\usepackage[utf8]{inputenc}
\usepackage{amsmath}
\usepackage{amssymb}
\usepackage{lipsum}
\usepackage{bm}
\usepackage{graphicx}
\usepackage{graphics}
\usepackage[dvipdf]{epsfig}
\usepackage{subcaption}
\usepackage{float}
\usepackage{wrapfig}
\usepackage{mathtools}
\usepackage{url}
\usepackage[dvipsnames]{xcolor}

\begin{document}

\title{Phase-field model for a weakly compressible soft layered material: \\
    morphological transitions on smectic-isotropic interfaces}

\author{Eduardo Vitral\,}
\email{evitral@unr.edu}
\affiliation{Department of Mechanical Engineering, 
    University of Nevada, \\ 1664 N. Virginia St., Reno, NV 89557, USA}
\author{Perry H. Leo}
\affiliation{Department of Aerospace Engineering and Mechanics,
    University of Minnesota, \\ 110 Union St. SE, Minneapolis, MN 55455, USA}
\author{Jorge Vi\~nals}
\affiliation{School of Physics and Astronomy,
    University of Minnesota, 116 Church St. SE, Minneapolis, MN 55455, USA}

\begin{abstract}
A coupled phase-field and hydrodynamic model is introduced to describe a two-phase, weakly compressible smectic (layered phase) in contact with an isotropic fluid of different density. A non-conserved smectic order parameter is coupled to a conserved mass density in order to accommodate non-solenoidal flows near the smectic-isotropic boundary arising from density contrast between the two phases. The model aims to describe morphological transitions in smectic thin films under heat treatment, in which arrays of focal conic defects evolve into conical pyramids and concentric rings through curvature dependent evaporation of smectic layers. The model leads to an extended thermodynamic relation at a curved surface that includes its Gaussian curvature, non-classical stresses at the boundary and flows arising from density gradients. The temporal evolution given by the model conserves the overall mass of the liquid crystal while still allowing for the modulated smectic structure to grow or shrink. A numerical solution of the governing equations reveals that pyramidal domains are sculpted at the center of focal conics upon a temperature increase, which display tangential flows at their surface. Other cases investigated include the possible coalescence of two cylindrical stacks of smectic layers, formation of droplets, and the interactions between focal conic domains through flow. 
\end{abstract}

\date{\today}

\maketitle


\section{Introduction}

Defect and texture engineering of soft matter~\cite{jangizehi2020defects,nielsen2020substrate} are promising design approaches for tuning the properties of materials by controlling the presence and spatial distribution of defects. Modulated soft phases, such as block copolymers~\cite{re:ruzette05}, smectic liquid crystals~\cite{degennes1995physics}, active and living matter~\cite{re:mitov17}, are of particular interest in this context since their broken translational symmetry is associated with lamellar patterns of uniaxial symmetry, which allows for a variety of topological defects. We focus here on smectic thin films which are known, under appropriate boundary conditions, to form focal conic domains: topological defects that arrange themselves in periodic arrays throughout the film, and form due to the hybrid molecular alignment between a substrate and the film's free surface.

\begin{figure}[ht]
	\centering
    \includegraphics[width=0.42\textwidth]{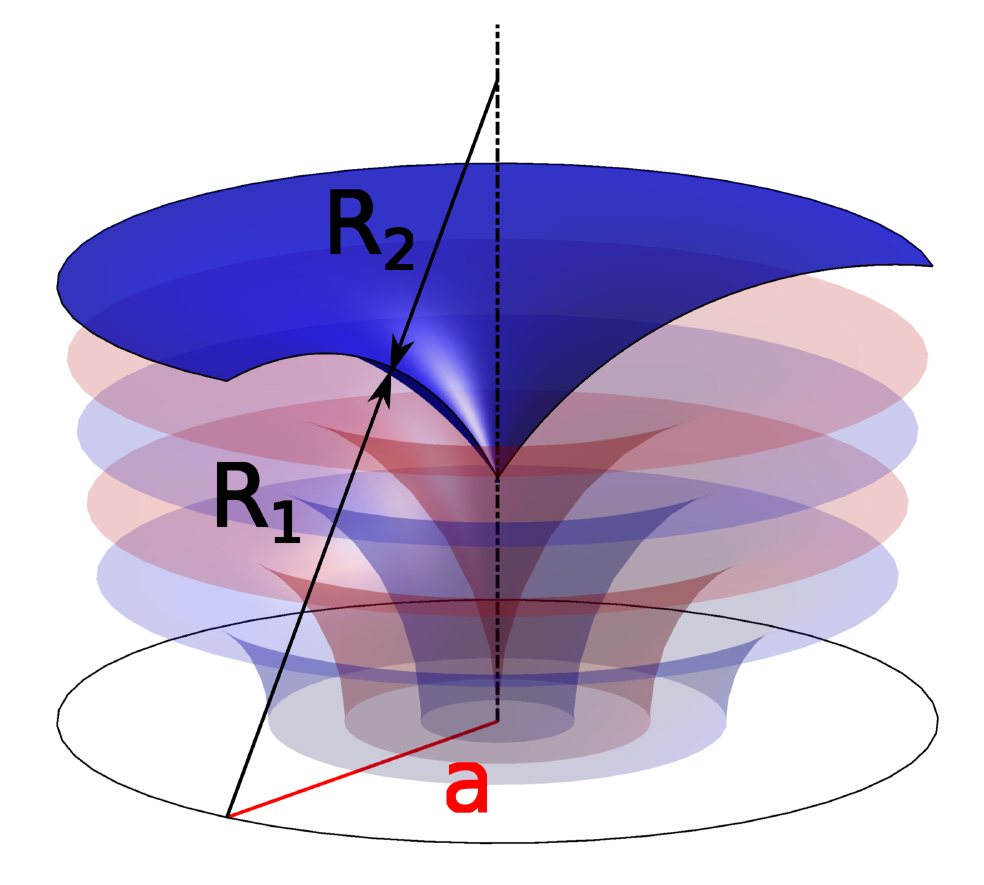}
    \caption{Structure of a toroidal FCD within a cylindrical region of radius \textsl{a}, presenting multiple molecular layers embedded in a smectic matrix of flat layers~\cite{kim2016controlling}. $R_1$ (with respective $c_1 > 0$) and $R_2$ ($c_2<0$) are the principal radii of a point on the external layer, so that the mean curvature $H$ changes from positive to negative as the core is approached, while the Gaussian curvature $G$ is negative everywhere.}
	\label{fig:fc-radius}
\end{figure}

The smectic phase of a liquid crystal is comprised of anisometric molecules that present collective orientational order along with a director axis $\mathbf{n}$, and are organized in periodically spaced molecular layers. When deposited on a substrate that induces tangential anchoring, smectic films in contact with air (with homeotropic molecular alignment at the smectic-air interface) may have their layers bend and form focal conic domains (FCDs)~\cite{shojaei2006role,guo2008controlling,kleman2009liquids,kim2018curvatures}, depending on the balance between elastic energy and surface anchoring~\cite{lavrentovich1994nucleation,kim2009confined}. These conical defects have been long known to be the equilibrium structure of the film, since the works of Friedel and Bragg in the  1920s and 1930s~\cite{friedel1922etats,bragg1934liquid}. The external layer forms a cusp-like macroscopic singularity at its center, whose neighborhood is a region of high mean and Gaussian curvatures. The mean curvature $H = (c_1+c_2)/2$ is the average of the principal curvatures $c_1 = 1/R_1$ and $c_2 = 1/R_2$ at a surface point (inverse of the principal radii), while the Gaussian curvature $G = c_1 c_2$ is the product of these curvatures. Figure~\ref{fig:fc-radius} presents a schematic for a toroidal FCD (axially symmetric) with multiple layers, defined within a cylindrical region of radius \textsl{a} embedded into a smectic matrix of flat layers. The principal radius $R_1$ of a point on a layer ($c_1 > 0$), associated with the hybrid alignment, has its origin at the bottom of the cylinder defined by \textsl{a}, whereas $R_2$ is antiparallel to $R_1$ ($c_2< 0$), and connects this point to the axis of symmetry of the FCD. One can observe that $R_2$ of the external molecular layer becomes very small as the core is approached.

Smectic films micropatterned with arrays of FCDs have been used as guides for colloidal dispersion~\cite{milette2012reversible,pratibha2010colloidal}, soft lithography~\cite{yoon2007internal,kim2010self}, and as templates for superhydrophobic surfaces~\cite{kim2014creation}. Nevertheless, there are still open problems associated with the formation and engineering uses of these defects. Recent findings have shown, for example, the existence of geometric memory in the nematic-smectic transition~\cite{gim2017morphogenesis,suh2019topological}, connecting the presence of boojum defects in the nematic to the nucleation of focal conics in the smectic, and have also shown that focal conics may even arise in the absence of hybrid alignment~\cite{selmi2017structures}. While the nucleation of defects in smectics has appealing functional applications, focal conics and dislocations can strongly dictate the structure of easily deformable thin films~\cite{coursault2016self}, and interfere in optical and conductivity properties~\cite{jangizehi2020defects}. Therefore, fine microstructural control is a key concern for soft material design.

A remarkable example is found in the experiments of Kim et al.~\cite{kim2016controlling,kim2018curvatures}, who have shown that smectic thin films presenting arrays of FCDs can undergo complex morphological transitions through sintering (i.e. reshaping of a smectic film at elevated temperatures for a certain amount of amount, with subsequent cooling). For particular sintering protocols, FCDs are sculpted through evaporation into unexpected patterns, which include conical pyramids and concentric rings. The observed configurations are controlled not only by the local mean curvature of the film surface (as classical thermodynamics would imply), but also by its Gaussian curvature~\cite{kim2016controlling}. This feature, together with the fact that the film is constantly evaporating and reshaping at elevated temperatures, are the main motivations for our study. We develop a three-dimensional phase-field framework that includes smectic elasticity in the film, interfacial thermodynamics and kinetics accounting for the effects of Gaussian curvature, and material compressibility at the smectic-air surface. 

The Oseen-Frank theory~\cite{oseen1933theory,frank1958liquid} of a bulk nematic phase naturally leads to incorporating the energy associated with small layer distortions as a function of a layer displacement variable in a bulk smectic~\cite{degennes1995physics,santangelo2005curvature}. Layer displacements along their normal direction are assumed parallel to the nematic director since their difference is a higher energy mode that relaxes quickly~\cite{santangelo2005curvature} (a constraint that may be abandoned with additional balance equations~\cite{capriz2001swelling}). This approach, alongside an explicit curvature elastic energy description~\cite{kleman2000curvature} adequate for surfaces with large mean and Gaussian curvatures, such as FCDs, have been successful in explaining why these defects are equilibrium configurations, in which layers are geometrically described as Dupin cyclides~\cite{friedel1922etats,schief2005nonlinear}. A nonlinear theory coupling the director to hydrodynamic flow was formulated by Brand and Pleiner~\cite{brand1980nonlinear}, which expanded on previous studies on flows and viscosities in mesophases~\cite{leslie1966some,martin1972unified,moritz1976nonlinearities}. A complete hydrodynamic theory for studying the smectic-isotropic transition was later proposed based on the Landau-de Gennes framework~\cite{brand2001macroscopic,mukherjee2001simple,abukhdeir2010edge}, written in terms of either the smectic layer displacement, the smectic complex order parameter, or the nematic $Q-$tensor. Small perturbations in a smectic are generally described by a scalar displacement normal to the layers. However, the layering can be also described by a complex amplitude, the leading order term in a Fourier series of the order parameter, with a phase that is a function of layer displacement, and a real amplitude that represents the strength of the local smectic order. Other coarse-grained models along these lines have also been proposed, based either on density deviations or local composition~\cite{poniewierski1991phase,linhananta1991phenomenological,pevnyi2014modeling,pezzutti2015smectic,xia2021structural}.

We have introduced a phase-field model for a smectic-isotropic system~\cite{vitral2019role,vitral2020model}, with a real order parameter $\psi$ that describes a modulated smectic phase in contact with an isotropic phase, and the diffuse interface between the two. By replacing discrete layers and describing two-phase interfaces by a smooth function, we allow tracking of arbitrarily distorted smectic planes, even while layers are breaking up or forming, as well as allowing two phase interfaces with cusps (such as in FCDs). We have chosen the energy of our phase-field model so that the Oseen-Frank energy for a smectic is recovered for small distortions; that is, the energy penalizes both deviations from the equilibrium layering spacing and bending of layers (splay of molecules). An asymptotic analysis of the model for a thin two phase interface reproduces the classical Gibbs-Thomson relation of interfacial thermodynamics, as well as corrections introduced by higher order curvatures and torsion. This extended thermodynamic relation at a distorted smectic-isotropic interface yields modified kinetic equations that account for the role of the Gaussian curvature and layering orientation on pattern formation, providing an explanation for the obervations by Kim et al.~\cite{kim2016controlling}. Numerical solutions of the governing equations reveal transitions from a FCD towards conical pyramids slightly away from coexistence (simulating a temperature increase in the experiments). 

The model of~\cite{vitral2019role} is limited by the assumption of incompressibility, and neglects advection of the modulated order parameter. That is, it strictly describes the diffusive evolution of an interface between a smectic and an isotropic phase of the same, constant, density. As our goal is to model an isotropic phase of lower density than the smectic, that is, an isotropic fluid (such as water or air) presenting small density of liquid crystal molecules, in~\cite{vitral2020model} we presented a way to incorporate a varying density field and hydrodynamics into the model. Smectic and isotropic phases of different density were considered, and fluid flows modeled along the lines of Cahn-Hilliard fluids~\cite{re:jasnow96,gurtin1996two} as extended by Lowengrub and Truskinovsky’s to quasi-incompressible fluids~\cite{lowengrub1998quasi} (see also \cite{lee2002modeling,abels2012thermodynamically,guo2017mass,gong2018fully,ding2007diffuse,shokrpour2018diffuse}). This assumption led to a density which smoothly varied from a uniform value inside the bulk of the smectic ($\rho_s$) to a uniform value in the bulk of the isotropic phase ($\rho_0$). However, the density and the smectic order parameter were not considered to be independent variables, but rather the local density was assumed to be a constitutive function of the local amplitude of the order parameter (quasi-incompressibility).  In this scenario, the balance of mass presents a strong constraint on how much the non-conserved order parameter can evolve. Therefore, the morphological transitions from FCDs we previously discussed~\cite{vitral2019role,kim2016controlling} are unable to take place in a quasi-incompressible smectic-isotropic system, so that the range of applicability of the model is limited.

In this work, we overcome these limitations by introducing a general model in which the smectic order parameter and the density are treated as independent variables, with model parameters that control how constrained is the smectic evolution, allowing for morphological transitions. We call it the \textit{weakly compressible smectic-isotropic model}.  We add a coupling (or penalty) term to the energy that limits deviations of the local density from the expected smectic and isotropic equilibrium densities in the bulk phases. The model enforces mass conservation, allows control on how strongly the conserved density affects the motion of the non-conserved smectic order parameter, and is shown to be numerically stable even in presence of a large density ratio between phases. This in sharp contrast with previously studied phase-field models for a modulated phase in contact with a $\psi = 0$ phase~\cite{thiele2013localized,knobloch2015spatial,espath2020generalized}, in which the order parameter $\psi$ was (i) non-conserved and free to evolve (e.g. Swift-Hohenberg dynamics) or (ii) conserved and restricted in its evolution (e.g. phase-field-crystal models). The weakly compressible smectic-isotropic model allows for non-conserved order parameter dynamics with a conserved density, as is physically the case. 

The paper is organized as follows. Section~\ref{sec:wcm} contains the model derivation, including the equations for the order parameter and balances of mass and momentum. We introduce a rotationally invariant energy which depends on a real order parameter $\psi$ and its gradients, with an additional penalty term for deviations from equilibrium density values. In Sec.~\ref{sec:ns} we discuss our numerical implementation based on a pseudo-spectral method. Section~\ref{sec:def} presents numerical results of growth and decay of planar smectic layers in order to show how the conserved density interacts with the non-conserved order parameter. In Sec.~\ref{sec:fcflow} we reconsider the evolution of FCDs, and show that morphological transitions from focal conics to conical pyramids or concentric rings can be obtained numerically from the weakly compressible model as long as the density coupling coefficient is not large. We also examine the flow at the surface of a FCD and a conical pyramid, and, based on the extended Gibbs-Thomson relation, we discuss how interfacial flows and stresses depend not only on the mean curvature but on the Gaussian curvature and layering orientation as well. Finally, in Sec.~\ref{sec:domain} we consider additional applications of the weakly compressible model, namely the coalescence of cylindrical stacks of smectic layers, formation of droplets from an initial smectic, and the interactions among neighboring FCDs in a smectic film.

\section{Weakly Compressible model}
\label{sec:wcm}

In this section we derive a diffuse interface model for a weakly compressible smectic phase in contact with an isotropic fluid when they have different equilibrium densities. A real order parameter representing the smectic layering is introduced~\cite{chaikin2000principles},
\begin{equation}
    \psi = \sum_n \frac{1}{2}[\bar{A}_n \,e^{in \mathbf{q}_0\cdot\mathbf{x}}+c.c.] \;,
\end{equation}
where $\mathbf{q}_0$ is a wave vector in the direction normal to the smectic layers, $q_0 = |\mathbf{q}_0|$ is its magnitude, $\lambda_0 = 2\pi/q_0$ is the characteristic period, and $\bar{A}_n$ is a complex amplitude. The order parameter $\psi$ is function of time $t$ and space $\mathbf{x} \in \mathbb{R}^3$. Near the transition, high harmonics are generally negligible, and smectic layering is well described by the approximate representation $\psi \approx \frac{1}{2}[\bar{A}\,e^{i\mathbf{q}_0\cdot\mathbf{x}}+c.c.]$. The complex amplitude has the form $\bar{A} = A\,e^{-iq_0 u}$, where $u(\mathbf{x},t)$ represents the displacement away from planar smectic layers, and $A$ is the real magnitude of the complex amplitude (the order parameter \textit{strength}).

We write the internal energy $\mathfrak{U}$ of the system in terms of the energy per unit mass $\mathfrak{u}$ and the mass density $\rho$, where $\mathfrak{u} = \mathfrak{u}(\mathfrak{s},\psi,\nabla \psi,\nabla^2\psi)$ and $\mathfrak{s}$ is the specific internal entropy. The functional dependence on $\nabla^2\psi$ does not appear for binary systems~\cite{gurtin1996two,lowengrub1998quasi}, but it is fundamental to model the smectic phase, as it makes the energy sensitive to layer distortions and curvatures. We also introduce in the energy an explicit coupling between the real amplitude $A$ of the modulated order parameter $\psi$, which is approximately constant in the bulk smectic and isotropic phases, and the density of the corresponding phase. This internal energy is written as
\begin{eqnarray}
    \mathfrak{U} &=& \int_\Omega \bigg\{\rho \mathfrak{u}\, + \frac{\zeta}{2}\Big[\rho-\rho_0-\kappa A\Big]^2\bigg\} d\mathbf{x} \; .
    \label{eq:energy}
\end{eqnarray}
The second term inside the integral penalizes density values that deviate from equilibrium values $\rho_{eq} = \rho_0+\kappa A$ in the smectic and isotropic phases, where $\zeta$, $\rho_0$ (the equilibrium density for the isotropic phase) and $\kappa$ are constants. In the limit of $\zeta \rightarrow \infty$, the density becomes constitutively governed by $A$, as $\rho = \rho_0 + \kappa A$, and the energy reduces to $\mathfrak{U} = \int_\Omega \rho\,\mathfrak{u}\,dx$. This is the quasi-incompressible limit that we previously studied in~\cite{vitral2020model}, and it imposes a strict constraint on the evolution of the non-conserved order parameter $\psi$. However, when $\zeta$ is finite, the density is an independent variable.

\subsection{Energy and entropy balances}

We first obtain the local form of the internal energy and entropy balances from the fist law of Thermodynamics, as detailed in~\cite{lowengrub1998quasi,vitral2020model}. The relations derived in this section are obtained in the absence of thermal radiation and heat flux though the boundary. When deriving the governing equations, we set no-flux boundary conditions: Neumann condition for the order parameter $\psi$ and zero normal fluid velocity $\mathbf{v}$ on the boundary, such that
\begin{eqnarray}
    \nabla \psi (\mathbf{x})  \cdot \mathbf{n} = \nabla \nabla^2 \psi (\mathbf{x})  \cdot \mathbf{n} =  0, \quad \mathbf{v}(\mathbf{x})\cdot \mathbf{n} = 0, \quad \mathbf{x} \in \partial \Omega.
\label{eq:bc-w}
\end{eqnarray}
Given the Neumann condition for $\psi$, the wave vector $\mathbf{q}_0$ is parallel to the boundaries, so that layers become perpendicularly anchored in these regions, allowing for focal conics to be created. 

We introduce $\mathbf{T}$ as the Cauchy stress tensor, and define the material time derivative of a vector $\mathbf{g}$ as $\dot{\mathbf{g}} = \partial_t\mathbf{g}+\mathbf{v}\cdot\nabla\mathbf{g}$. The first law is then stated as
\begin{eqnarray}
    \frac{d}{dt}\int_\Omega\bigg\{\rho \mathfrak{u}\, + \frac{\zeta}{2}\Big[\rho-\rho_0-\kappa A\Big]^2 + \frac{\rho|\mathbf{v}|^2}{2}\bigg\}d\mathbf{x}
    = \int_{\partial\Omega}\bigg[\mathbf{T}\mathbf{n}\cdot\mathbf{v}+(\boldsymbol\xi\cdot\mathbf{n})\dot{\psi}\bigg]dS \;,
\end{eqnarray}
where the surface integral is the rate of work done on the surface of the system, and the volume integral includes both internal energy and kinetic energy.

Given the balances of linear momentum and mass,
\begin{equation}
    \rho\dot{\mathbf{v}}=\nabla\cdot\mathbf{T} \quad,\quad 
    \dot{\rho}+\rho\nabla\cdot\mathbf{v}=0 \;,
\end{equation}
the local form of the balance of internal energy~\cite{gurtin2010mechanics} becomes
\begin{eqnarray}
    \rho\dot{\mathfrak{u}}-\kappa\zeta\Big(\rho-\rho_0-\kappa A\Big)\dot{A} &=& \bigg\{\mathbf{T}+\frac{\zeta}{2}\bigg[\rho^2-(\rho_0+\kappa A)^2\bigg]\bigg\}:\nabla \mathbf{v} + \nabla \cdot (\boldsymbol\xi\,\dot{\psi}) \;.
    \label{eq:ebal-local-w}
\end{eqnarray}
Here $\boldsymbol\xi$ is a microstress (also known as a generalized force), whose origin lies in the theory of microforces, a generalization of order parameter theories such as Ginzburg-Landau and Cahn-Hilliard~\cite{hutter2020coleman,espath2020generalized,vitral2020mesoscale,duda2021coupled}. Similarly to the balance of linear momentum, we can write the balance of microforces in terms of the microforce $\pi$ as
\begin{equation}
    \nabla\cdot\boldsymbol\xi + \pi = 0 \;.
\end{equation}

Since the energy density is of the form $\mathfrak{u} = \mathfrak{u}(\mathfrak{s},\psi,\nabla \psi,\nabla^2\psi)$, from partial differentiation the local balance of entropy can be derived from Eq.~(\ref{eq:ebal-local-w}). Note that $\mathfrak{u}$ is an energy per unit mass, independent of the density $\rho$. Therefore, by the chain rule
\begin{equation}
    \dot{\mathfrak{u}} =
        \frac{\partial \mathfrak{u}}{\partial \mathfrak{s}}\dot{\mathfrak{s}} 
        + \frac{\partial \mathfrak{u}}{\partial \psi}\dot{\psi}
        + \frac{\partial \mathfrak{u}}{\partial \nabla\psi}\cdot\dot{\overline{\nabla\psi}}
        + \frac{\partial \mathfrak{u}}{\partial \nabla^2\psi}\dot{\overline{\nabla^2\psi}} \; .
        \label{eq:chain}
\end{equation}
By accounting for the boundary conditions from Eq.~(\ref{eq:bc-w}), the gradient terms appearing in $\rho\dot{\mathfrak{u}}$ can be rewritten as
\begin{eqnarray}
 \rho \frac{\partial \mathfrak{u}}{\partial \nabla\psi}\cdot\dot{\overline{\nabla\psi}} &=& 
 \rho\frac{\partial \mathfrak{u}}{\partial \nabla\psi}\cdot \nabla\dot{\psi} 
 - \rho \nabla\psi \otimes \frac{\partial \mathfrak{u}}{\partial\nabla\psi} : \nabla\mathbf{v} \; ,
\end{eqnarray}
and also
\begin{eqnarray}
    \nonumber
    \rho \frac{\partial \mathfrak{u}}{\partial \nabla^2\psi}\dot{\overline{\nabla^2\psi}} &=&
    \rho \frac{\partial \mathfrak{u}}{\partial \nabla^2\psi}\nabla^2\dot{\psi} 
    - \rho \frac{\partial \mathfrak{u}}{\partial \nabla^2\psi}\nabla^2\mathbf{v}\cdot\nabla\psi
    -2\rho \frac{\partial \mathfrak{u}}{\partial \nabla^2\psi}\nabla\mathbf{v} : \mathbf{D}\psi 
   \\[2mm] &=&
    -\nabla\bigg(\rho\frac{\partial \mathfrak{u}}{\partial\nabla^2\psi}\bigg)\cdot\nabla\dot{\psi}
    + \bigg[ \nabla\psi\otimes\nabla\bigg(\rho \frac{\partial \mathfrak{u}}{\partial\nabla^2\psi}\bigg)
    - \rho \frac{\partial \mathfrak{u}}{\partial\nabla^2\psi} \mathbf{D}\psi \bigg] : \nabla\mathbf{v} \; ,
\end{eqnarray}
where $\mathbf{D}$ stands for $\nabla\nabla$ (i.e. $``\partial_i\partial_j "$), so that $\mathbf{D}\psi$ is a second order tensor. 

Given that the temperature $\theta = \partial \mathfrak{u}/ \partial \mathfrak{s}$ and that the real amplitude of the order parameter has a simple dependency $A = A(\psi)$, by substituting Eq.~(\ref{eq:chain}) into Eq.~(\ref{eq:ebal-local-w}) we obtain the following local balance of entropy:
\begin{eqnarray}
    \nonumber
    \rho \theta \dot{\mathfrak{s}} &=&
        \bigg\{ \mathbf{T} + \frac{\zeta}{2}\Big[\rho^2-(\rho_0+\kappa A)^2 \Big]\mathbf{I}
        + \rho \nabla\psi \otimes \frac{\partial \mathfrak{u}}{\partial\nabla\psi}
        -\nabla\psi\otimes\nabla
        \bigg(\rho\frac{\partial \mathfrak{u}}{\partial \nabla^2\psi}\bigg)
        +\rho\frac{\partial \mathfrak{u}}{\partial \nabla^2\psi}\mathbf{D}\psi
    \bigg\} : \nabla \mathbf{v} \\[2mm]
    && +\bigg[ \boldsymbol\xi -\rho \frac{\partial \mathfrak{u}}{\partial\nabla\psi} + \nabla\left(\rho\frac{\partial \mathfrak{u}}{\partial \nabla^2 \psi}\right
        )\bigg] \cdot \nabla \dot{\psi} + \bigg[\kappa\zeta\Big(\rho-\rho_0-\kappa A\Big)\frac{\partial A}{\partial\psi}-\rho\frac{\partial \mathfrak{u}}{\partial \psi}+\nabla\cdot\boldsymbol\xi\bigg]\dot{\psi} \; .
\label{eq:sbal-w}
\end{eqnarray}

\vspace{5mm}

In equilibrium, we observe from Eq.~(\ref{eq:sbal-w}) that the balance of microforces is satisfied for
\begin{eqnarray}
  \boldsymbol\xi &=& \rho \frac{\partial \mathfrak{u}}{\partial\nabla\psi} - \nabla\left(\rho\frac{\partial \mathfrak{u}}{\partial \nabla^2 \psi}\right) \;,
  \\[2mm]
  \pi&=& \kappa\zeta\Big(\rho-\rho_0-\kappa A\Big)\frac{\partial A}{\partial\psi}-\rho\frac{\partial \mathfrak{u}}{\partial \psi} \;.
\end{eqnarray}
The terms in square brackets proportional to $\dot{\psi}$ and $\nabla\dot{\psi}$ in Eq.~(\ref{eq:sbal-w}) are both related to variations of $\psi$ and can be grouped together. Accounting for the boundary conditions, we obtain the final form of the local entropy balance
\begin{eqnarray}
    \nonumber
    \rho \theta \dot{\mathfrak{s}} &=& 
     \bigg\{ \mathbf{T} + \frac{\zeta}{2}\Big[\rho^2-(\rho_0+\kappa A)^2 \Big]\mathbf{I}
        +\rho \nabla\psi \otimes \frac{\partial \mathfrak{u}}{\partial\nabla\psi}
        -\nabla\psi\otimes\nabla
        \bigg(\rho\frac{\partial \mathfrak{u}}{\partial \nabla^2\psi}\bigg)
        +\rho\frac{\partial \mathfrak{u}}{\partial \nabla^2\psi}\mathbf{D}\psi
    \bigg\} : \nabla \mathbf{v} \\[2mm]
    && +\bigg[\kappa\zeta\Big(\rho-\rho_0-\kappa A\Big)\frac{\partial A}{\partial\psi}-\rho\frac{\partial \mathfrak{u}}{\partial \psi}
    +\nabla\cdot\bigg(\rho \frac{\partial \mathfrak{u}}{\partial\nabla\psi}\bigg)
    -\nabla^2\bigg(\rho\frac{\partial \mathfrak{u}}{\partial \nabla^2 \psi}\bigg) \bigg] \dot{\psi}
    \; .
\label{eq:sbal2-w}
\end{eqnarray}
Note that the microstress is not explicit in this form of the entropy balance.

\subsection{Governing equations}

Constitutive relations and governing equations are now derived by imposing strict requirements on the entropy production $\dot{\mathfrak{s}}$, as in the Coleman-Noll procedure~\cite{coleman1963thermodynamics,hutter2020coleman}. The Clausius-Duhem inequality states that every admissible thermomechanical process must satisfy $\dot{\mathfrak{s}} \geq 0$, which has direct implications for Eq.~(\ref{eq:sbal2-w}). We start by splitting the stress into a sum of reversible and dissipative parts, $\mathbf{T} = \mathbf{T}^R + \mathbf{T}^D$, so that the reversible part $\mathbf{T}^R$ can be derived from Eq.~(\ref{eq:sbal2-w}) in the limit of zero entropy production, while dissipative parts are obtained by enforcing positive entropy production.

The reversible stress $\mathbf{T}^R$, is obtained from the expression inside the brackets associated with the rate $\nabla\mathbf{v}$ in the limit $\dot{\mathfrak{s}} = 0$. We find
\begin{eqnarray}
    \mathbf{T}^R &=& -\frac{\zeta}{2}\Big[\rho^2-(\rho_0+\kappa A)^2 \Big]\mathbf{I}
        - \rho \nabla\psi \otimes \frac{\partial \mathfrak{u}}{\partial\nabla\psi}
             +\nabla\psi\otimes\nabla
             \bigg(\rho\frac{\partial \mathfrak{u}}{\partial \nabla^2\psi}\bigg)
             -\rho\frac{\partial \mathfrak{u}}{\partial \nabla^2\psi}\mathbf{D}\psi \; .
        \label{eq:rev-stress-w}
\end{eqnarray}
Note that non-classical stress terms appear in Eq.~(\ref{eq:rev-stress-w}), not only from the dependence of $\mathfrak{u}$ on $\nabla^2\psi$, but also from the density coupling through $\zeta$. Generally, we will expect the latter to be small for the bulk smectic and isotropic phases, since $\rho$ tends to approach $\rho_0 + \kappa A$ in the bulk due to energy minimization. However, this term can potentially become large for compressible flows near the smectic-isotropic interface.

The generalized chemical potential $\mu$, which is the the thermodynamic conjugate to $\psi$, $\mu = \delta \mathfrak{U} / \delta \psi$, appears in Eq.~(\ref{eq:rev-stress-w}) inside the brackets multiplying $\dot{\psi}$. That is,
\begin{eqnarray}
  \mu &=&
          - \kappa\zeta\Big(\rho-\rho_0-\kappa A\Big)\frac{\partial A}{\partial\psi}
          + \rho\frac{\partial \mathfrak{u}}{\partial \psi}
    -\nabla\cdot\bigg(\rho \frac{\partial \mathfrak{u}}{\partial\nabla\psi}\bigg)
    +\nabla^2\bigg(\rho\frac{\partial \mathfrak{u}}{\partial \nabla^2 \psi}\bigg) \; .
    \label{eq:chemicalp-w}
\end{eqnarray}
For a slowly relaxing variable such as $\psi$, which is not a conserved quantity, the associated dynamic equation~\cite{pleiner1996hydrodynamics} is given by a quasi-current $Z$, and takes the form 
\begin{equation}
    \partial_t\psi + \mathbf{v}\cdot\nabla\psi + Z = 0\;.
    \label{eq:psydyn}
\end{equation}
While reversible motion requires $\dot{\mathfrak{s}} = 0$ in Eq.~(\ref{eq:sbal2-w}), the generalized chemical potential $\mu$ from Eq.~(\ref{eq:chemicalp-w}) is arbitrary, and reversible motion has $\dot{\psi} = 0$. The implication is that $Z$ has no reversible part, and is purely dissipative.

Irreversible currents are obtained by requiring a positive entropy production $\dot{\mathfrak{s}} \geq 0$. From Eq.~(\ref{eq:sbal2-w}), $\dot{\psi}$ must be proportional to the chemical potential $\mu$ times a mobility constant $\Gamma$, so that
\begin{equation}
    Z = \Gamma\mu \;.
\end{equation}
The irreversible stress $\mathbf{T}^D$ when contracted with $\nabla\mathbf{v}$ in Eq.~\ref{eq:sbal2-w} should result in a positive quantity to satisfy the Clausius-Duhem inequality. This implies that $\mathbf{T}^D = \boldsymbol\eta:\nabla\mathbf{v}$, where $\boldsymbol\eta$ is a fourth order viscosity tensor. For simplicity, we will restrict our analysis for a Newtonian fluid, with a dissipative stress of the form
\begin{equation}
    \mathbf{T}^D = \eta(\nabla\mathbf{v}+\nabla\mathbf{v}^\top)+\lambda(\nabla\cdot\mathbf{v})\mathbf{I}\;,
    \label{eq:dstress}
\end{equation}
where $\eta$ and $\lambda$ are the first and second coefficient of viscosity, respectively. The coefficient $\lambda$ is important for compressibility effects, as it controls the magnitude of the longitudinal part of the flow. We note that the dissipative stress for a general uniaxial phase can be written in terms of five independent viscosity coefficients~\cite{degennes1995physics}, as we argued in~\cite{vitral2020model}. While an extension to such a dissipative stress is straightforward in the model, these coefficients are poorly characterized experimentally, so that we focus exclusively on the simpler form of Eq.~(\ref{eq:dstress}). Further, we consider $\eta$ and $\lambda$ to be homogeneous throughout the smectic-isotropic system, so that another possible extension of the model would consider viscosity contrast between the phases.

From the constitutive relations for the stress $\mathbf{T}$ and the quasi-current $Z$, we find that the balance of mass, the balance of linear momentum, and the dynamic equation for the order parameter, which govern the weakly compressible smectic-isotropic system, have the following form
\begin{eqnarray}
    \dot{\rho} &=& 
    -\rho\nabla\cdot\mathbf{v} \; ,
    \label{eq:wcom-bm}
    \\[2mm]
    \rho\dot{\mathbf{v}} &=&
    \nabla\cdot\Big(\mathbf{T}^R + \mathbf{T}^D\Big) \; ,
    \label{eq:wcom-blm}
    \\[2mm]
    \dot{\psi} &=& -\Gamma \mu \;.
    \label{eq:wcom-psi}
\end{eqnarray}
Boundary conditions are specified by Eqs.~(\ref{eq:bc-w}), the reversible stress $\mathbf{T}^{R}$ is defined in Eq.~(\ref{eq:rev-stress-w}), the dissipative stress $\mathbf{T}^{D}$ in Eq.~(\ref{eq:dstress}) and the generalized chemical potential $\mu$ in Eq.~(\ref{eq:chemicalp-w}).

\subsection{Energy density of the smectic phase}

The specific energy $\mathfrak{u}$ we use allows for coexistence of isotropic and smectic phases, and remains rotationally invariant. It is given by
\begin{eqnarray}
    \mathfrak{u}(\psi,\nabla^2\psi) &=& \frac{1}{2}\bigg\{ \epsilon \psi^2 
    + \alpha\left[ \left(\nabla^2+q_0^2 \right)\psi \right]^2            
    - \frac{\beta}{2} \psi^{4} + \frac{\gamma}{3} \psi^{6}\;\bigg\},
    \label{eq:energy-density}
\end{eqnarray}
where $\psi = 0$ represents the isotropic phase, and a periodic $\psi$ the smectic phase. The coefficients $\alpha$, $\beta$ and $\gamma$ are three constant, positive parameters, $\epsilon$ is a bifurcation parameter that describes the distance away from the smectic-isotropic transition, and $q_{0}$ is approximately the smectic layering wave number. Hence, this energy penalizes deviations from the preferred equidistant layering~\cite{napoli1999smectic}. The associated quintic Swift-Hohenberg equation has been used in the past to describe modulated phases~\cite{sakaguchi1996stable,sakaguchi1998localized,burke2007snakes} in coexistence with an isotropic ($\psi=0$) phase, and describes the diffusive relaxation of the order parameter $\psi$ driven by energy minimization. The sixth order polynomial in $\psi$ leads to a triple well energy potential function, whose minima represent the smectic and isotropic phases. By fixing $\beta$ and $\gamma$, the relative height between these wells can be controlled through $\epsilon$, such that smectric-isotropic coexistence occurs at $\epsilon_{c} = 27 \beta^{2}/160 \gamma$, when both phases present the same energy density. For $\epsilon > \epsilon_{c}$, $\psi = 0$ (isotropic) becomes the equilibrium phase, whereas for $\epsilon < \epsilon_{c}$, a modulated phase (smectic) $\psi \approx \frac{1}{2} (A\,e^{i{\bf q}\cdot{\bf x}} + c.c.)$ is the equilibrium one. The amplitude $A_s$ of the bulk smectic is given by~\cite{sakaguchi1996stable}
\begin{equation}
    A_s^2 = \frac{3\beta+\sqrt{9\beta^2-40\epsilon\gamma}}{5\gamma} \;,
\end{equation} 
which is relevant when choosing the constants $\kappa$ and $\rho_0$ for a specific density ratio. For small layer perturbations away from planarity, the energy reduces to the classical Oseen-Frank form of the smectic energy, where the elastic moduli for compression of layers and splay of molecules can be written as a function of $\alpha$~\cite{vitral2019role}.

By substituting this definition of specific energy into the chemical potential from Eq.~(\ref{eq:chemicalp-w}), we obtain
\begin{eqnarray}
    \mu &=&  - \kappa\zeta\Big(\rho-\rho_0-\kappa A\Big)\frac{\partial A}{\partial\psi}+\rho\Big[\epsilon\psi +\alpha q_0^2(\nabla^2+q_0^2)\psi-\beta\psi^3+\gamma\psi^5\Big] 
    + \alpha\nabla^2\Big[\rho(\nabla^2+q_0^2)\psi\Big]\;.
    \label{eq:chemicalp2-w}
\end{eqnarray}

An issue concerning the actual computation of $\mu$ from Eq.~(\ref{eq:chemicalp2-w}) is that while we are able to numerically extract the amplitude from $\psi$, the dependence of $A$ as a function of $\psi$ is unknown, and so is its derivative with respect to $\psi$. In this work, since the order parameter $\psi$ has an asymptotic sinusoidal form, we compute the amplitude $A$ through $A = (\psi^2+q_0^{-2}|\nabla\psi|^2)^{1/2}$. By accounting for this dependency on $\nabla\psi$, we then write the chemical potential $\mu = \delta \mathfrak{U} / \delta \psi$ as,
\begin{eqnarray}
  \nonumber
  \mu &=&  - \kappa\zeta\Big(\rho-\rho_0-\kappa A\Big)\frac{\psi}{A}+\kappa\zeta q_0^{-2}\nabla\cdot\bigg[(\rho-\rho_0-\kappa A)\frac{\nabla\psi}{A} \bigg]
  \\[2mm]&&
            +\rho\Big[\epsilon\psi +\alpha q_0^2(\nabla^2+q_0^2)\psi-\beta\psi^3+\gamma\psi^5\Big] 
    + \alpha\nabla^2\Big[\rho(\nabla^2+q_0^2)\psi\Big]\;.
    \label{eq:chemicalp3-w}
\end{eqnarray}

We also neglect inertia compared to viscous effects, so that by using the definition of the energy density in Eq.~(\ref{eq:energy-density}), the balance of linear momentum, Eq.~(\ref{eq:wcom-blm}), becomes
\begin{eqnarray}
    \nonumber
  \mathbf{0} &=&  -\frac{\zeta}{2}\nabla\Big[\rho^2-(\rho_0+\kappa A)^2 \Big]
        + \alpha\nabla^2\Big[\rho(\nabla^2+q_0^2)\psi\Big]\nabla\psi        
  \\[2mm]
    &&
       -\alpha\rho(\nabla^2+q_0^2)\psi\nabla^2\nabla\psi
       + \eta \nabla^2\mathbf{v} 
       + (\eta + \lambda)\nabla(\nabla\cdot\mathbf{v}) \; .
    \label{eq:wcom-stokes}
\end{eqnarray}
By defining the modified chemical potential $\bar{\mu}$ as
\begin{eqnarray}
    \nonumber
  \bar{\mu} &=&  
    \rho\Big[\epsilon\psi +\alpha q_0^2(\nabla^2+q_0^2)\psi-\beta\psi^3+\gamma\psi^5\Big] 
    + \alpha\nabla^2\Big[\rho(\nabla^2+q_0^2)\psi\Big]\;,
    \label{eq:chemicalp4-w}
\end{eqnarray}
we can rewrite Eq.~(\ref{eq:wcom-stokes}) as,
\begin{eqnarray}
  \mathbf{0} &=&  -\frac{\zeta}{2}\nabla\Big[\rho^2-(\rho_0+\kappa A)^2 \Big]
        + \bar{\mu}\nabla\psi - \rho\nabla\mathfrak{u}
       + \eta \nabla^2\mathbf{v} 
       + (\eta + \lambda)\nabla(\nabla\cdot\mathbf{v}) \; .
    \label{eq:wcom-stokes2}
\end{eqnarray}
The quantity $\bar{\mu}\nabla\psi$ is known as the osmotic force \cite{fielding2003flow}, and is exactly the forcing term that appears for an incompressible smectic-isotropic system, where both phases have the same density~\cite{vitral2019role}. Therefore, $-\rho\nabla \mathfrak{u}$ is a force that originates from compressibility effects. 

Dimensionless variables are introduced similarly to~\cite{vitral2020model}. Let $U$ and $L$ represent characteristic scales for the velocity and length, and $\tilde{\rho}$ and $\tilde{\mu}$ represent typical values for $\rho$ and $\mu$ in the modulated phase. We then perform the non-dimensionalization by defining $\mathbf{v}^* = \mathbf{v}/U$, $\mathbf{x}^* = \mathbf{x}/L$, $t^* = U t/L$,  $\rho^* = \rho/\tilde{\rho}$, $\mu^* = \mu/\tilde{\mu}$ and $\psi^* = \psi\tilde{\mu}/\tilde{\rho}U^2$. The dimensionless constants one finds are $\kappa^* = \kappa / \tilde{\rho}$, $\Gamma^* = \Gamma \tilde{\mu}^2 L / \tilde{\rho} U^3$, $\zeta^* = \zeta \tilde{\rho}/U^2$, $\eta^* = \eta / \tilde{\rho}U L$ and $\lambda^* = \lambda / \tilde{\rho}U L$. Note that $\eta^*$ is the inverse Reynolds number, and $\zeta^*$ is a dimensionless number that controls the ratio of $\psi$ currents arising from density gradients compared to $\psi$ owing to local curvatures. 
This can be seen by taking the ratio of terms in Eq.~(\ref{eq:chemicalp2-w}) in dimensionless form:
\begin{equation}
    \frac{\kappa\zeta\delta\rho}{\bar{\mu}}\,\frac{\partial A}{\partial \psi}
    = \zeta^* \bigg(\frac{\kappa^*\delta\rho^*}{\bar{\mu}^*}\,\frac{\partial A}{\partial \psi^*}\bigg) \;,
\end{equation}
where $\delta\rho = \rho-\rho_{eq}$ and the chemical potential $\bar{\mu}$ is proportional to local curvatures through the Gibbs-Thomson relation, see Eqs.~(\ref{eq:gibbs-thomson}) and (\ref{eq:gt-perp}). Hence, when $\zeta$ is small, the evolution of $\psi$ is driven primarily by the chemical potential $\bar{\mu}$.

By dropping the star notation from variables and constants, we here summarize the complete set of dimensionless governing equations for the weakly compressible smectic-isotropic fluid system, including balance of mass, balance of linear momentum, and order parameter equation,
\begin{eqnarray}
  \label{eq:bm-ge-w}
  \dot{\rho}
  &=& -\rho\nabla\cdot\mathbf{v}\; ,
  \\[4mm]
   \label{eq:blm-ge-w}
  \mathbf{0}
  &=& -\frac{\zeta}{2}\nabla\Big[\rho^2-(\rho_0+\kappa A)^2 \Big]
      +\bar{\mu}\nabla\psi -\rho\nabla \mathfrak{u} 
      + \eta \nabla^2\mathbf{v} 
      + (\eta + \lambda)\nabla(\nabla\cdot\mathbf{v}) \; ,
  \\[4mm]
  \label{eq:psi-ge-w}
  \dot{\psi} &=& -\Gamma \mu \; ,
\end{eqnarray}
with the generalized chemical potential $\mu$ given by  Eq.~(\ref{eq:chemicalp3-w}), modified chemical potential $\bar{\mu}$ given by Eq.~(\ref{eq:chemicalp4-w}), and energy density $\mathfrak{u}$ given by Eq.~(\ref{eq:energy-density}). Note that $\zeta$ in Eq.~(\ref{eq:blm-ge-w}) plays a role in setting diffuseness of the density across the interface.  For small $\zeta$ the interface is more diffuse and the penalty for deviations of $\rho$ from equilibrium is small, while for larger $\zeta$ the variation in $\rho$ across the interface becomes sharper and the system approaches quasi-incompressibility.

\section{Numerical algorithm}
\label{sec:ns}

The governing equations~(\ref{eq:bm-ge-w})-(\ref{eq:psi-ge-w}), with boundary conditions specified in Eq.~(\ref{eq:bc-w}), are integrated with a pseudo-spectral method in three spatial dimensions, in which linear and gradient terms are computed in Fourier space and nonlinear terms in real space. A regular cubic grid is used with linear spacing $\Delta x = 2\pi/(n_w q_0)$, where $n_w$ is the number of grid points per base wavelength. We have developed a custom C++ code (\textit{smaiso-wcomp}) based on the parallel FFTW library and the standard MPI passing interface for parallelization, which is publicly avilable~\cite{smaiso-wcomp}. In order to accommodate the boundary conditions, we use both the Discrete Cosine Transform of ($\psi$, $\rho$) and the Discrete Sine Transform of ($\nabla\psi$, $\mathbf{v}$).

We compute the order parameter amplitude $A$ by $A = (\psi^2+q_0^{-2} |\nabla\psi|^2)^{1/2}$.  While this approximation gives us an adequate value of $A$ in regions where the smectic layers are well formed and only weakly distorted, it becomes noisier on the interface, and also in regions where layers are highly distorted or break up. Therefore in our numerical calculations we smooth the computed amplitude with a Gaussian filter in Fourier space, given by the operator $F_{\omega} = \textrm{exp}(-\omega^2 q^2/2)$, where $q$ is the local wavenumber and $\omega$ the filtering radius, chosen as $1/q_0$.

\subsection{Order parameter equation}

The numerical scheme for integrating Eq.~(\ref{eq:psi-ge-w}) has been detailed in~\cite{vitral2020model}, and here we summarize it. Due to the variable density multiplying the RHS of Eq.~(\ref{eq:psi-ge-w}), it cannot be dealt with in the same form as the uniform density case~\cite{vitral2019role}, for which the split in linear and nonlinear parts was immediate. Instead, we follow a scheme previously employed for phase-field models with variable mobility~\cite{zhu1999coarsening,badalassi2003computation}. First, we split Eq.~(\ref{eq:psi-ge-w}) as $\partial_t\psi = \Gamma(\rho L \psi + N)$ with
\begin{eqnarray}
    L &=& -\Big[\epsilon + (\nabla^2+q_0^2)^2\Big] \;,
    \label{eq:linear-w}
    \\[2mm]
    \nonumber
    N &=&  \kappa\zeta\Big(\rho-\rho_0-\kappa A\Big)\frac{\psi}{A}-\kappa\zeta q_0^{-2}\nabla\cdot\bigg[(\rho-\rho_0-\kappa A)\frac{\nabla\psi}{A}\bigg]
    \\[2mm] &&
    -2\alpha\nabla\rho\cdot(\nabla^2+q_0^2)\nabla\psi
    -\alpha\nabla^2\rho\,(\nabla^2+q_0^2)\psi+\beta\psi^3-\gamma\psi^5
    -\Gamma^{-1}\mathbf{v}\cdot\nabla\psi  \;,
    \label{eq:nlinear-w}
\end{eqnarray}
where $L$ is a linear operator, and $N$ is a collection of nonlinear terms. We split the density as $\rho \rightarrow \rho_m +(\rho-\rho_m)$, where $\rho_m = \frac{1}{2}(\rho_s + \rho_0)$. Here, $\rho_s$ is the density of the smectic bulk, which can be obtained from the system parameters by $\rho_s = \kappa A_s + \rho_0$, where $A_s$ is the amplitude solution of the sinusoidal phase. The term associated with $\rho_m$ can be treated implicitly, and $(\rho-\rho_m)$ is treated explicitly, with a choice of $\rho_m$ that satisfies $|\rho -\rho_m| \leq \rho_m$. We treat the linear term $L$ implicitly, while the nonlinear term $N$ is treated with a second order Adams-Bashforth scheme. In Fourier space, the order parameter $\psi_q$ for the new time step $n+1$ is computed from
\begin{eqnarray}
    (3/2-\Delta t\, \Gamma\rho_m L )\psi_q^{n+1}
    \,=\,  (2-\Delta t\, \Gamma\rho_m L)\psi_q^{n}-\frac{1}{2}\psi_q^{n-1}
    +\frac{\Delta t\, \Gamma}{2}(3N_q^n-N_q^{n-1}) \; .
    \label{eq:method2-w}
\end{eqnarray}
Numerical verification and stability of this scheme have been studied in~\cite{vitral2020model}, where we found that for $n_w = 8$ a time step of $\Delta t = 1\cdot 10^{-3}$ or less was necessary to guarantee stability.

\subsection{Velocity decomposition}

The weakly compressible model requires additional considerations for computing the velocity and density as compared to~\cite{vitral2020model}. A Helmholtz decomposition the velocity field $\mathbf{v}$ is introduced,
\begin{eqnarray}
  \mathbf{v} &=& \nabla\Phi + \nabla\times\mathbf{P} \;,
  \label{eq:helmholtz}
\end{eqnarray}
where $\Phi$ is a scalar potential and $\mathbf{P}$ is a vector potential. By substituting this decomposition of the velocity into the balance of linear momentum from Eq.~(\ref{eq:blm-ge-w}), we obtain
\begin{eqnarray}
  \mathbf{0} &=&  -\frac{\zeta}{2}\nabla\Big[\rho^2-(\rho_0+\kappa A)^2 \Big]
       + \mathbf{f}   
       + \eta \nabla^2\nabla\times\mathbf{P} 
       + (2\eta + \lambda)\nabla\nabla^2\Phi \; ,
       \label{eq:blm-helm}
\end{eqnarray}
where $\mathbf{f} = \bar{\mu}\nabla\psi-\rho\nabla \mathfrak{u}$. Since the gradient of the density is computed numerically at every time step, one can avoid computing the gradient of the energy density $\mathfrak{u}$ by adding $\rho \mathfrak{u}$ inside the gradient from the first term in Eq.~(\ref{eq:blm-helm}) and rewriting $\mathbf{f}$ as $\bar{\mathbf{f}} = \bar{\mu}\nabla\psi + \mathfrak{u}\nabla \rho$.

The solenoidal field can be obtained from the transverse part of Eq.~(\ref{eq:blm-helm}). By eliminating irrotational terms through an orthogonal projection (and, consequently, modulations due to the layering), we compute $\nabla\times\mathbf{P}$ in Fourier space from
\begin{eqnarray}
     (\nabla\times\mathbf{P})_q
     &=& \frac{1}{\eta\, q^2}\bigg(\mathbf{I} - 
     \frac{\mathbf{q}\otimes\mathbf{q}}{q^2} \bigg)\mathbf{f}_q \;.
     \label{eq:transv}
\end{eqnarray}
Similarly, the longitudinal component of Eq.~(\ref{eq:blm-helm}) eliminates the solenoidal terms, and allows us to compute $\nabla\Phi$. By substituting Eq.~(\ref{eq:transv}) into Eq.~(\ref{eq:blm-helm}), we obtain 
\begin{eqnarray}
     (\nabla \Phi)_q
     &=& \frac{1}{(2\eta +\lambda)\, q^2}\bigg\{\frac{\zeta\,\mathbf{q}}{2}
     \Big[\rho^2-(\rho_0+\kappa A)^2 \Big]_q + 
     \frac{\mathbf{q}\otimes\mathbf{q}}{q^2}\,\mathbf{f}_q \bigg\} \;.
     \label{eq:irrot}
\end{eqnarray}
Note that the density does not change in the scale of the smectic layering modulations, and also that the force $\mathbf{f}$ contains both resonant terms on the same scale as the amplitude, and modes $\pm 2 i q_0$ or higher. In order to avoid spurious oscillations in the irrotational flow along smectic layers, we dampen contributions from the higher order frequencies by applying the filter $F_{\omega} = \textrm{exp}(-\omega^2 q^2/2)$ to the $\mathbf{f}_q$ term in Eq.~(\ref{eq:irrot}), with a filtering radius $\omega = 1/q_0$.

\subsection{Balance of mass}

The balance of mass from Eq.~(\ref{eq:bm-ge-w}) can be easily integrated by a multistep method such as Adam-Bashforth. Here, we compute the density at the new time step $n+1$ through
\begin{eqnarray}
    \rho^{n+1} &=& \rho^n -
    \Delta t \bigg[
    \frac{3}{2}\big(\mathbf{v}^n\cdot\nabla\rho^n
    +\rho^n \nabla\cdot\mathbf{v}^n\big)
             -\frac{1}{2}\big(\mathbf{v}^{n-1}\cdot\nabla\rho^{n-1}
             +\rho^{n-1} \nabla\cdot\mathbf{v}^{n-1}\big) \bigg] \;,
\end{eqnarray}
where the divergence of the velocity is computed from $\nabla\cdot\mathbf{v} = \nabla^2\Phi$.

\section{Compressibility effects on order parameter diffusion}
\label{sec:def}

One of the main goals of the weakly compressible model is to explore how the energy penalty for  deviations in density from equilibrium in Eq.~(\ref{eq:energy}) affects the evolution of the non-conserved order parameter $\psi$.  We have shown that the dimensionless $\zeta$ controls the ratio between order parameter currents arising from density gradients and curvatures. Now we investigate its role on the motion of $\psi$ given by Eq.~(\ref{eq:psi-ge-w}).

Assume first we have a region of constant density and amplitude $A$. By setting $\Gamma = 1$, as in our numerical computations, the order parameter equation~(\ref{eq:psi-ge-w}) can be written as
\begin{eqnarray}
  \dot{\psi} &=&  \frac{\kappa\zeta}{A}\Big(\rho-\rho_0-\kappa A\Big)(1-q_0^{-2}\nabla^2)\psi
            -\rho\Big[\epsilon\psi +\alpha(\nabla^2+q_0^2)^2\psi-\beta\psi^3+\gamma\psi^5\Big] 
   \;.
    \label{eq:psi-const}
\end{eqnarray}
The first term on the RHS is linear under this scenario. Since the order parameter asymptotic solution is \hbox{$\psi \approx A \,\textrm{cos}(\mathbf{q}_0\cdot\mathbf{x})$}, this implies that $(1-q_0^{-2}\nabla^2)\psi \approx 2\psi$. By using this approximation  we can regroup the terms in Eq.~(\ref{eq:psi-const}) as
\begin{eqnarray}
\nonumber
  \dot{\psi} &\approx&  -[\rho\epsilon-2\frac{\kappa\zeta}{A}(\rho - \rho_0 -\kappa A)]\psi
            -\rho\Big[\alpha(\nabla^2+q_0^2)^2\psi-\beta\psi^3+\gamma\psi^5\Big]
   \\[2mm]
   &=& -\rho\epsilon_{e}\psi
   -\rho\Big[\alpha(\nabla^2+q_0^2)^2\psi-\beta\psi^3+\gamma\psi^5\Big] \;.
    \label{eq:psi-const2}
\end{eqnarray}
Therefore, the term proportional to $\zeta$ can be viewed as a correction to the bifurcation parameter $\epsilon$, resulting in an effective bifurcation parameter $\epsilon_e$.

We first investigate the stability of our algorithm by computing a stationary configuration at coexistence with planar smectic-isotropic interfaces, as well as the evolution near coexistence to evaluate numerically the effect of $\epsilon_e$.  The governing equations are integrated with a time step $\Delta t = 0.001$ and grid spacing $\Delta x = \pi/4$ (8 points per wavelength). These equilibrium configurations have been obtained by setting  $q_0 = 1$, $\beta = 2$, $\gamma = 1$, $\nu = 1$, $\lambda = 1$, $\zeta = 1$, an isotropic equilibrium density $\rho_0 = 0.5$ and $\kappa = 0.3727$, so that the equilibrium density of the smectic is $\rho_s = 1$ (density ratio $\rho_s:\rho_0 = 2:1$). 

\begin{figure}[ht]
	\centering
    \begin{subfigure}[b]{0.4\textwidth}
    \includegraphics[width=\textwidth]{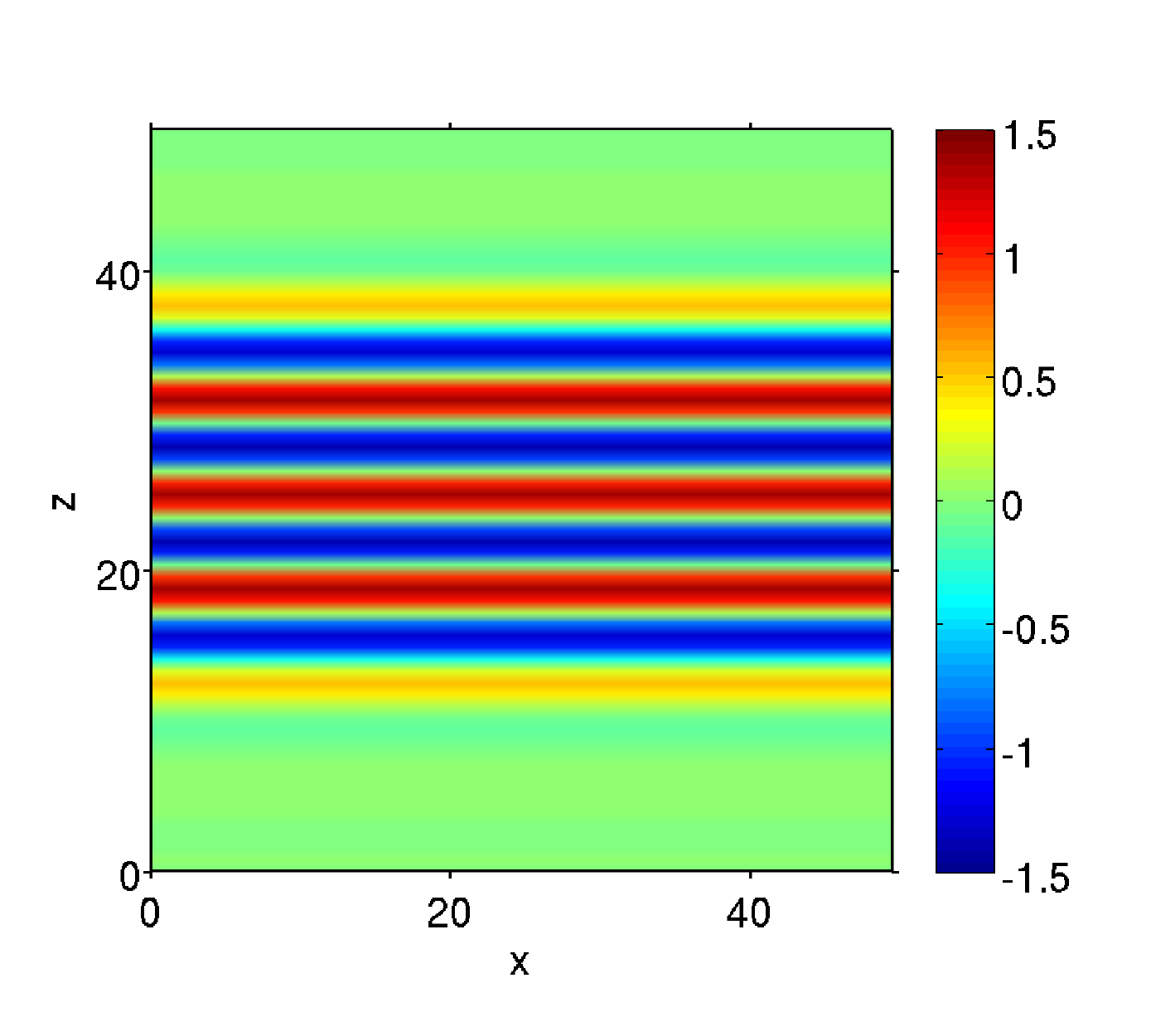}
    \caption{$\psi$}
    \end{subfigure}
    \begin{subfigure}[b]{0.4\textwidth}
    \includegraphics[width=\textwidth]{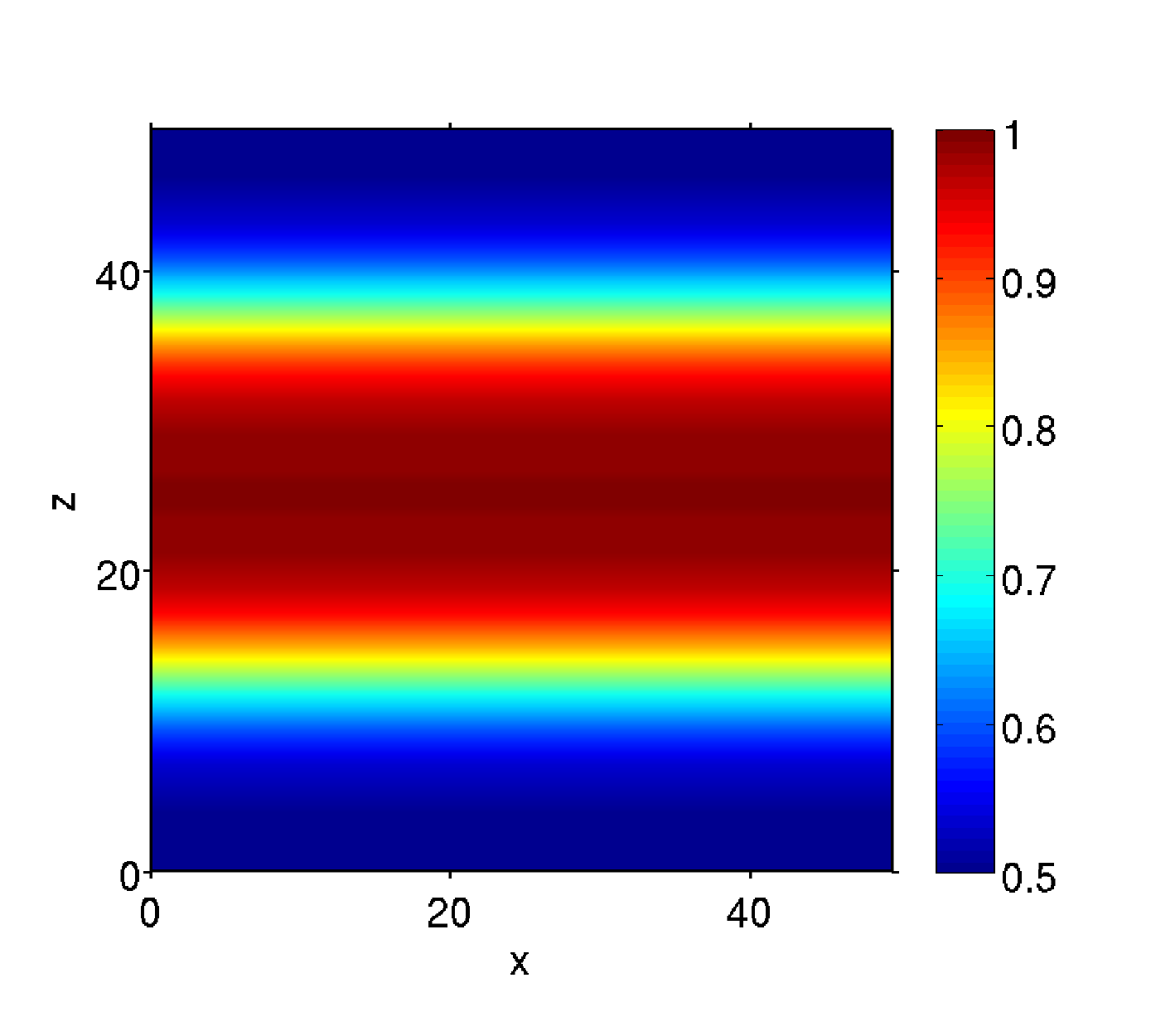}
    \caption{$\rho$}
    \end{subfigure}
    \caption{Middle cross section $xy$ of a stack of flat smectic layers with normal in the $z$ direction, showing the order parameter field $\psi$ and density field $\rho$ used as initial condition. Simulations employing this initial conditions use parameters $\beta = 2$, $\gamma = 1$, $\nu = 1$, $\lambda = 1$, $\rho_0 = 0.5$ and $\kappa = 0.3727$. For these values, the coexistence parameter is $\epsilon_c = 0.675$.}
	\label{fig:flat-ic}
\end{figure}

Consider a stack of planar smectic layers in contact with an isotropic phase, where the layer normal is in the $\hat{z}$ direction. When both phases are at coexistence ($\epsilon = \epsilon_c$), the density profile is $\rho = \rho_0 + \kappa A$, as illustrated in Fig.~\ref{fig:flat-ic}. This figure shows the $xz$ cross-section of a cubic computational domain with $N = 64^3$ nodes at coexistence with $\epsilon_c = 0.675$.  The velocity is $\mathbf{v} = \mathbf{0}$, and the interface is stationary with $\partial_t\psi = 0$. If $\epsilon \neq \epsilon_{c}$ there is a direct coupling between the local density difference relative to the stationary case and the order parameter. This coupling term generally opposes interfacial motion.

\begin{figure}[ht!]
	\centering
    \begin{subfigure}[b]{0.24\textwidth}
        \includegraphics[width=\textwidth]{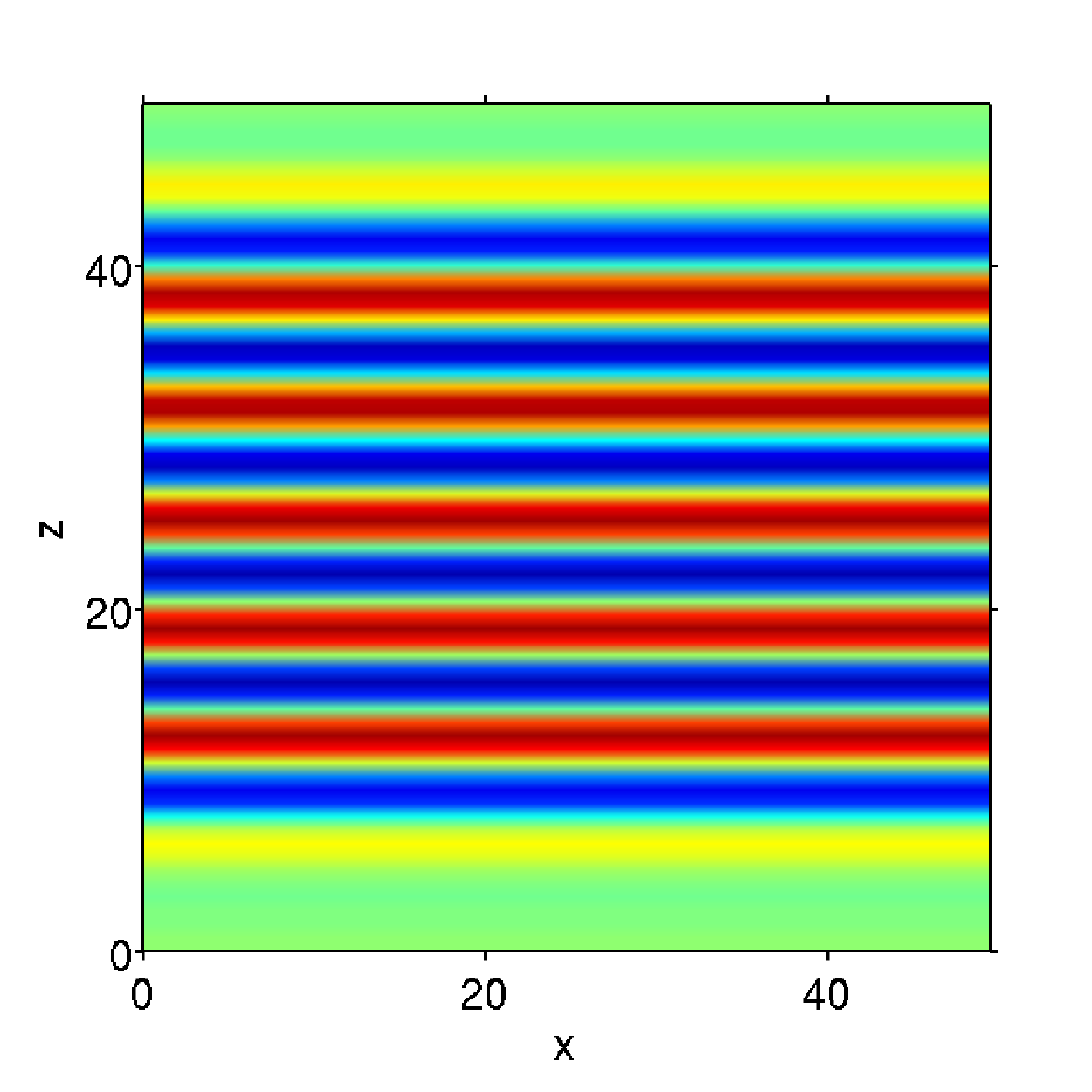}
        \includegraphics[width=\textwidth]{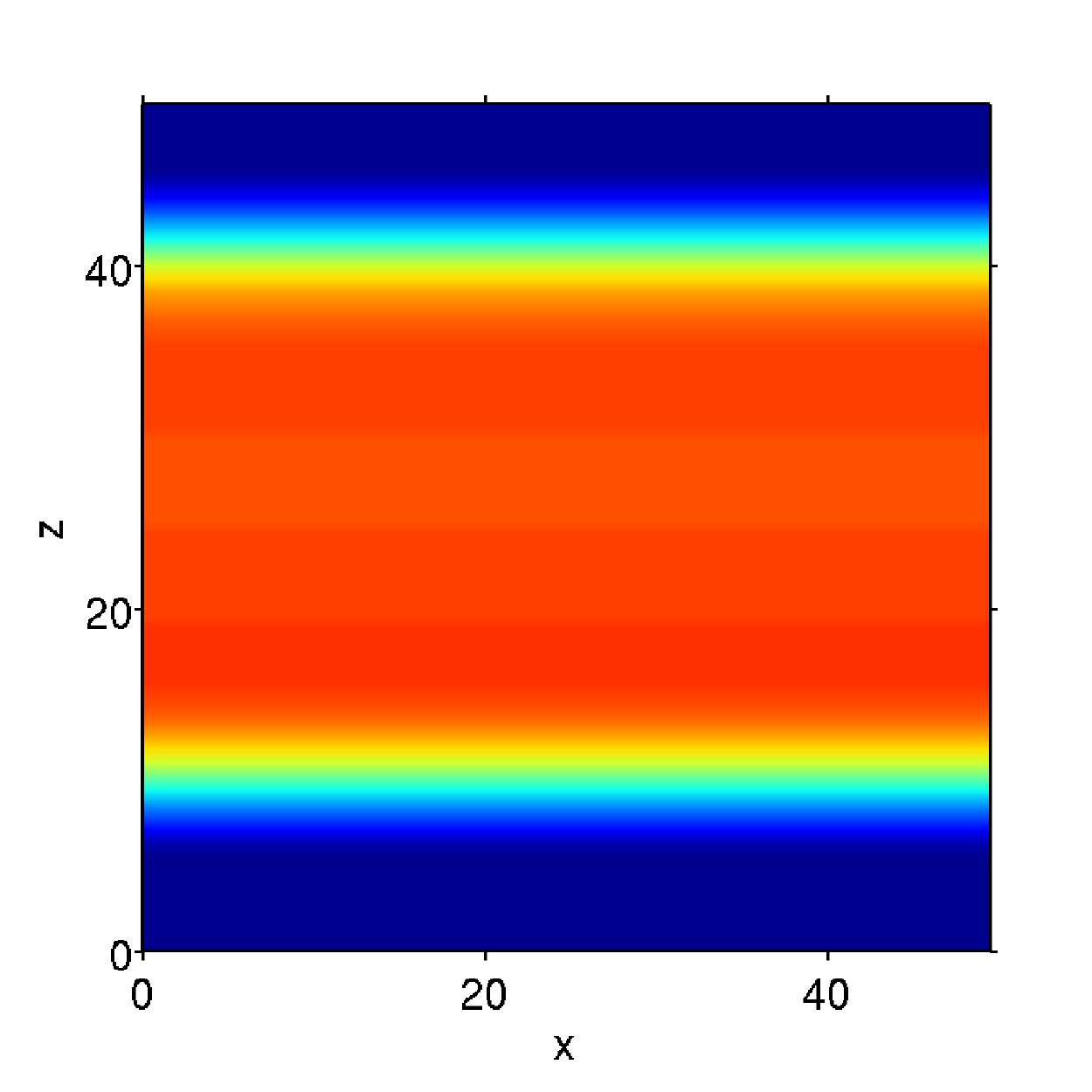}
        \caption{$\epsilon = 0.5,\,\zeta = 1$}
        \label{fig:flat-param-a}
    \end{subfigure}
    \begin{subfigure}[b]{0.24\textwidth}
        \includegraphics[width=\textwidth]{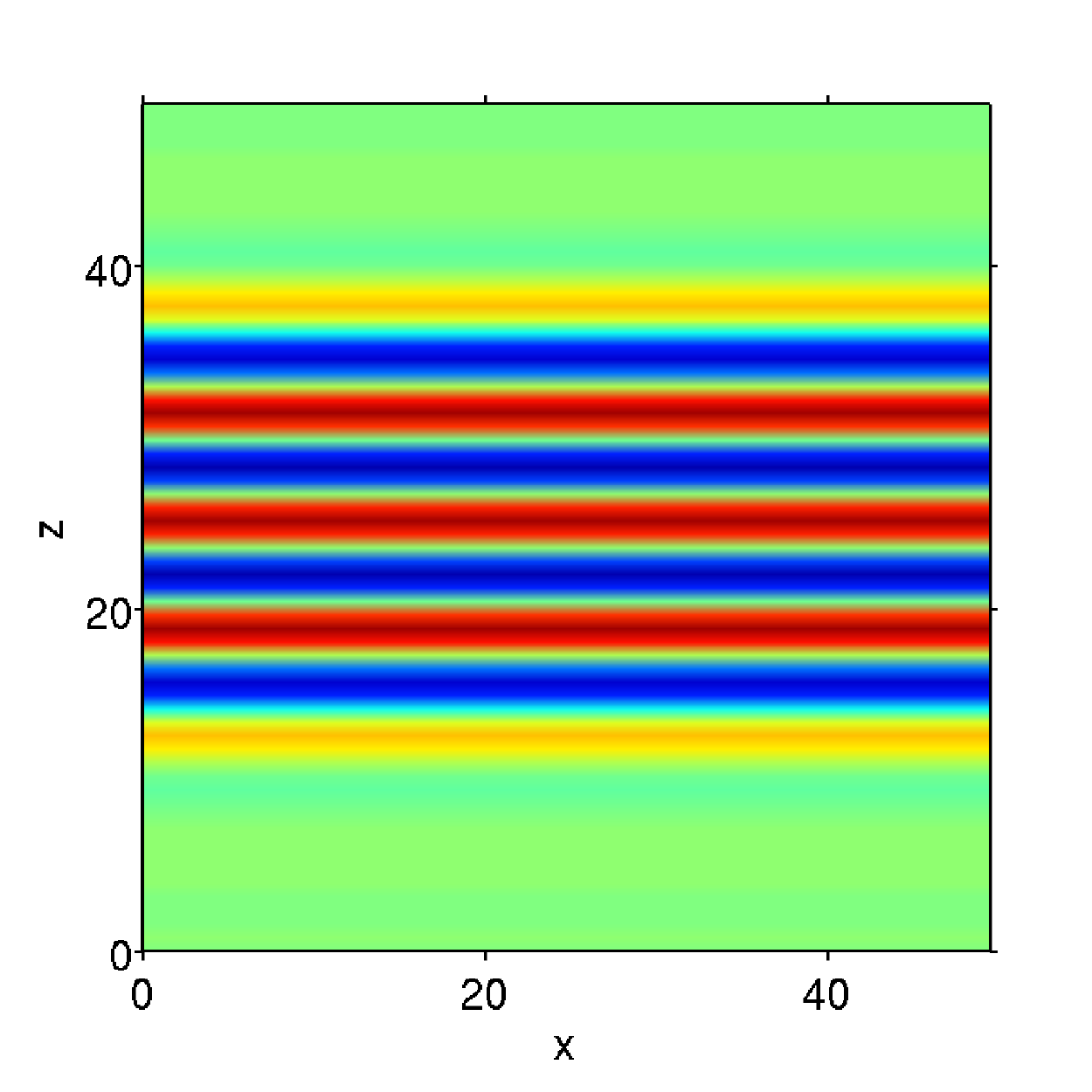}
        \includegraphics[width=\textwidth]{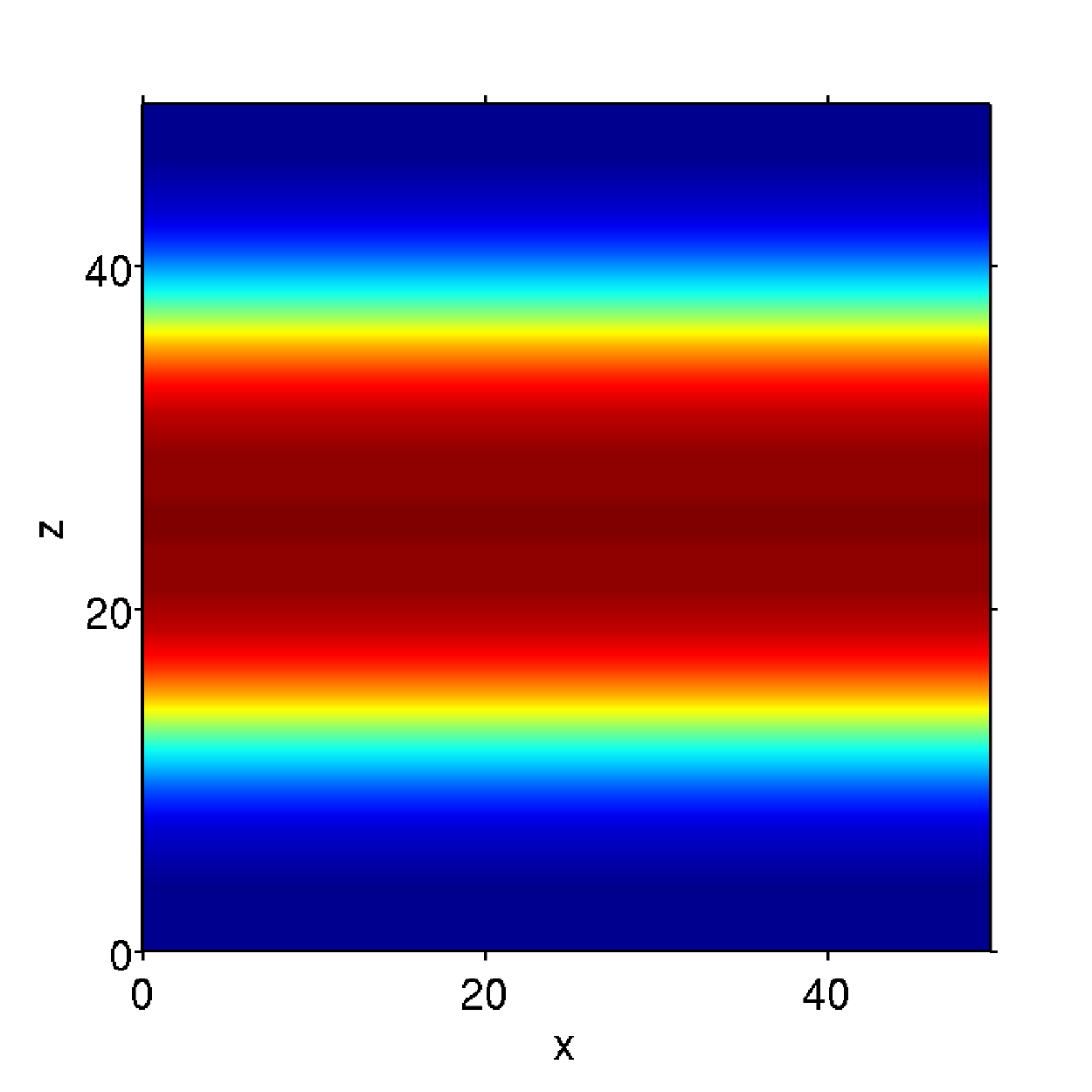}
        \caption{$\epsilon = 0.5,\,\zeta = 100$}
        \label{fig:flat-param-b}
    \end{subfigure}
    \begin{subfigure}[b]{0.24\textwidth}
        \includegraphics[width=\textwidth]{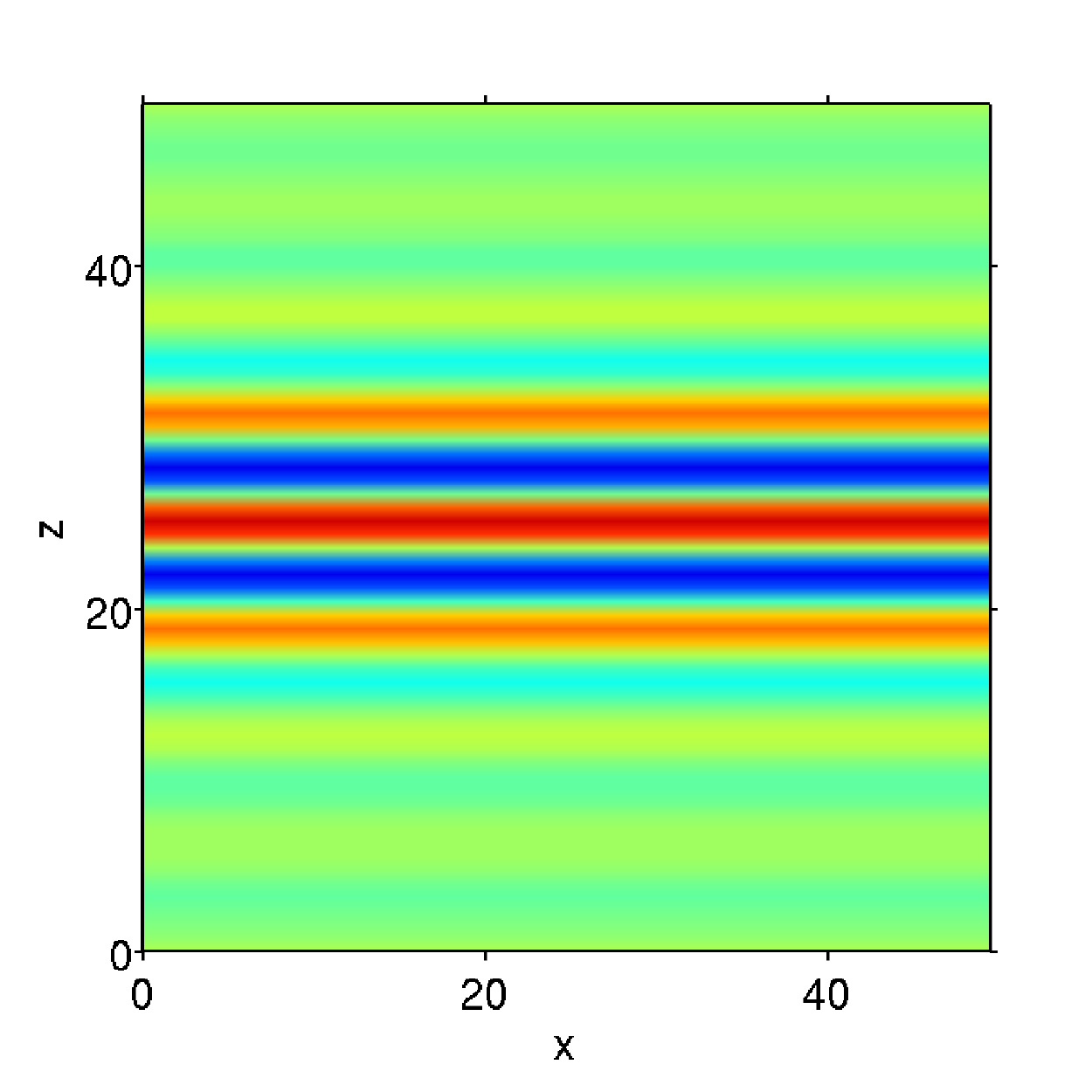}
        \includegraphics[width=\textwidth]{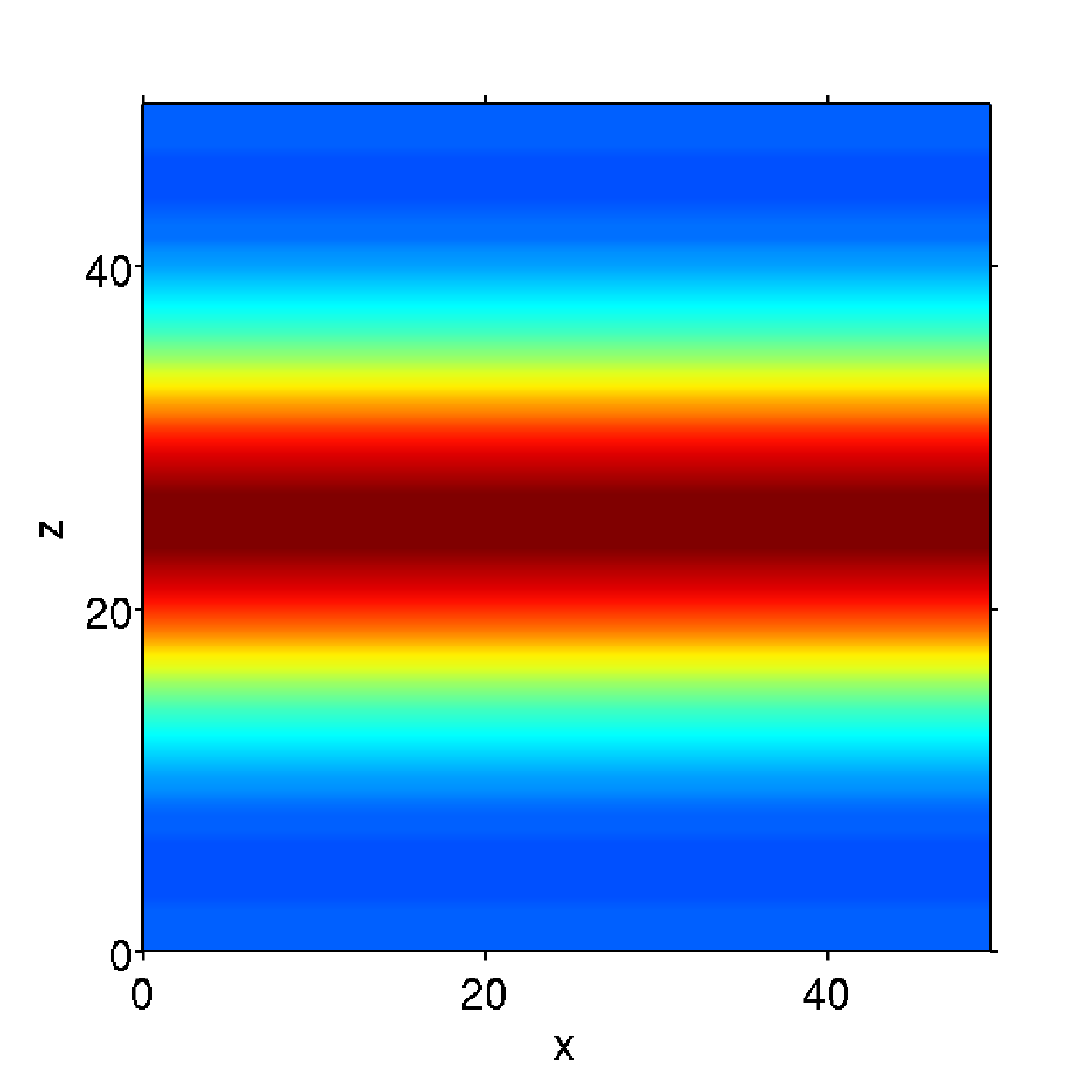}
        \caption{$\epsilon = 0.8,\,\zeta = 1$}
        \label{fig:flat-param-c}
    \end{subfigure}
    \begin{subfigure}[b]{0.24\textwidth}
        \includegraphics[width=\textwidth]{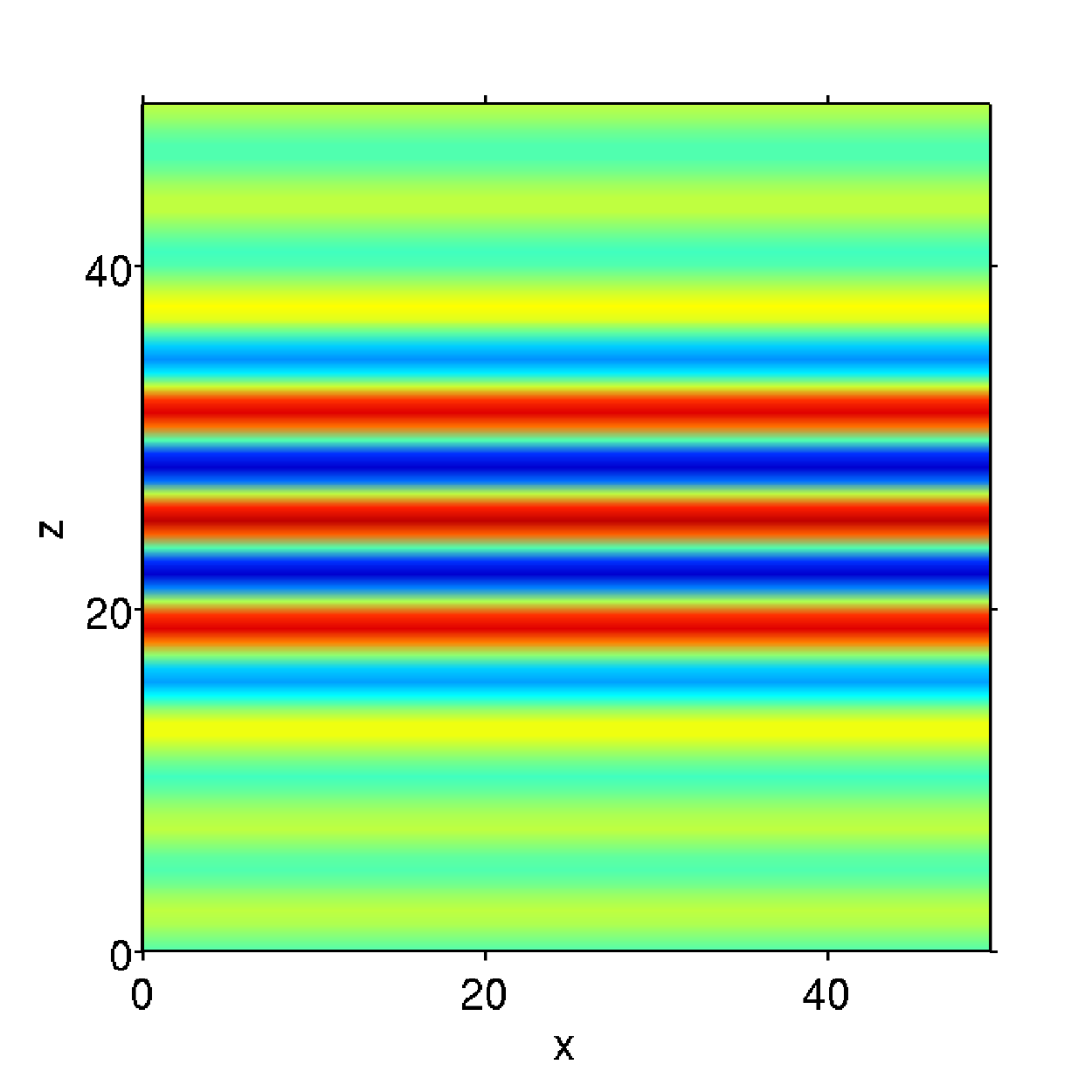}
        \includegraphics[width=\textwidth]{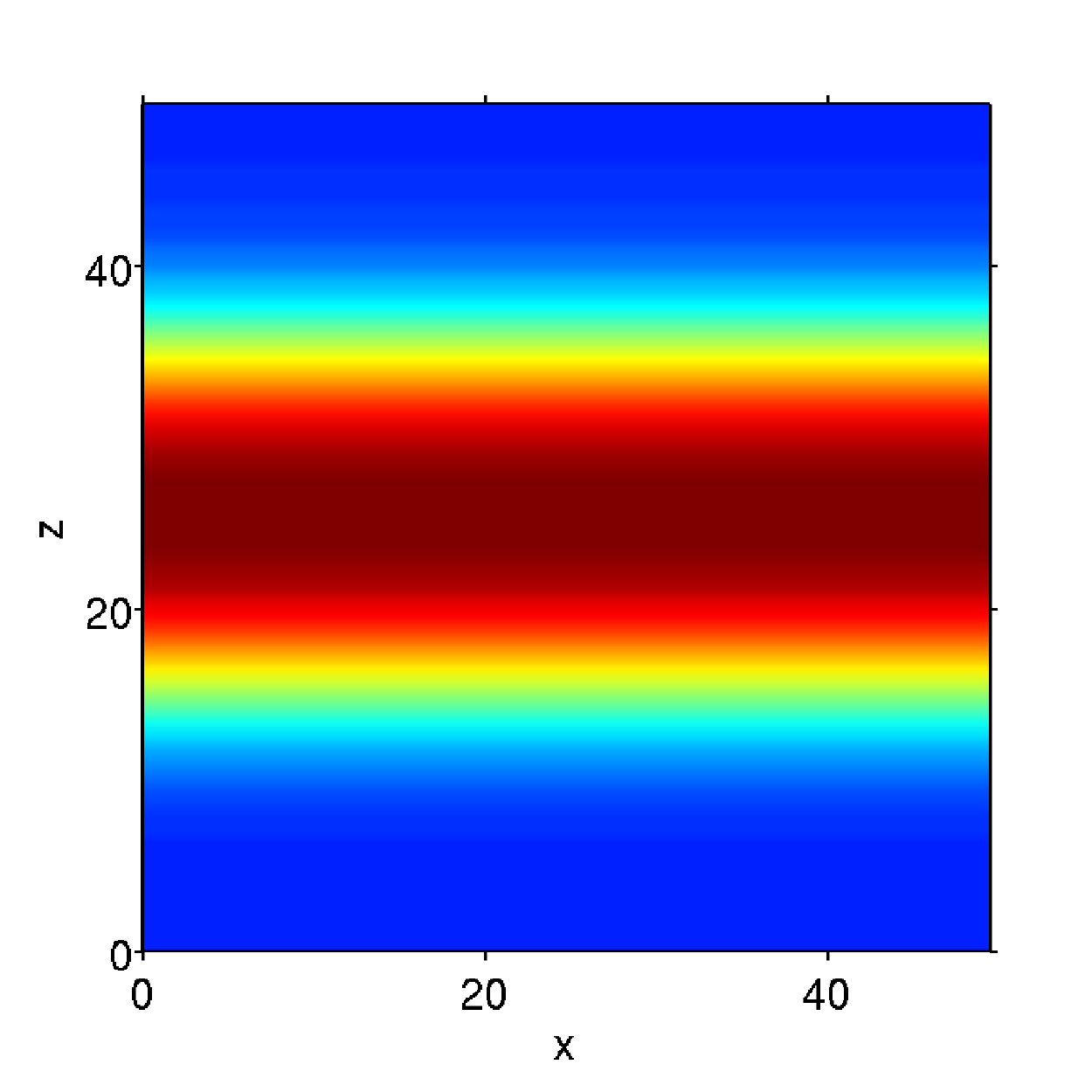}
        \caption{$\epsilon = 0.8,\,\zeta = 100$}
        \label{fig:flat-param-d}
    \end{subfigure}
    \caption{Order parameter (top) and density (bottom) configuration at time $t = 90$, starting from the initial condition in Fig. \ref{fig:flat-ic}. For $\epsilon = 0.5$, the smectic phase is energetically favored, while for $\epsilon  = 0.8$, the isotropic phase is energetically favored. For $\zeta = 1$ the interface moves to grow the energetically favored phase, while for $\zeta = 100$ the interface does not move when $\epsilon = 0.5$ and moves only slightly when $\epsilon = 0.8$.}
	\label{fig:flat-param}
\end{figure}

Starting from this initial configuration, Fig.~\ref{fig:flat-param-a} shows the evolving configuration at time $t = 90$ when $\epsilon$ is set to $0.5$ (within the smectic region of the phase diagram), and coupling constant $\zeta =1$. The initial smectic has grown, and the region of high density has also spread, with its value reduced from the initial $\rho_s = 1$ (so as to conserve mass). We note that the smectic is only able to grow significantly due to the fact that $\rho_s:\rho_0 = 2:1$ so that there are enough molecules in the isotropic phase to allow growth.  Smectic growth was not observed for larger density ratios such as $\rho_s:\rho_0 = 100:1$ when $\zeta = 1$. 

Figure~\ref{fig:flat-param-b} shows the evolution of the same initial configuration and $t=90$ but for $\zeta = 100$.  In this case the smectic region and the density remain almost the same as in the initial condition shown in Fig.~\ref{fig:flat-ic}. The reason why further growth is not observed can be explained by the effective bifurcation parameter $\epsilon_e$ in Eq.~(\ref{eq:psi-const2}). If the smectic grows and the smectic-isotropic interface advances into regions of lower density, we argue that $\epsilon_e$ increases over $\epsilon$ in such regions. Hence when $\zeta$ is large enough, the effective bifurcation parameter can increase even beyond the coexistence value $\epsilon_c$, which prevents any growth of the smectic.

Similarly, we show in Figs.~\ref{fig:flat-param-c} and \ref{fig:flat-param-d} the case $\epsilon = 0.8$ (within the isotropic region of the phase diagram) for $\zeta = 1$ and $100$, at $t = 90$.  Consistent with the above discussion, we see some shrinkage of the smectic while increasing the density of the isotropic phase (satisfying the balance of mass) when $\zeta=1$.  However when $\zeta=100$ there is significantly slower evolution of the smectic because of the strong coupling between the order parameter and the density. This can also be explained in terms of the effective bifurcation parameter $\epsilon_e$, as in this case if the smectic tries to shrink it creates a region where $\epsilon_e$ decreases from $\epsilon$.

\section{Morphological evolution and flows in focal conic domains}
\label{sec:fcflow}

The outer boundary of a FCD in smectic films has a mean curvature $H$ that changes sign from positive away from the center, to negative near the center. At that point there is a macroscopic singularity in the form of a cusp. Near the cusp, the principal curvatures become very large, which can lead to interesting interfacial stresses and flows. Experiments have shown that the FCD undergoes complex morphological changes under temperatures changes, mediated by evaporation of smectic layers. In this section, we study the velocity field in a smectic-isotropic system presenting layers bent in a focal conic configuration. We also revisit the morphological transitions studied in~\cite{vitral2019role}, which were previously investigated within a purely diffusional model (governed by the order parameter equation only), and for uniform density. First, we show that for a certain range of parameters, the transition from focal conic defects to conical pyramids or concentric rings is also observed in the weakly compressible model with density contrast. We then investigate how the velocity field changes during these morphological transitions, revealing the roles of Gaussian curvature $G$ and layer orientation at the interface on flow fields.

\subsection{Flows in focal conics at coexistence}

We analyze the velocity fields obtained at coexistence $\epsilon_c = 0.675$ for different values of $\zeta$ and density ratios, using an initial condition consisting of bent layers forming a focal conic (see Fig.~\ref{fig:fc-cp} left panel). The computational domain is a cubic cell with $N = 256^3$ grid points, with approximately 8 points per wavelength and $q_0 = 1$, so that $L_x = 200$, $L_y = 200$ and $L_z = 200$. Parameters used are $\beta = 2$, $\gamma = 1$, and $\nu = \lambda = 1$. We will compare the velocity field $\mathbf{v}$ obtained when $\zeta$ is large and the system approaches quasi-incompressibility with that obtained when $ \zeta$ is small.

\begin{figure}[ht]
	\centering
    \begin{subfigure}[b]{0.46\textwidth}
    \includegraphics[width=\textwidth]{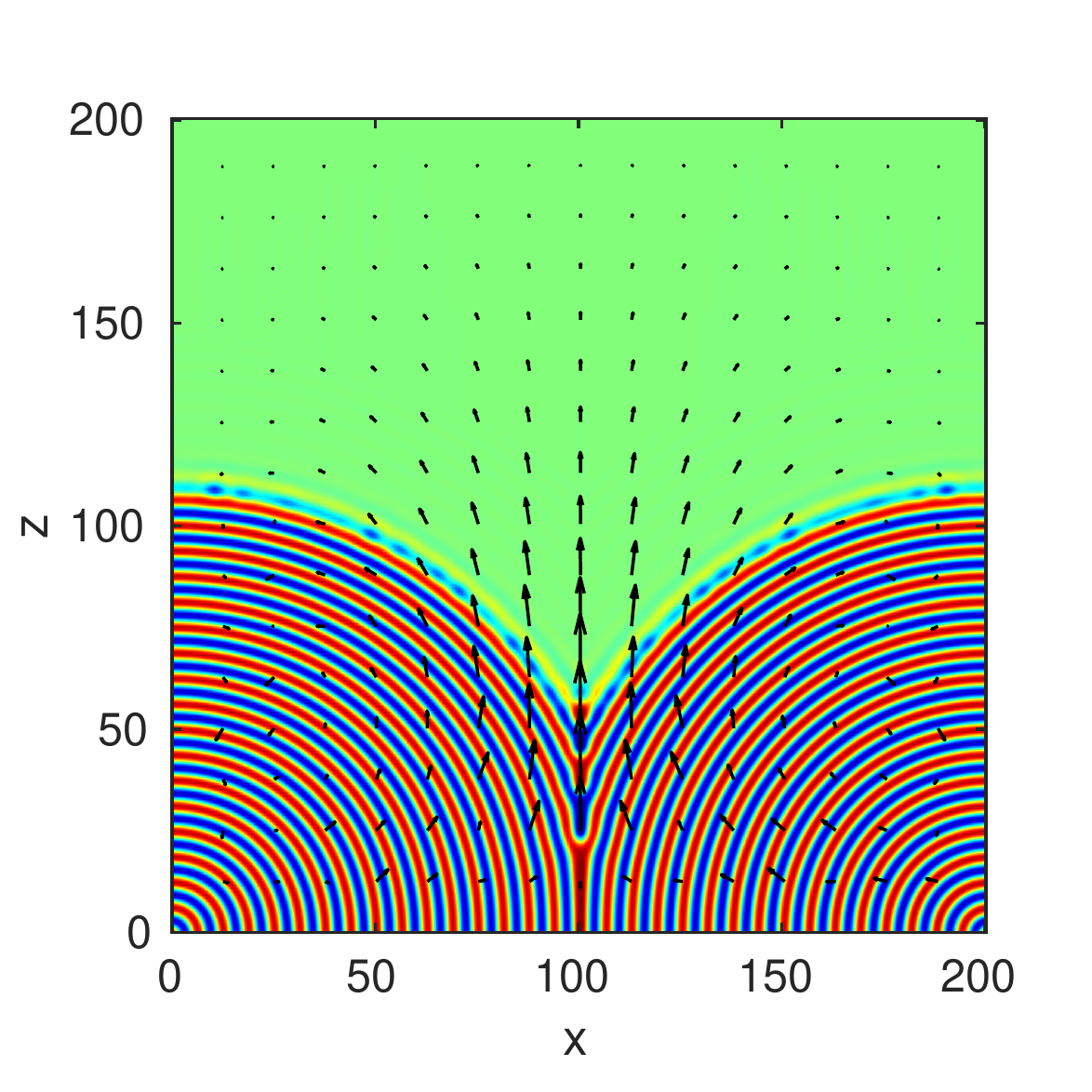}
    \caption{$\zeta = 100$}
    \end{subfigure}
    \begin{subfigure}[b]{0.46\textwidth}
    \includegraphics[width=\textwidth]{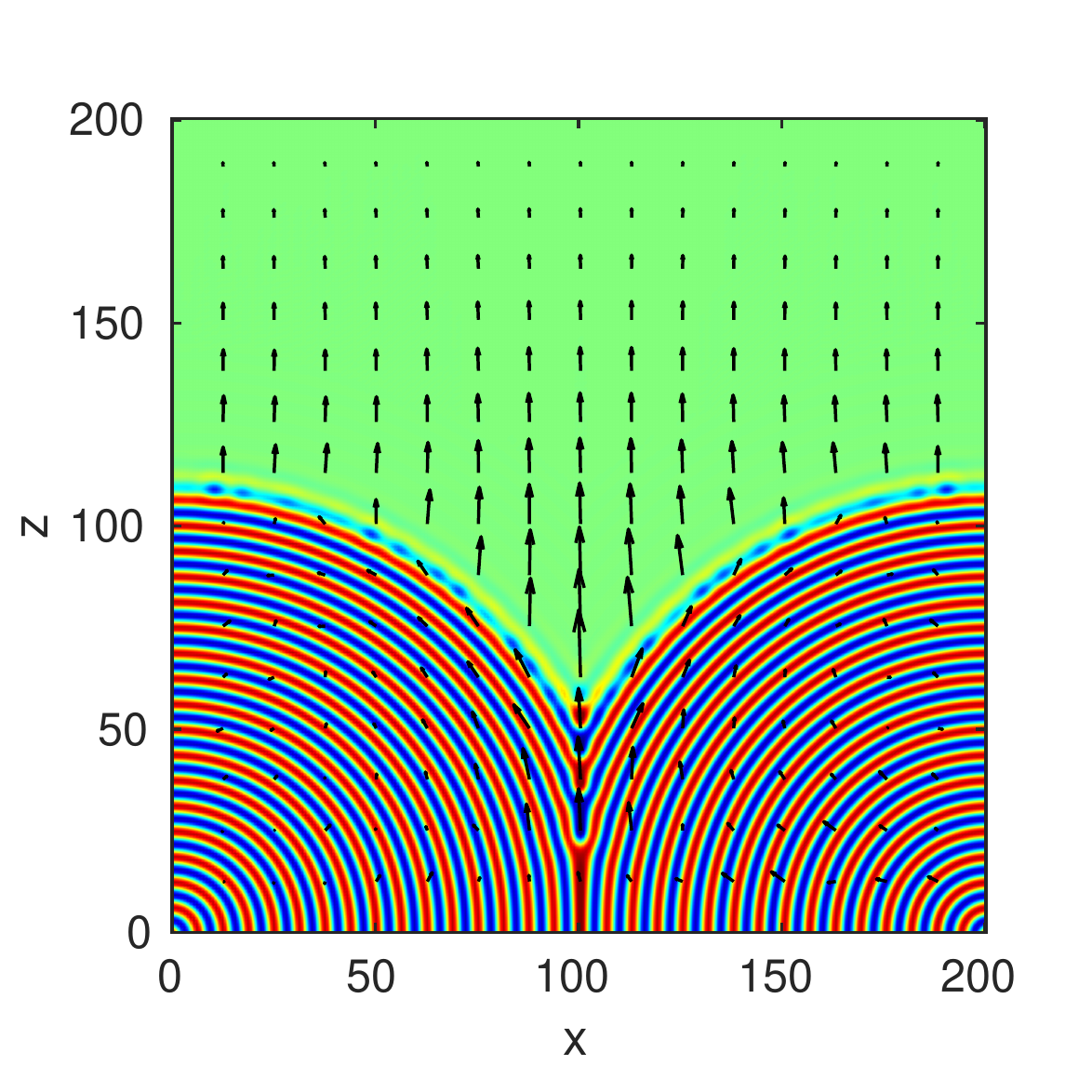}
    \caption{$\zeta = 0.01$}
    \end{subfigure}
    \caption{Comparison between the transient fluid flow $\mathbf{v}$ on smectic-isotropic fluid system for different $\zeta$, at an early time $t = 4$, where both phases have the same bulk density ($\kappa = 0$ and $\rho_0 = 1$). Background color is the order parameter $\psi$. We use $N = 256^3$, $\Delta t = 1\times 10^{-3}$, and parameters $q_0=1$, $\eta=1$, $\epsilon = 0.675$ (coexistence), $\alpha=1$, $\beta=2$ and $\gamma=1$.}
	\label{fig:fcflow1}
\end{figure}

As a reference, we begin with the case with $\kappa = 0$, so $\rho_s = \rho_0 = 1$. Figure~\ref{fig:fcflow1} shows $\mathbf{v}$ at time $t = 4$ at the mid section of the domain, $y = L_y/2$, for $\zeta = 100$ and $\zeta = 0.01$. As expected, for $\zeta = 100$ the flow behaves similarly to an incompressible uniform density system~\cite{vitral2020model}.  Vortices appear at the smectic-isotropic interface, where the flow moves outward from the smectic in regions of negative mean curvature, and inward towards the smectic in regions of positive mean curvature. The flow is governed by local curvatures since no significant density deviation from $\rho = 1$ is observed due to the large value of $\zeta$. The term $\bar{\mu}\nabla\psi$ determines $\mathbf{v}$ in Eq.~(\ref{eq:wcom-stokes2}), where the local curvatures appear from the difference in chemical potential $\mu$ between a planar and a curved interface. This difference is given by the Gibbs-Thomson equation for the case of layers parallel to the interface~\cite{vitral2019role}:
\begin{eqnarray}
    \delta\mu\Delta A &=& 2H\sigma_h + (4H^2-2G)\sigma_b - 2H(3G-4H^2)\sigma_t \;.
    \label{eq:gibbs-thomson}
\end{eqnarray}
Here $\Delta A$ is the difference in amplitude between the two phases, $\sigma_h$ is the surface tension, $\sigma_b$ the interface bending coefficient, and $\sigma_t$ the interface torsion coefficient. These coefficients can be directly computed from the model parameters, see~\cite{vitral2019role}.

For surfaces where the leading order term proportional to $H$ dominates Eq.~(\ref{eq:gibbs-thomson}), the difference in chemical potential $\delta\mu$ is positive for $H > 0$, implying in a thermodynamic force $\bar{\mu}\nabla\psi$ pointing towards the smectic at the interface, and the opposite when $\delta\mu$ is negative. Based on Eq.~(\ref{eq:wcom-stokes2}), this corroborates with the numerical results discussed for Fig.~\ref{fig:fcflow1}(a), where we also notice that the flow is stronger in regions where the magnitude $H$ is high. However, when $\zeta$ is reduced to $\zeta = 0.01$, the flow in the isotropic phase at $t = 4$ simply points away from the interface, whereas inside the smectic the flow also moves up, and curves near the interface. The reason is that for small $\zeta$ the density has more freedom to deviate from the equilibrium values $\rho_0$ and $\rho_s$, particularly at the interface.  This creates density gradients that interfere with the resulting flow structure.

\begin{figure}[ht]
	\centering
    \begin{subfigure}[b]{0.46\textwidth}
    \includegraphics[width=\textwidth]{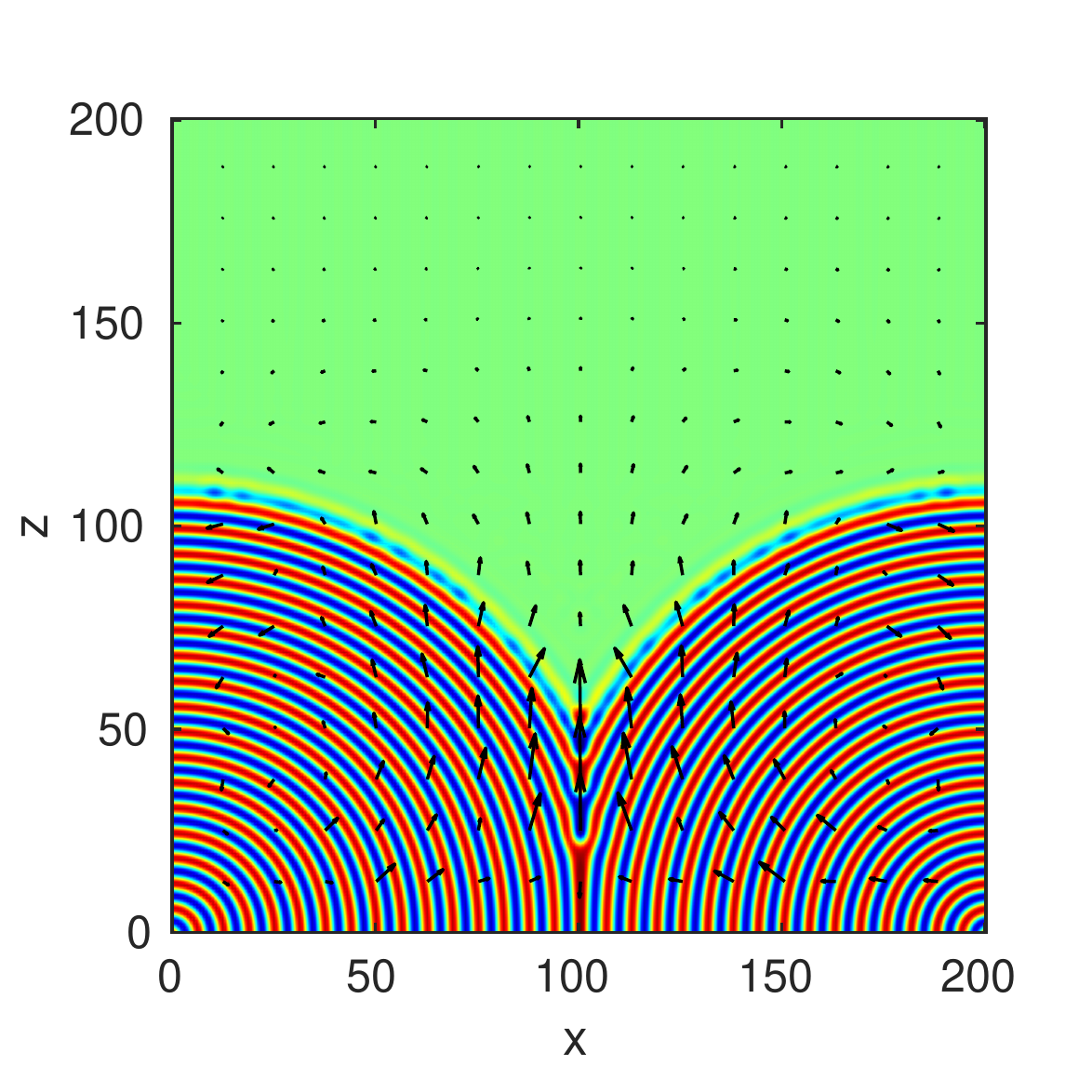}
    \caption{$\zeta = 100$}
    \end{subfigure}
    \begin{subfigure}[b]{0.46\textwidth}
    \includegraphics[width=\textwidth]{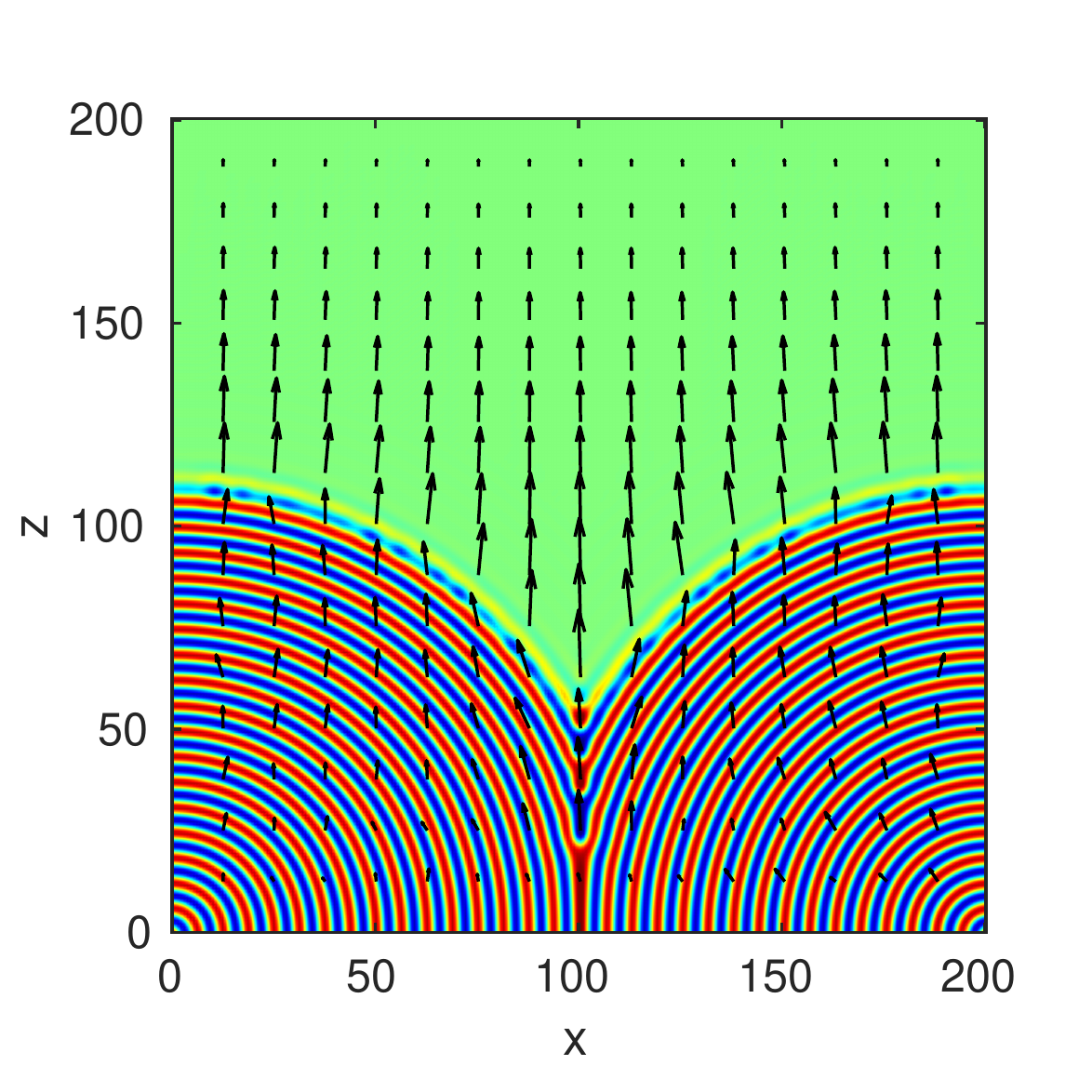}
    \caption{$\zeta = 0.01$}
    \end{subfigure}
    \caption{Comparison between the transient fluid flow $\mathbf{v}$ on smectic-isotropic fluid system for different $\zeta$, at an early time $t = 4$, with $\rho_s:\rho_0 = 2:1$ density ratio between bulk phases ($\kappa = 0.3727$ and $\rho_0 = 0.5$). Background color is the order parameter $\psi$. We use $N = 256^3$, $\Delta t = 1\times 10^{-3}$, and parameters $q_0=1$, $\eta=1$, $\epsilon = 0.675$ (coexistence), $\alpha=1$, $\beta=2$ and $\gamma=1$.}
	\label{fig:fcflow2}
\end{figure}

We now increase the density contrast to $\rho_s:\rho_0 =  2:1$, using $\kappa =  0.3727$ and $\rho_0 = 0.5$. Figure~\ref{fig:fcflow2} shows numerical results for $\mathbf{v}$ at time $t = 4$ for the cases of $\zeta = 100$ and $\zeta = 0.01$. When compared to Fig.~\ref{fig:fcflow1}(a), the velocity field for $\zeta = 100$ presents smaller vortices closer to the smectic region, which quickly decay away from the interface in the isotropic phase. This is characteristic of quasi-incompressibility, since for large $\zeta$ the density is constant almost everywhere except at the sharp interface. For the case of $\zeta = 0.01$, the velocity points upwards throughout the smectic-isotropic system, implying that the density gradient dominates the orientation of the flow. This result is similar to Fig.~\ref{fig:fcflow1}(b), although for Fig.~\ref{fig:fcflow2}(b) the flow in the smectic is stronger; this happens owing to the combination of the larger density gradient for $\rho_s:\rho_0 =  2:1$ with the small value of $\zeta$.  In this case the density at the interface becomes more diffuse, enhancing the longitudinal flow that moves from regions of high to low density.
 
\begin{figure}[ht!]
	\centering
    \begin{subfigure}[b]{0.4\textwidth}
    \includegraphics[width=\textwidth]{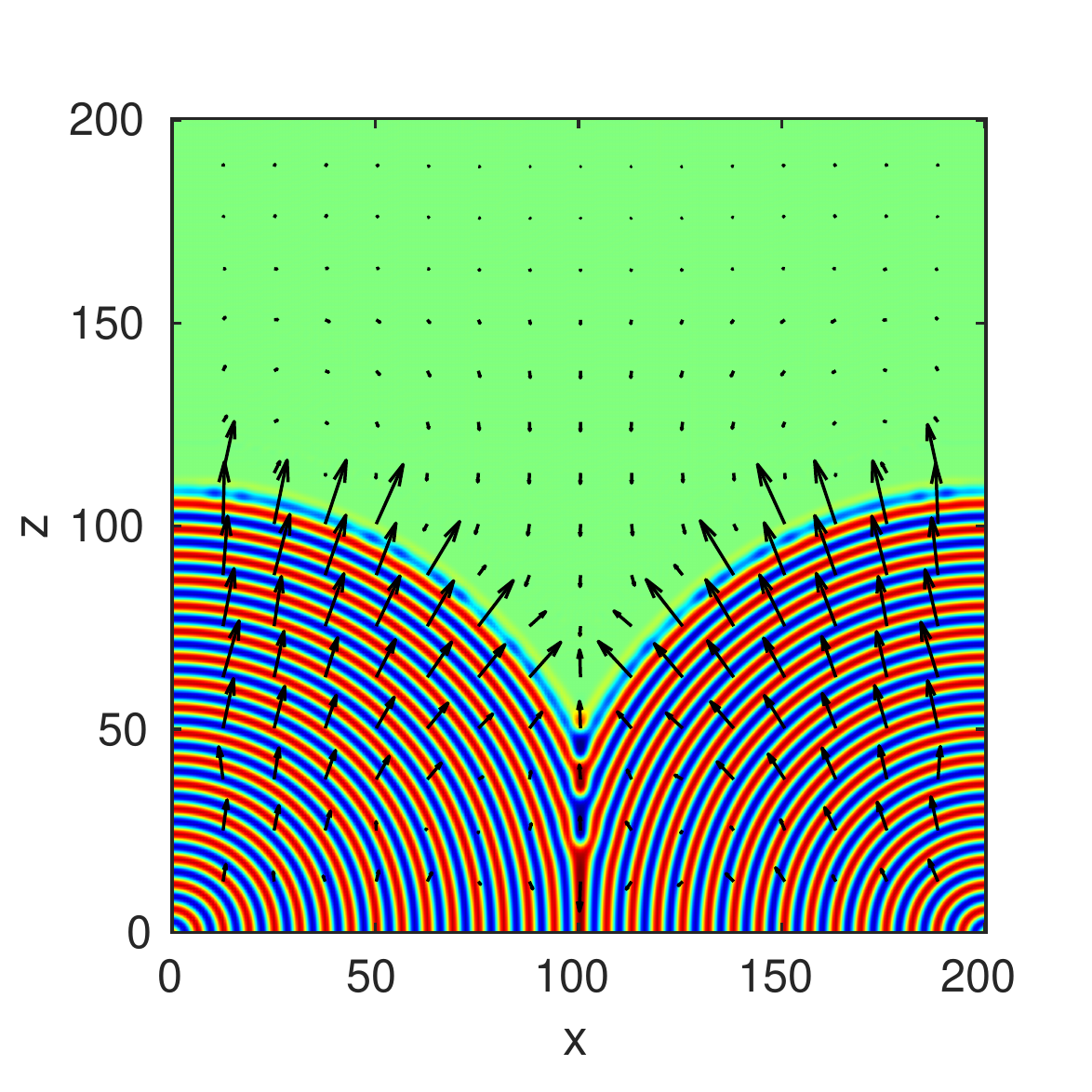}
    \caption{$t = 4$, $\zeta = 100$}
    \label{fig:fcflow3-a}
    \end{subfigure}
    \begin{subfigure}[b]{0.4\textwidth}
    \includegraphics[width=\textwidth]{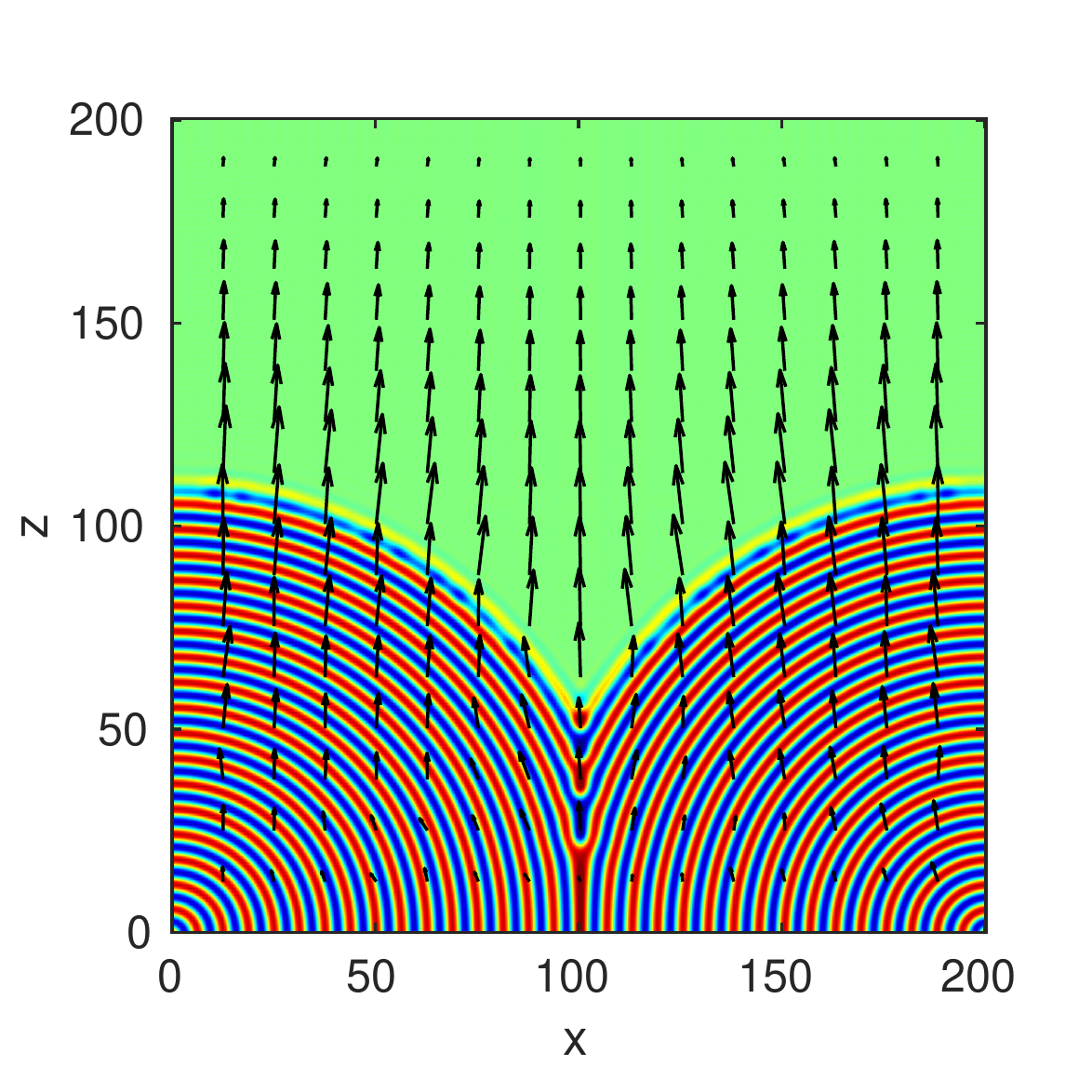}
    \caption{$t = 4$, $\zeta = 0.01$}
    \label{fig:fcflow3-b}
\end{subfigure}
    \begin{subfigure}[b]{0.4\textwidth}
    \includegraphics[width=\textwidth]{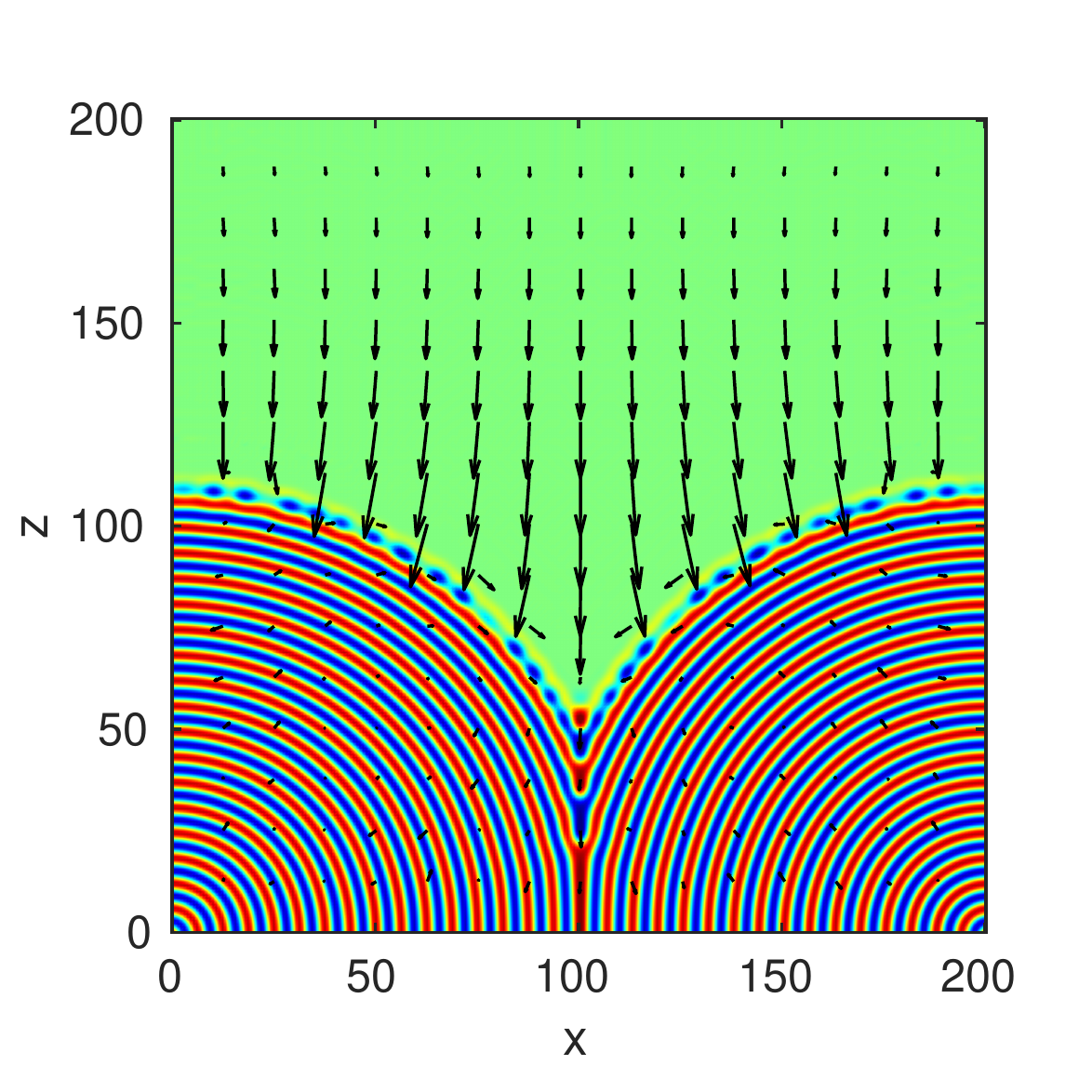}
    \caption{$t = 25$, $\zeta = 100$}
    \label{fig:fcflow3-c}
    \end{subfigure}
    \begin{subfigure}[b]{0.4\textwidth}
    \includegraphics[width=\textwidth]{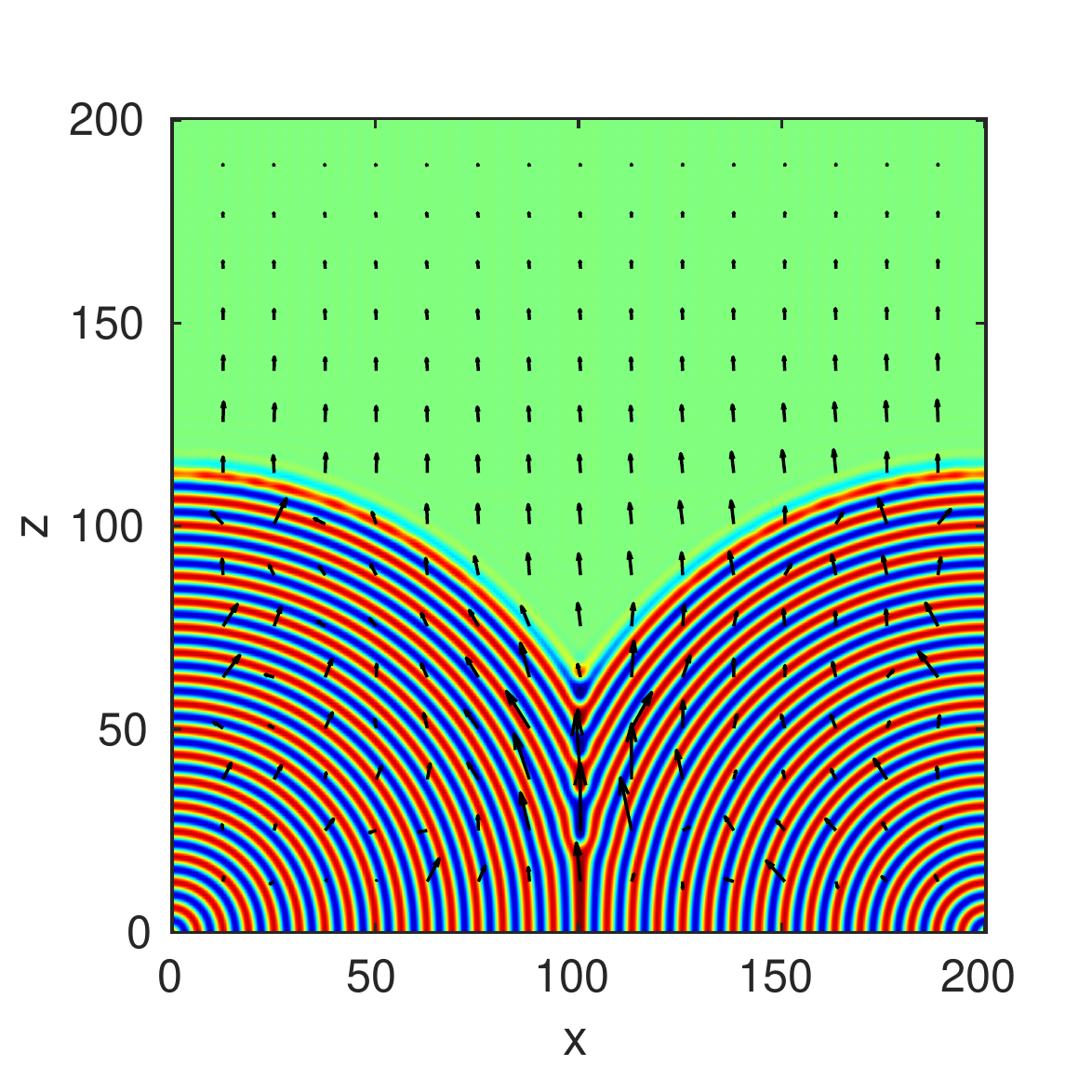}
    \caption{$t = 25$, $\zeta = 0.01$}
    \label{fig:fcflow3-d}
    \end{subfigure}
    \caption{Comparison between the transient fluid flow $\mathbf{v}$ on smectic-isotropic fluid system for different $\zeta$, at times $t = 4$ and $t = 25$, with $\rho_s:\rho_0 = 100:1$ density ratio between bulk phases ($\kappa = 0.7379$ and $\rho_0 = 0.01$). Background color is the order parameter $\psi$. We use $N = 256^3$, $\Delta t = 1\times 10^{-3}$, and parameters $q_0=1$, $\eta=1$, $\epsilon = 0.675$ (coexistence), $\alpha=1$, $\beta=2$ and $\gamma=1$.}
	\label{fig:fcflow3}
\end{figure}

On increasing the density ratio to $\rho_s:\rho_0 = 100:1$, by using $\kappa = 0.7379$ and $\rho_0 = 0.01$, some significant changes in the velocity are observed for large $\zeta$. Figure~\ref{fig:fcflow3-a} shows that for $\zeta = 100$ at $t = 4$, the flow in the smectic becomes dominated by the potential part of the velocity, so that it points radially outward from the layers in the direction of the density gradient. The velocity in the isotropic phase is smaller in magnitude, and points towards the smectic. This happens because for large $\zeta$ the density gradient becomes large at the interface owing to the large density ratio. We also show in Fig.~\ref{fig:fcflow3-c} the transient flow for $\zeta = 100$ at a later time, $t = 25$.  The flow in the smectic has mostly disappeared, so the only significant flow is in the isotropic phase towards the smectic and tangential at the interface. Since the density of the isotropic phase is small, the mass flux in this case becomes very small in magnitude and decreases with time.

For $\zeta = 0.01$ Fig.~\ref{fig:fcflow3-b} shows that the velocity at $t = 4$ points upwards as in Fig.~\ref{fig:fcflow2}(b). In Fig.~\ref{fig:fcflow3-d}, we see that the velocity at $t = 25$ is still primarily pointing upwards, and that significant growth of smectic layers at the interface has taken place when since $t = 4$.  Since the energy penalty for deviations of the preferred bulk density is small for a small $\zeta$, the mass flux $\rho\mathbf{v}$ from the bulk smectic towards the interface leads to the growth of layers. That is, growth of the smectic by mass flow may occur, albeit at the cost of reducing the bulk smectic density in order to satisfy overall mass conservation.

\subsection{Flows in focal conics under thermal sintering}

\begin{figure}[ht]
	\centering
    \begin{subfigure}[b]{0.4\textwidth}
    \includegraphics[width=\textwidth]{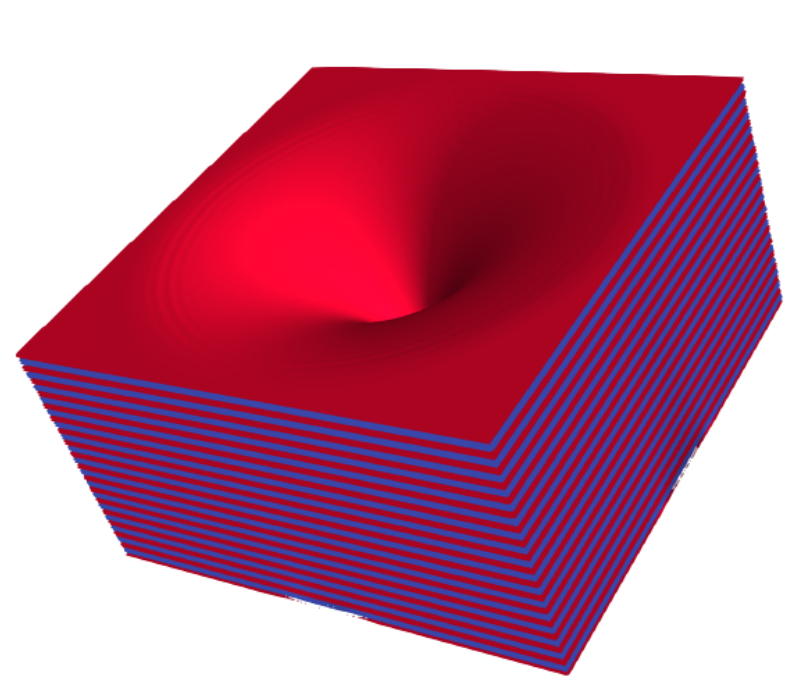}
    \end{subfigure}
    \hspace{10mm}
    \begin{subfigure}[b]{0.4\textwidth}
    \includegraphics[width=\textwidth]{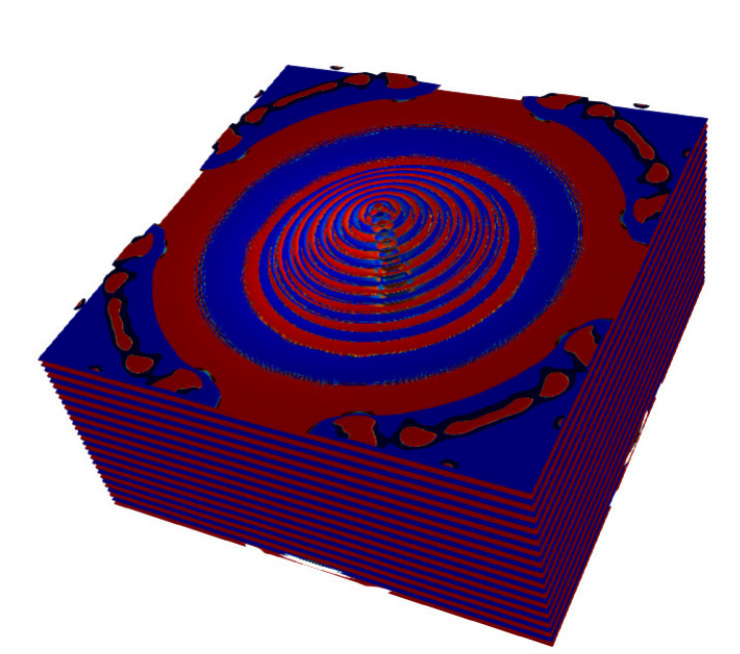}
    \end{subfigure}
    \caption{Focal conic (initial condition) transforms into a conical pyramid ($t = 80$) through evaporation of layers. Red and blue planes represent $+1$ and $-1$ values of $\psi$, respectively. Parameters used are $\epsilon = 0.8$ (favoring the isotropic phase), $\beta = 2$, $\gamma = 1$, $\nu = 100$, $\lambda = 100$, $\rho_0 = 0.5$, $\kappa = 0.3727$ ($\rho_s = 1$), and $\zeta = 0.01$.}
	\label{fig:fc-cp}
\end{figure}

Transitions in smectic thin films from focal conic domains to conical pyramids through sintering have been experimentally observed by Kim et al.~\cite{kim2016controlling,kim2018curvatures}. By increasing the value of $\epsilon$ (favoring the isotropic phase) we can simulate a heat treatment of a smectic film similar to the sintering experiments. We set as initial condition the configuration of Fig.~\ref{fig:fc-cp} (left), with smectic layers bent in a focal conic configuration, in a domain with $N = 256^3$ grid points, so that $L_x = 200$, $L_y = 200$ and $L_z = 200$. Parameter values used are $\beta = 2$, $\gamma = 1$, $\nu = \lambda = 100$, $\rho_0 = 0.5$, $\rho_s = 1$ and $\zeta = 0.01$. We set $\epsilon = 0.8$, so that the initial smectic region should slowly evaporate at the interface with the isotropic phase. We choose a small value for $\zeta$ in order to allow the transition to take place. If $\zeta$ is too large the focal conic morphology is unable to change significantly since the motion of $\psi$ becomes restricted due to the balance of mass. Figure~\ref{fig:fc-cp} shows the initial focal conic (left) and the resulting conical pyramid (right) at time $t = 80$, after a number of layers have evaporated away from the core, sculpting a pyramidal structure that is more resilient to evaporation than the original focal conic. Also, the velocity field for a focal conic under $\epsilon = 0.8$ initially points away from the smectic, similarly to Fig.~\ref{fig:fcflow2}(b).

Note that the edges of the smectic layers in the pyramidal region become exposed at the interface. As discussed in~\cite{vitral2019role}, the equation describing interfacial thermodynamics for this perpendicular alignment of layers with respect to the interface is different from the classical Gibbs-Thomson equation found in literature, even at leading order. The difference in chemical potential between planar and curved interfaces in this case is given by
\begin{eqnarray}
  \delta\mu \Delta A &=& \bigg[\frac{1}{2}\nabla^2_s H + 2H(H^2 - G)\bigg] \frac{\sigma_h}{q_0^2} \;,
  \label{eq:gt-perp}
\end{eqnarray}
where the combination of curvatures in the RHS is similar to Willmore-type flows~\cite{willmore1996riemannian}. Therefore, we consider the effect of layer alignment at the interface on flow.

For a small dimensionless $\zeta$, local curvatures dominate the motion of $\psi$. A small $\zeta$ also leads to a very diffuse density variation at the interface, which evolves slowly compared to the order parameter. Based on the balance of linear momentum in Eq.~(\ref{eq:blm-ge-w}), under these conditions the velocity at the interface of a conical pyramid is governed by the force $\bar{\mu}\nabla\psi$. We have previously shown~\cite{vitral2019role} that conical pyramids are surfaces for which the RHS of Eq.~(\ref{eq:gt-perp}) is approximately zero, and present values of $H^2$ and $G$ that are locally close to each other along most of the surface. That is, conical pyramids in this context are analogous to Willmore surfaces~\cite{willmore1996riemannian}, for which the Willmore flow is zero. This has an important implication to the normal force balance at the interface, as we now show.

When layers are perpendicular to the interface, $\nabla\psi$ has both a tangential and a normal component at this interface, where the normal component is the same as the gradient of the amplitude $\nabla A$ at this interface (which only acts on the slow spatial scale~\cite{vitral2020spiral}). Since Eq.~(\ref{eq:gt-perp}) is derived from the amplitude equation that governs the motion of $A$, we find that $\delta\mu \Delta A$ corresponds to the asymptotic form of the force $\bar{\mu}\nabla\psi$ in the normal direction to the curved interface. 

\begin{figure}[ht]
	\centering
    \includegraphics[width=0.45\textwidth,height=0.42\textwidth]{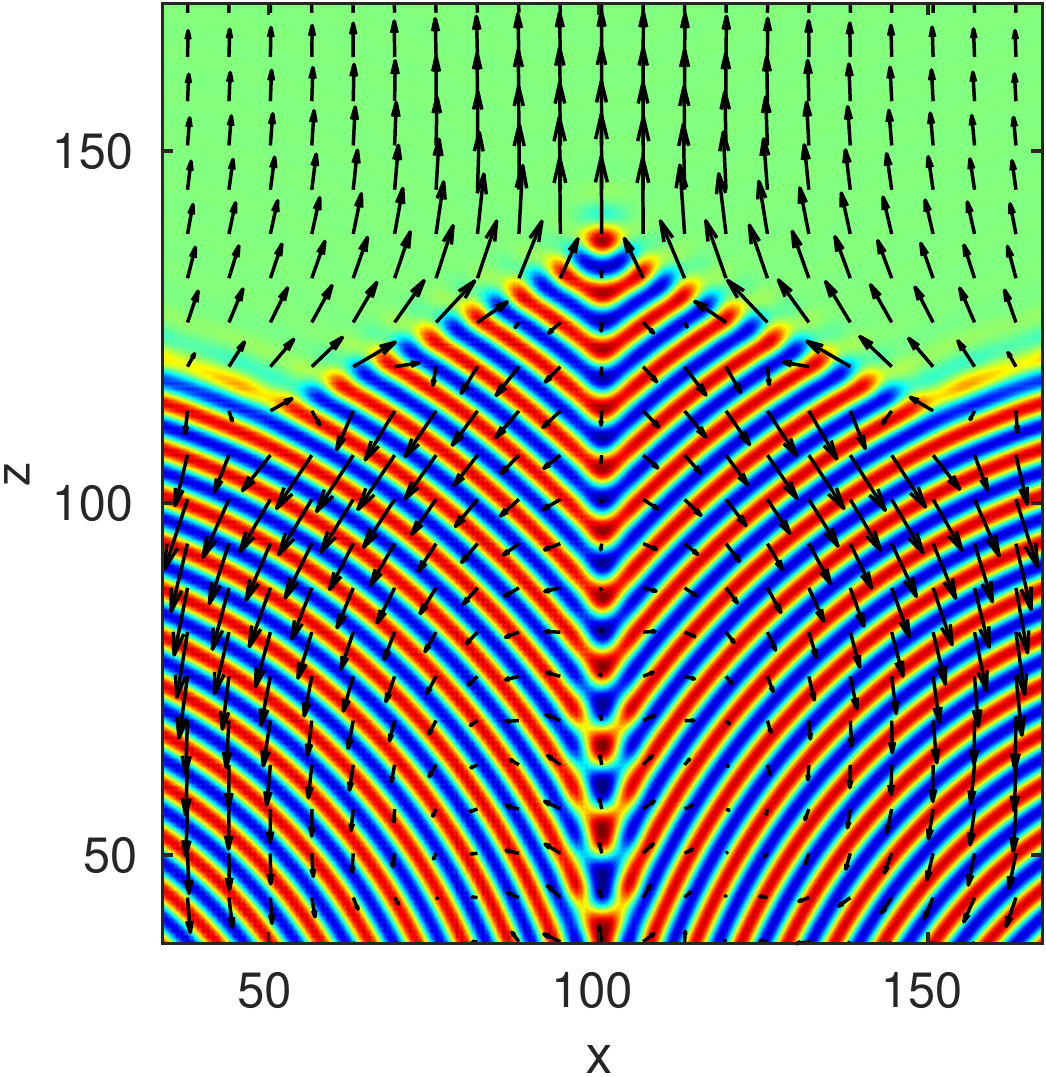}
    \caption{Order parameter field for a conical pyramid (in cross-section) at $t = 80$ obtained from an initial focal conic configuration. The velocity field $\mathbf{v}$ is also plotted, showing that the velocity becomes tangential to the interface of the pyramid. Parameters are $\epsilon = 0.8$ (favoring the isotropic phase), $\beta = 2$, $\gamma = 1$, $\nu = 100$, $\lambda = 100$, $\rho_0 = 0.5$, $\kappa = 0.3727$, and $\zeta = 0.01$.}
	\label{fig:cp-flow}
\end{figure}

The fact that the RHS of Eq.~(\ref{eq:gt-perp}) becomes very small for a conical pyramid then suggests that the normal velocity $\mathbf{v}\cdot\mathbf{n}$ at the interface is negligible when compared to tangential flow. In Fig.~\ref{fig:cp-flow}, we show the middle cross section of the $\psi$ field for a conical pyramid alongside the velocity field. As expected, the numerical result reveals that the flow indeed becomes tangential on the surface of the pyramid, reinforcing that mean and Gaussian curvatures balance in such a way that the RHS of Eq.~(\ref{eq:gt-perp}) and normal forces become negligible on this surface.

\section{Domain interactions in smectic-isotropic systems}
\label{sec:domain}

We consider two applications of the weakly compressible smectic-isotropic model that involve interactions between two domains. The first involves examining the coalescence of neighboring cylindrical stacks of smectic layers, which is motivated by experiments in freely-suspended smectic films. The second is an investigation of the interactions between two FCDs, once again motivated by the experimental observations from Kim et al.~\cite{kim2016controlling,kim2018curvatures}.

\subsection{Coalescence of smectic cylindrical domains}

Domain coalescence driven by capillarity is a widely studied phenomenon in fluid dynamics. For two infinite isotropic fluid cylinders with the same radius, Hopper~\cite{hopper1984coalescence,hopper1993coalescence} developed an exact theory based on the assumption of creeping planar viscous flow, parameterizing the coalescence in the plane through a 1-parameter family of closed inverse ellipses of constant area. This theory has been shown to be qualitatively consistent with a bridge growing between two smectic islands~\cite{shuravin2019coalescence,nguyen2020coalescence}, although with a slower temporal evolution. In a thin smectic-A liquid crystal equilibrium, which is freely suspended, islands are regions with more smectic layers than the embedding film, resulting in smectic cylinders enveloped by outer layers, bounded by edge dislocation loops. The line tension arising from the dislocations is argued to drive the initial bridge expansion in the coalescence process. In contrast to ordinary fluids, Nguyen et al.~\cite{nguyen2020coalescence,nguyen2011smectic} showed that permeation through the molecular layers of the merging islands is an additional channel for dissipative motion, not included in Hopper's model, and it may be responsible for the observed slow coarsening dynamics. Coalescence of smectic islands~\cite{shuravin2019coalescence,nguyen2020coalescence} and holes~\cite{dolganov2020coalescence} have been observed in freely-suspended smectic films (FSSF) with layers parallel to the surface of the film, with thicknesses that can range from two layers up to thousands of layers.

\begin{figure}[ht]
	\centering
    \begin{subfigure}[b]{0.4\textwidth}
    \includegraphics[width=\textwidth]{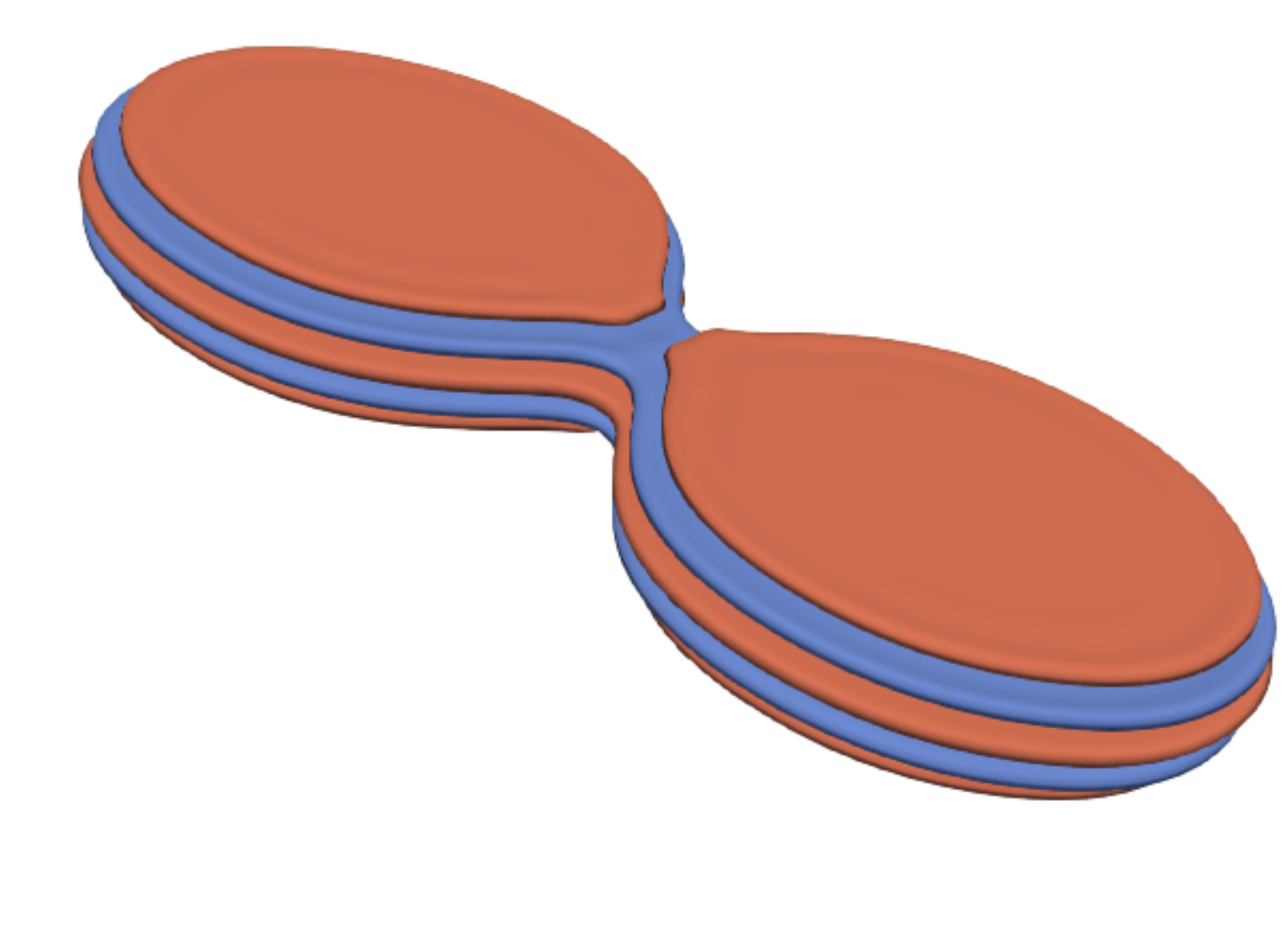}
    \caption{$t = 2.5$}
    \end{subfigure}
    \hspace{10mm}
    \begin{subfigure}[b]{0.4\textwidth}
    \includegraphics[width=\textwidth]{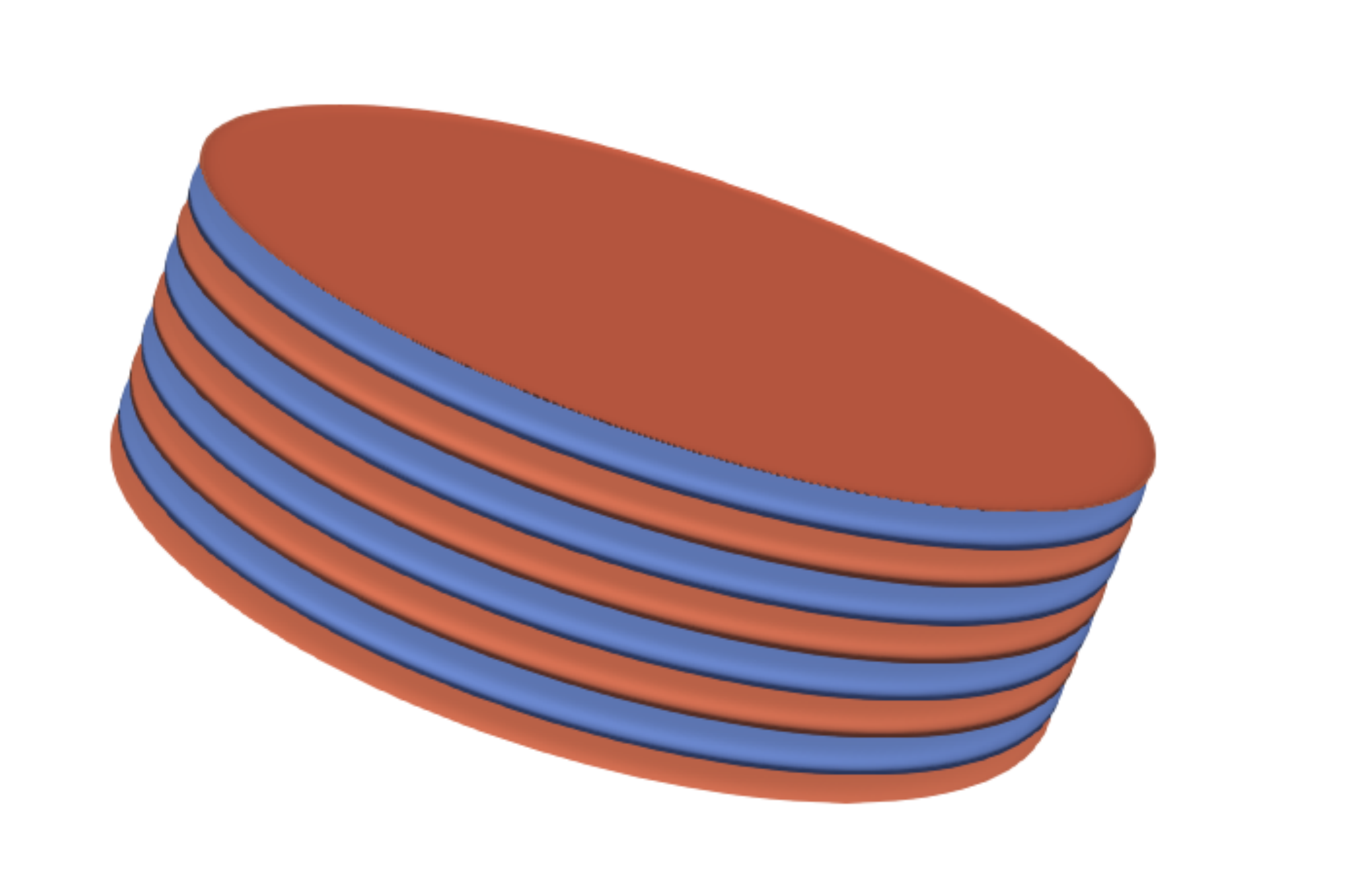}
    \caption{$t = 200$}
    \end{subfigure}
    \caption{Coalescence of smectic cylinders inside a box with $L_x = L_y = 200$ and $L_z  = 25$. Besides coalescence, the number of layers of the final cylinder grow with respect to the initial ones. Parameters are $\epsilon = 0.3$ (smectic region), $\zeta = 100$, $\beta = 2$, $\gamma = 1$, $\nu = 1$, $\lambda = 1$, $\rho_s = 1$, and $\rho_0 = 0.5$ ($\kappa = 0.3727$).}
	\label{fig:cls3d}
\end{figure}

Here, we study numerically a related configuration consisting of two smectic cylinders, which are initially touching, surrounded by an isotropic phase of different density. While this geometry is much simpler than a FSSF, it connects to Hopper's work on the coalescence of two fluid cylinders, and also shows the role played by irrotational flow in our model. As smectic is a uniaxial phase, we expect to observe qualitatively different features in the interaction between two smectic cylinders when compared to isotropic fluids. We consider a $N = 256^2\times32$ grid as the computational domain, so that $L_x = 200$, $L_y = 200$ and $L_z = 25$. The domain contains two smectic cylinders, each with radius $R_0 = 37.5$ and five layers, embedded in the middle of an isotropic domain. Parameters used are $\beta = 2$, $\gamma = 1$, $\nu = \lambda = 1$, $\rho_0 = 0.5$, $\kappa = 0.3727$ (2:1 density ratio) and $\zeta = 100$. We choose $\epsilon = 0.3$, deep in the smectic state (as $\epsilon_c = 0.675$), so that the smectic is energetically favored. At the same time, we take $\zeta = 100$ to be high enough to guarantee a weak conservation of $\psi$, a necessary condition for smectic coalescence to take place (if $\zeta$ is too small, $\psi$ would grow and occupy the whole domain with $\epsilon = 0.3$ and a 2:1 density ratio). Figure~\ref{fig:cls3d} shows the two smectic cylinders creating a bridge between each other at time $t = 2.5$, and also the resulting stationary cylinder at time $t = 200$ after coalescence occurs. Note that in addition to a two-dimensional coalescence process in the plane of the smectic layers, we observe a number of layers growing in the normal direction $\hat{z}$, from five initial layers to eight in the resulting cylinder (filling the whole $L_z$). 

\begin{figure}[ht]
	\centering
    \begin{subfigure}[b]{0.3\textwidth}
        \includegraphics[width=\textwidth]{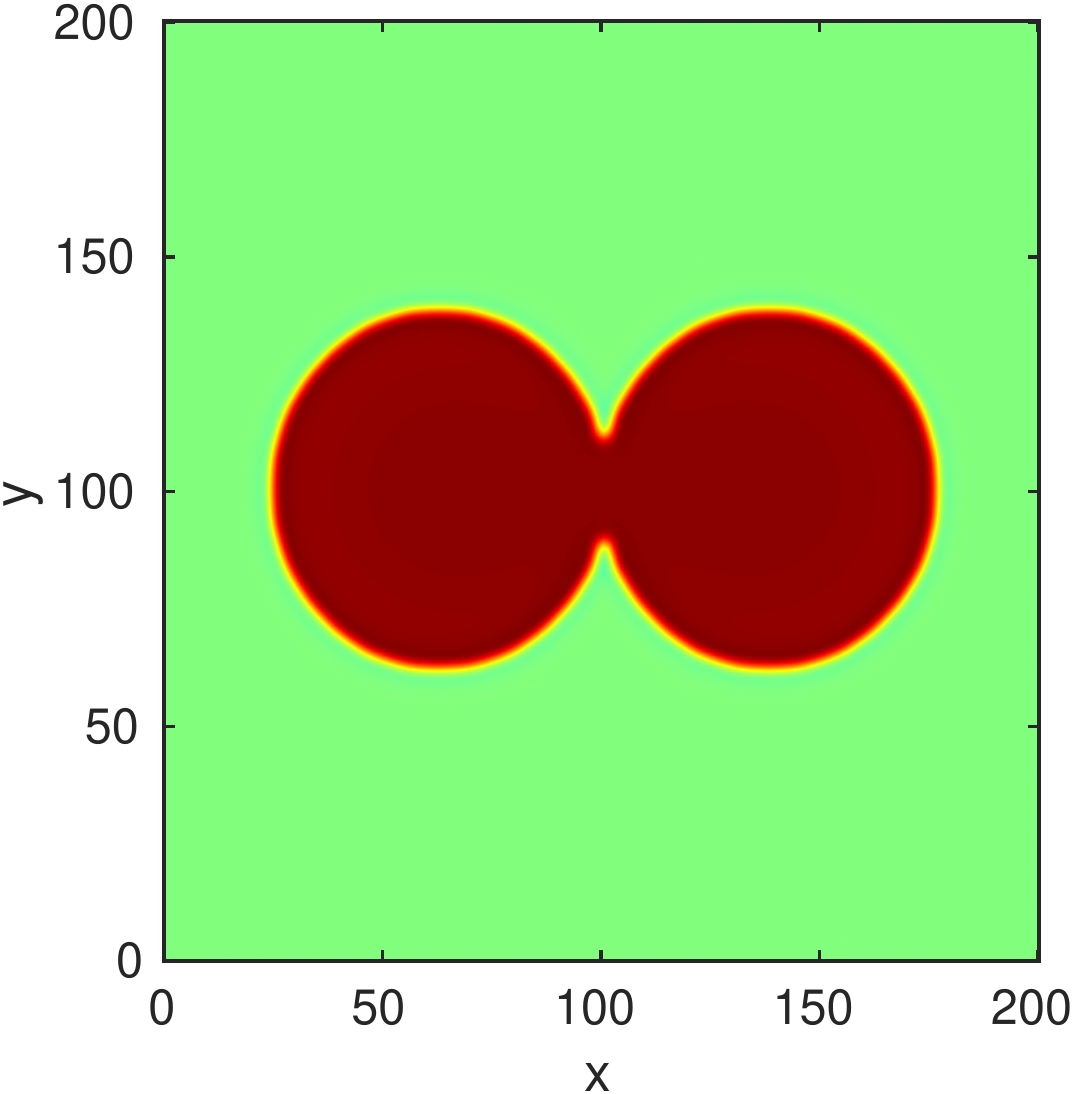}
        \caption{$t = 0.5$}
    \end{subfigure}
    \begin{subfigure}[b]{0.3\textwidth}
        \includegraphics[width=\textwidth]{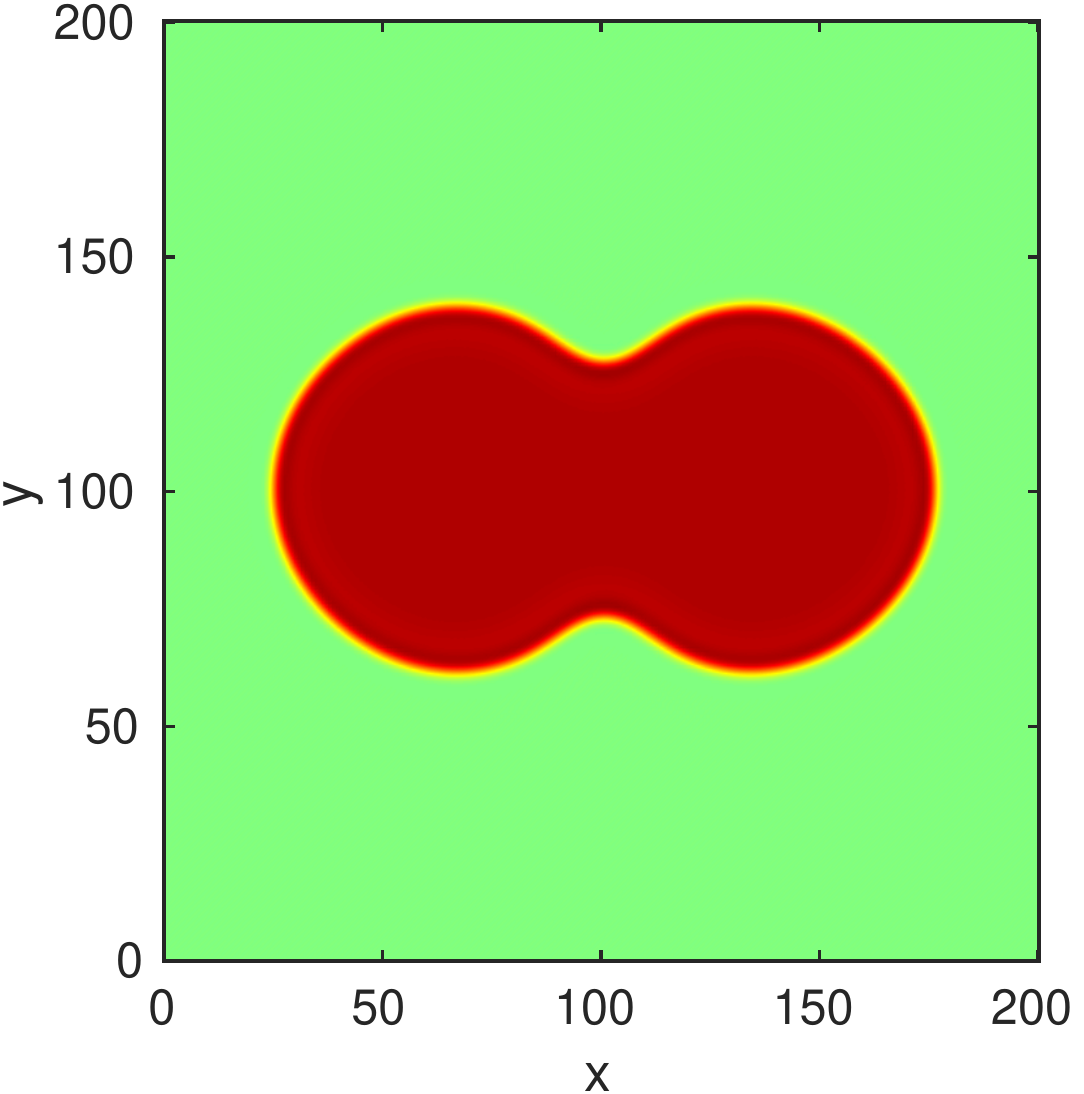}
        \caption{$t = 10$}
    \end{subfigure}
    \begin{subfigure}[b]{0.3\textwidth}
        \includegraphics[width=\textwidth]{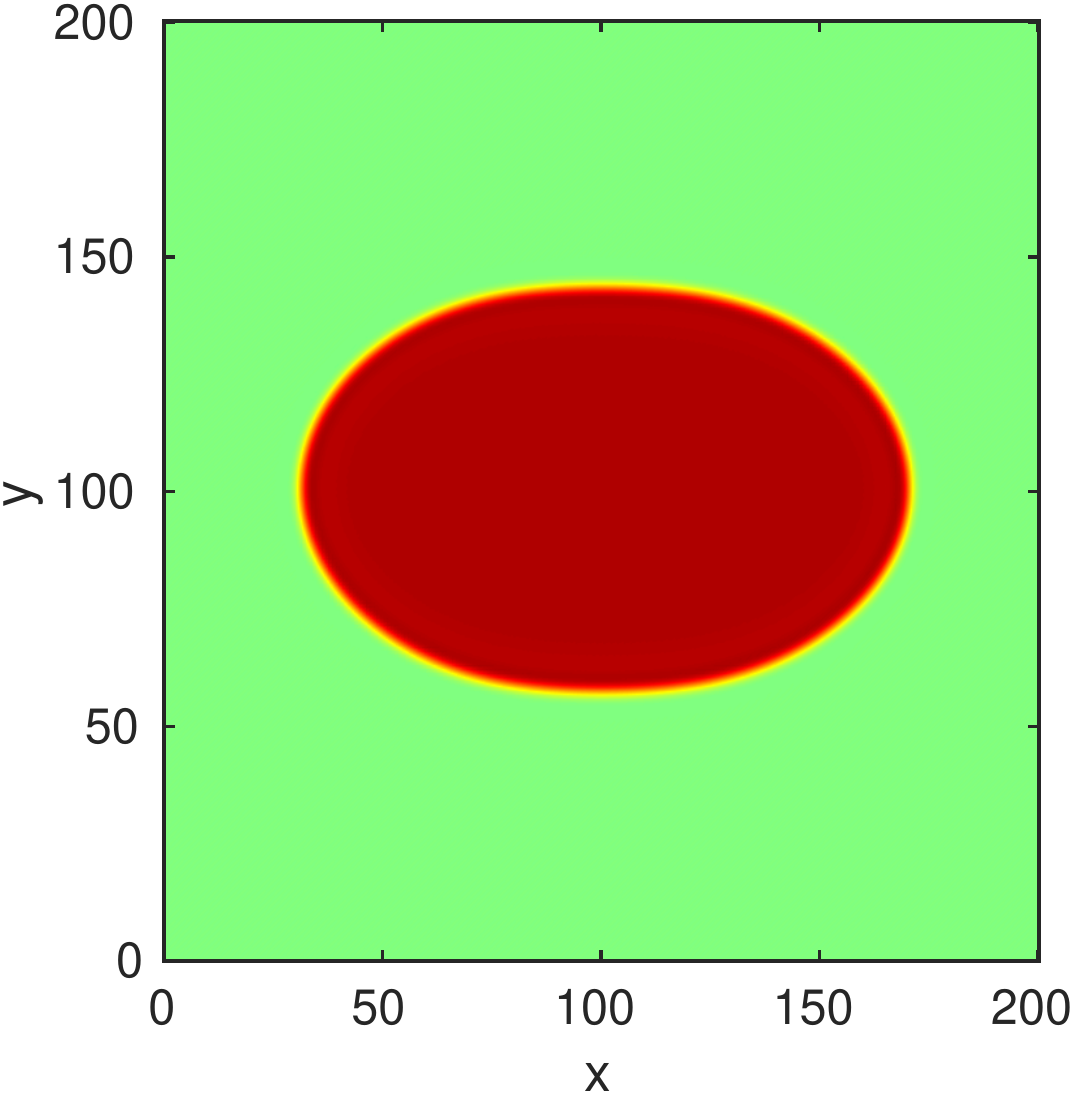}
        \caption{$t = 25$}        
    \end{subfigure}
    \caption{Order parameter at the mid height $xy$ plane of a box with $L_x = L_y = 200$ and $L_z  = 25$ showing coalescence of smectic cylinders, at times $t = 0.5$, $t = 10$ and $t = 50$. Initial conditions present two parallel smectic cylinders with layers filling the whole height $L_z$. Parameters are $\epsilon = 0.3$ (smectic region), $\zeta = 100$, $\beta = 2$, $\gamma = 1$, $\nu = 1$, $\lambda = 1$, $\rho_s = 1$, and $\rho_0 = 0.5$ ($\kappa = 0.3727$).}
    \label{fig:cls2d}    
\end{figure}

In order to make contact with Hopper's theory for infinite fluid cylinders, we focus on the two-dimensional dynamics on the plane of the stacks by changing the initial condition from five to eight smectic layers, so that both cylinders occupy the entire $z$ range of the domain. This initial condition is intended to mimic Hopper's case of infinite cylinders (and is also closer to the smectic islands constrained by enveloping layers observed in experiments). Figure~\ref{fig:cls2d} shows the evolution of order parameter on the midplane ($L_z/2$) of the box: due to the large $\zeta$, the area defined by $\psi$ shows little growth as it evolves. That is, as $\zeta$ and the density ratio gets larger, the area becomes more closely conserved in its evolution. The midplane density evolves in the same way, since the motion of $\psi$ is tied to $\rho$ for large $\zeta$.

While capillarity driven coalescence is well understood for objects such as droplets or cylinders of isotropic fluids, the modulated nature of a smectic leads to several significant differences.  First, the interface between the cylinder and the isotropic phase is composed of layers oriented perpendicular to the interface. According to the equations derived in~\cite{vitral2019role}, shown here in Eq.~(\ref{eq:gt-perp}), diffusion driven interface motion of the smectic for such orientation goes as $v_n \sim -H^{3}$ for zero Gaussian curvature (as in a cylinder), which by itself leads to coalescence at a slower rate than the classical motion by mean curvature. The smectic motion also depends on the evolution of the density, $\dot{\rho}$, governed by Eq.~(\ref{eq:bm-ge-w}), which is proportional to $-\nabla\cdot\mathbf{v}$. Therefore, in the coalescence of smectic cylinders, a competition exists between the diffusional motion of the order parameter and mass transport.

\begin{figure}[ht]
	\centering
    \begin{subfigure}[b]{0.4\textwidth}
        \includegraphics[width=\textwidth]{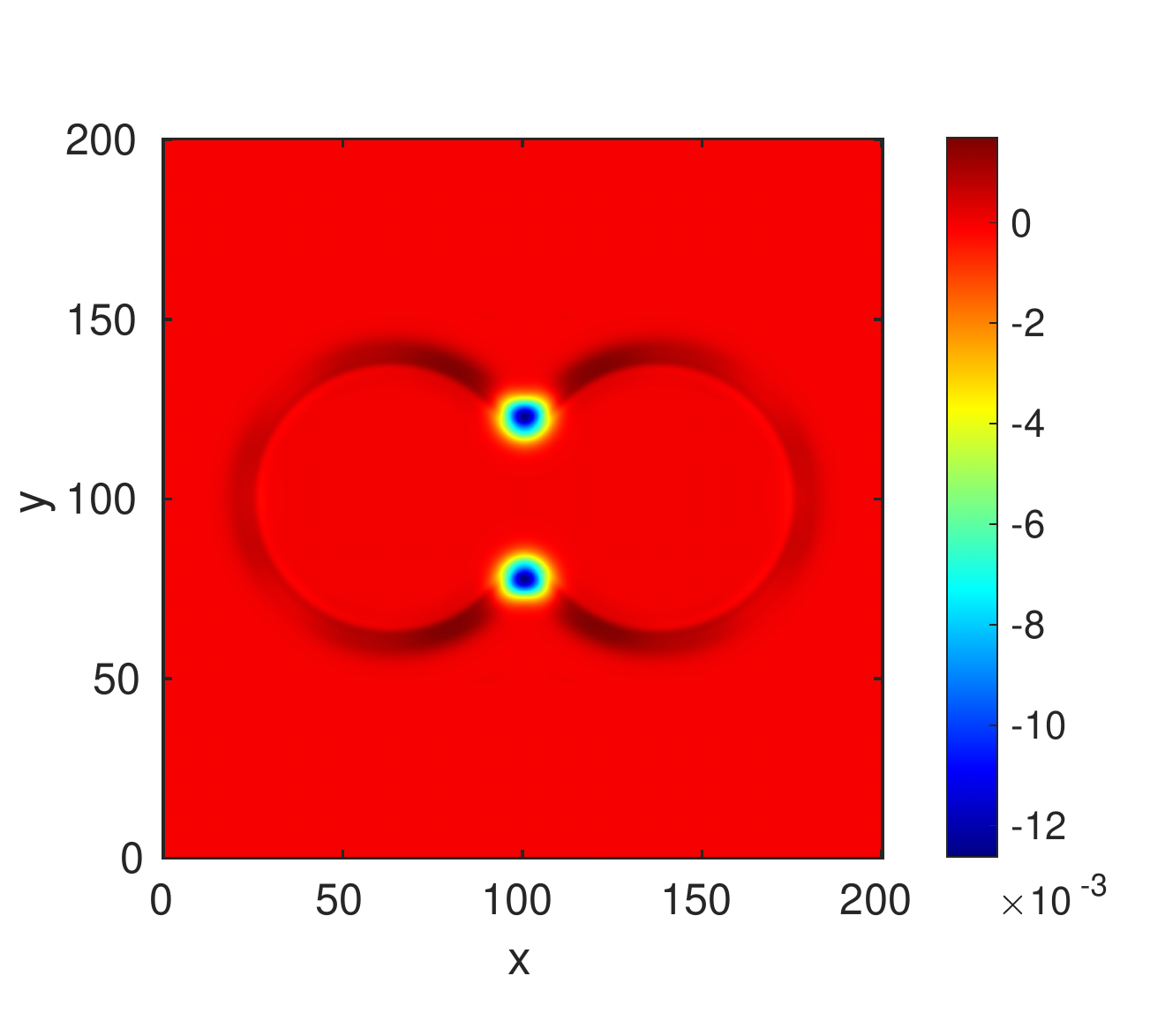}
    \end{subfigure}
    \hspace{10mm}
    \begin{subfigure}[b]{0.4\textwidth}
        \includegraphics[width=\textwidth]{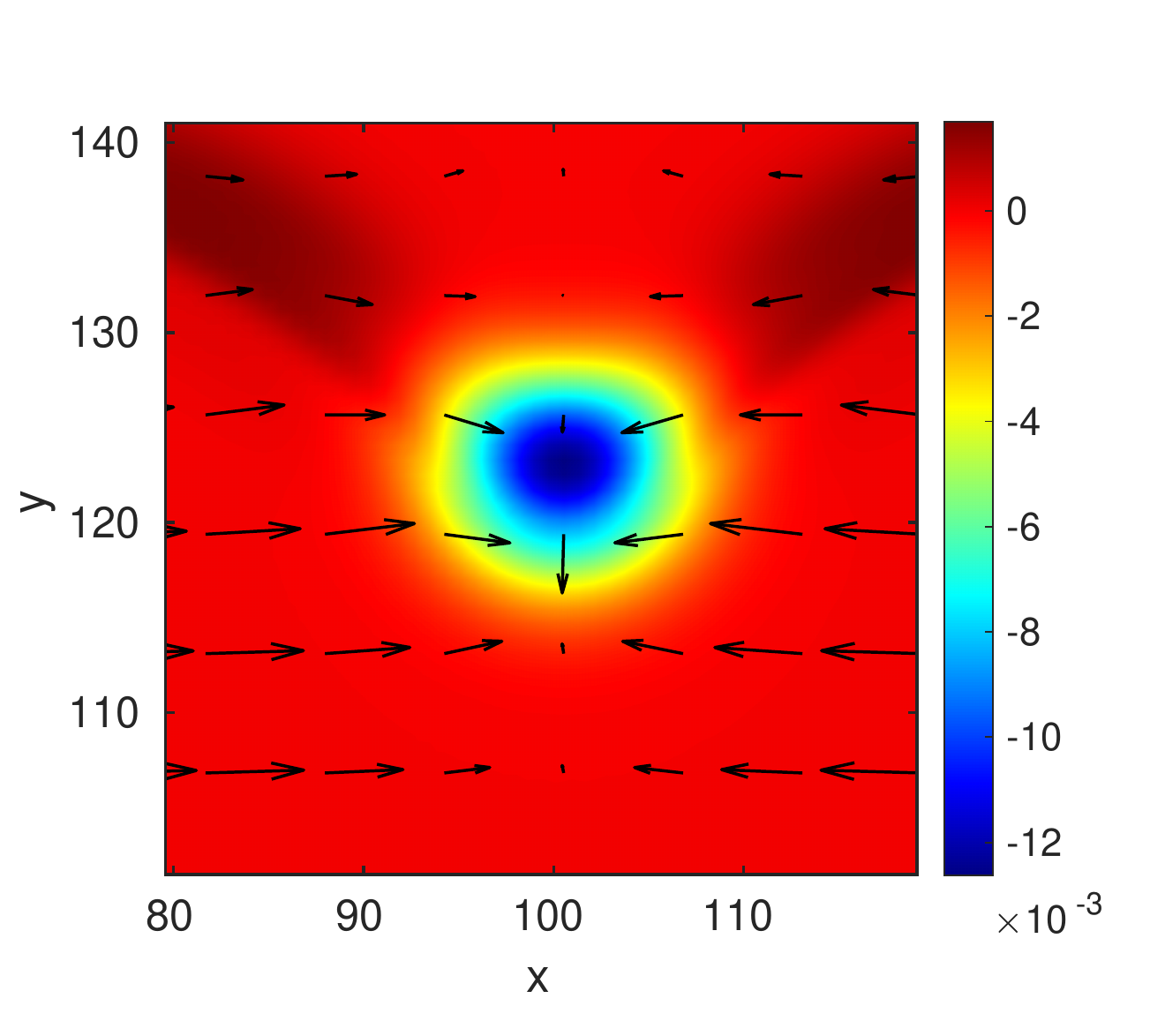}
    \end{subfigure}
    \caption{Divergence of the velocity $\nabla\cdot\mathbf{v}$ at the mid height $xy$ plane of a box with $L_x = L_y = 200$ and $L_z  = 25$ at time $t = 5$. Initial conditions and parameters are the same as in Fig. \ref{fig:cls2d}. The right figure is a close up of the bridge region, showing that the velocity field $\mathbf{v}$ moves from regions of positive $\nabla\cdot\mathbf{v}$ ($H > 0$) towards regions
    of negative $\nabla\cdot\mathbf{v}$ ($H < 0$).}
    \label{fig:divv}    
\end{figure}

Recall that the divergence of the velocity depends on the irrotational flow, $\nabla\cdot\mathbf{v} = \nabla^2\Phi$, given by Eq.~(\ref{eq:irrot}) in Fourier space. The first term on the RHS of Eq.~(\ref{eq:irrot}) is proportional to the normal $\mathbf{n}$ at the interface when $\rho \neq \rho_0 + \kappa A$. Since $\nabla\cdot\mathbf{n} = 2H$, this implies that $\dot{\rho}$ is proportional to the negative of the mean curvature, so that there is a mean curvature driven flow of mass from regions of positive to negative curvature. To verify this, we plot the divergence of the velocity for time $t = 5$ in Fig.~\ref{fig:divv}. Regions of positive $\nabla\cdot\mathbf{v}$ correspond to an interface of positive mean curvature, while regions of negative $\nabla\cdot\mathbf{v}$ have an interface of negative mean curvature. Figure~\ref{fig:divv} shows the enlarged bridge region, and the velocity field $\mathbf{v}$. During coalescence, the flow in the left cylinder moves toward the right, and vice-versa, and at the bridge the flow moves inwards in the region of negative mean curvature, as expected from the coalescence process. 

\begin{figure}[ht]
	\centering
    \includegraphics[width=0.44\textwidth]{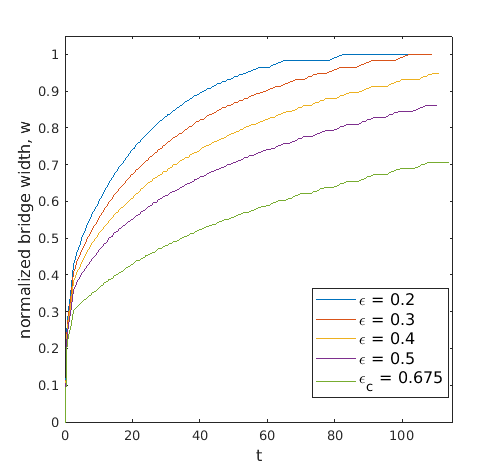}
    \caption{Bridge width as a function of time, for simulations employing different values of $\epsilon$. Domain size, initial conditions and parameters are the same as in Fig. \ref{fig:cls2d}.}
	\label{fig:bridge}
\end{figure}

In our numerical calculations involving smectic cylinders spanning the domain in $z$, we observe coalescence deep inside the smectic region from $\epsilon = 0.2$ up to the coexistence point \mbox{$\epsilon_c = 0.675$}. In Fig.~\ref{fig:bridge} we plot the normalized bridge length between the smectic stacks as a function of time for different values of $\epsilon$. Coalescence occurs faster as we decrease the bifurcation parameter from the coexistence value, that is, as we decrease the energy of the smectic in comparison to the isotropic phase (corresponding to a decrease in temperature). Qualitatively these curves and the numerical evolution of the order parameter resemble the bridge width evolution shown in experiments and Hopper's theoretical model. The phase-field model also allows for more intricate order parameter morphologies, so that a possibility for future work is to simulate FSSF and study the role of permeation on coalescence.

\begin{figure}[ht]
	\centering
    \begin{subfigure}[b]{0.4\textwidth}
        \includegraphics[width=\textwidth]{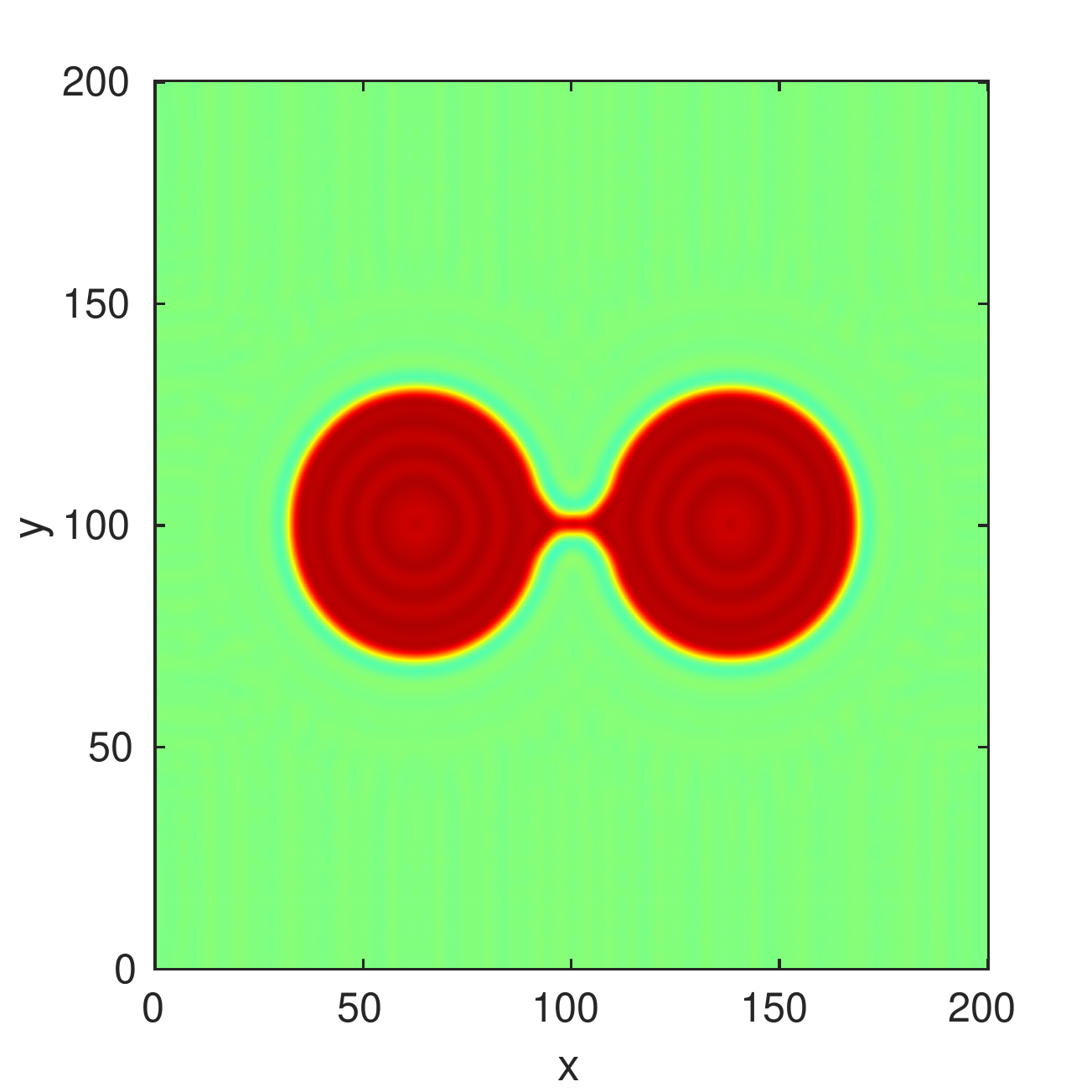}
        \caption{$t = 25$}        
    \end{subfigure}
    \hspace{10mm}
    \begin{subfigure}[b]{0.4\textwidth}
        \includegraphics[width=\textwidth]{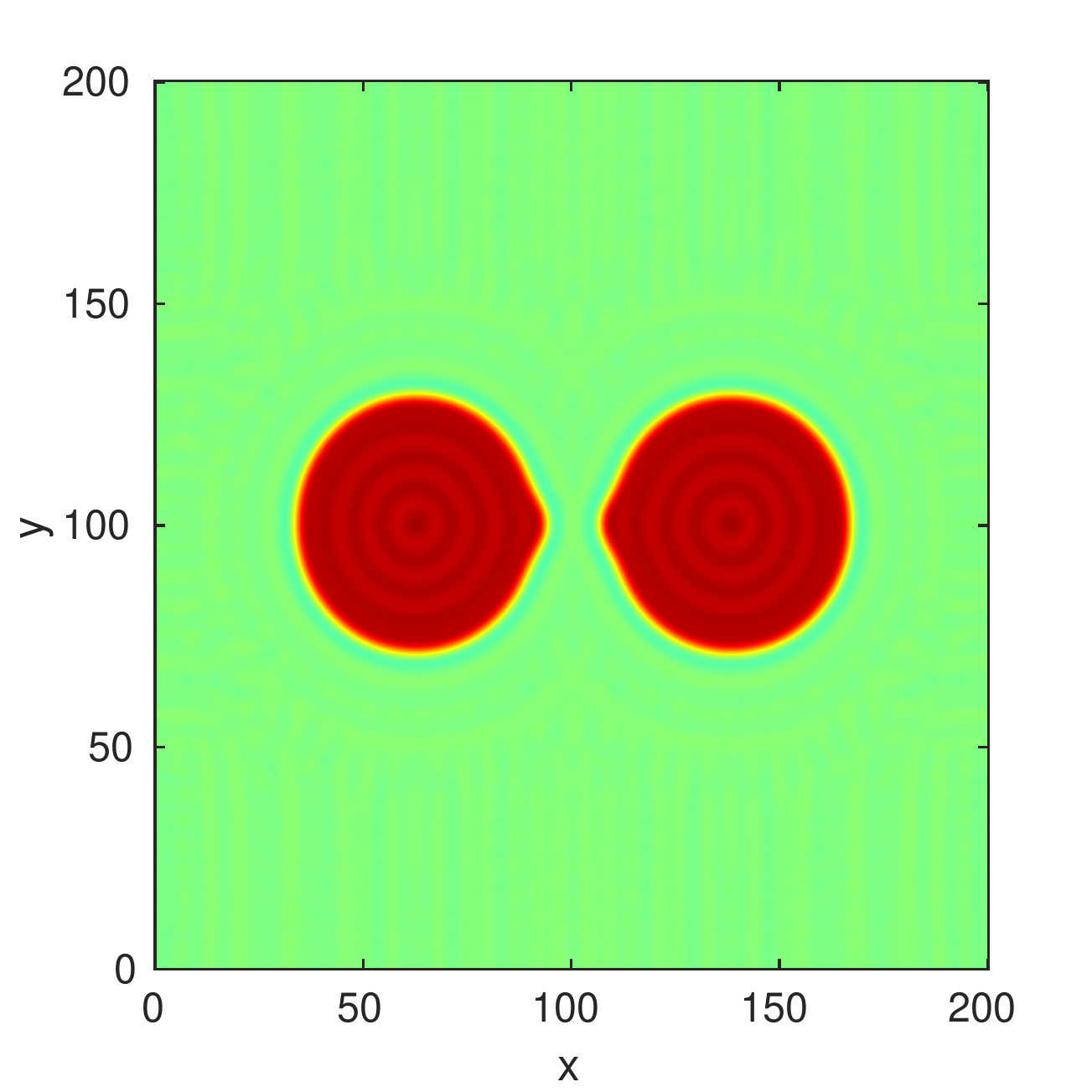}
        \caption{$t = 35$}
    \end{subfigure}
    \caption{Order parameter on the mid $xy$ plane of a domain with $L_x = L_y = 200$ and $L_z  = 50$ showing separation of smectic cylinders, at times $t = 2.5$ and $t = 25$. Initial conditions are two smectic tangential cylinders with approximately five layers. Parameters are $\epsilon = 0.675$ (coexistence), $\zeta = 100$, $\beta = 2$, $\gamma = 1$, $\nu = 1$, $\lambda = 1$, $\rho_s = 1$, and $\rho_0 = 0.5$ ($\kappa = 0.3727$).}
    \label{fig:sep}    
\end{figure}

\begin{figure}[ht]
	\centering
    \begin{subfigure}[b]{0.4\textwidth}
        \includegraphics[width=\textwidth]{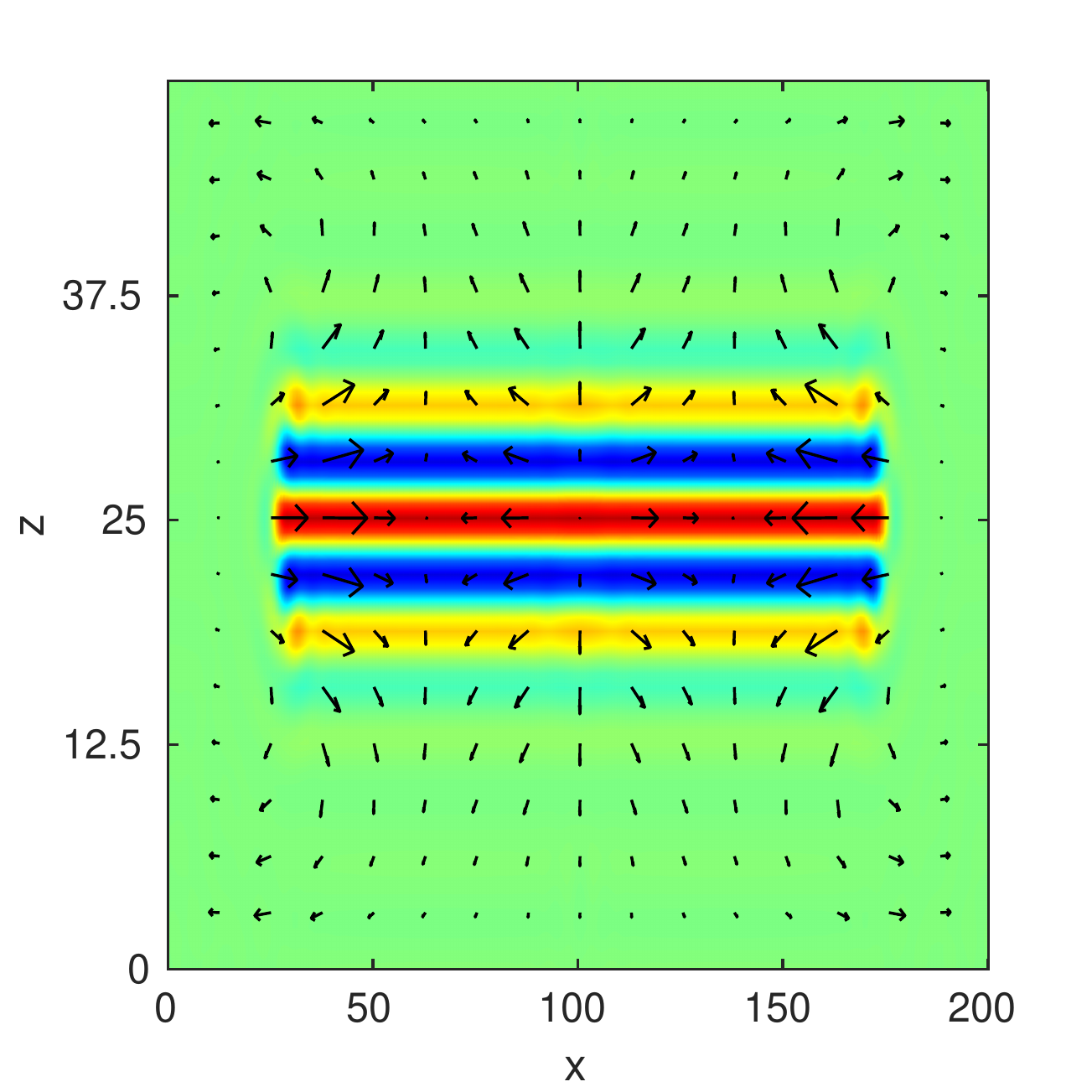}
        \caption{$t = 2.5$}
    \end{subfigure}
    \hspace{10mm}
    \begin{subfigure}[b]{0.4\textwidth}
        \includegraphics[width=\textwidth]{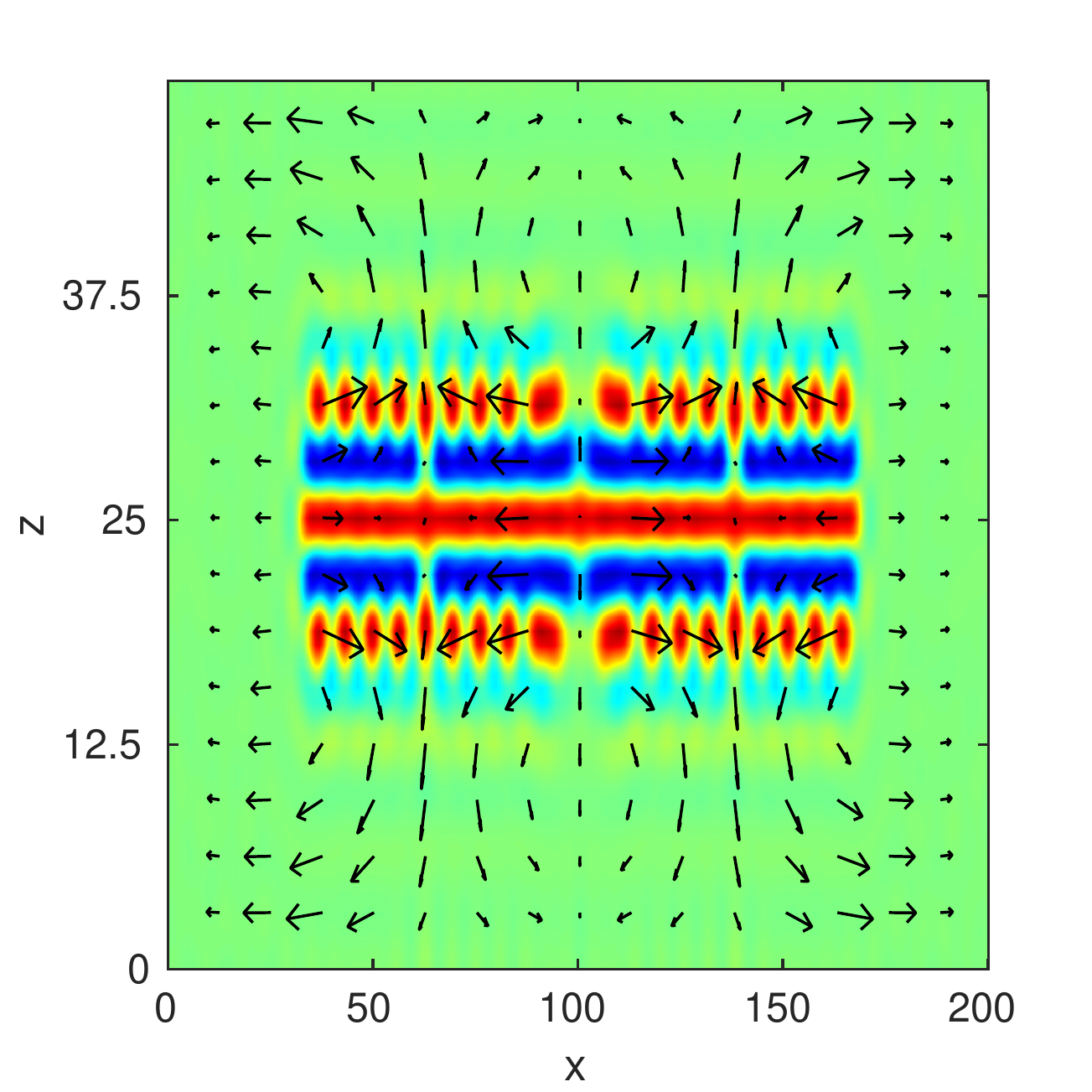}
        \caption{$t = 25$}
    \end{subfigure}
    \caption{Order parameter and velocity field on the mid $xz$ plane for the same calculation of Fig.~\ref{fig:sep}. The $z$ axis has been multipled by a factor of four for better visualizing the flow. Left: two smectic cylinder in contact at $x = 100$, and time $t = 2.5$. Right: smectic cylinders are almost separating due to the flow, time $t = 25$.}
    \label{fig:sep-flow}    
\end{figure}

Interestingly, if the initial smectic cylinders do not span the computational domain in the $z$-direction, coalescence is only observed once $\epsilon$ becomes sufficiently small. For example, using a domain with $N = 256^2\times64$, and the same parameters as above, smectic domains of approximately five layers did not coalesce for $\epsilon$ close to $\epsilon_c = 0.675$. From plots of the midplane order parameter in Fig.~\ref{fig:sep}, we observe that after an initial thin bridge is formed (at $t = 25$), the two stacks move apart, until the bridge breaks (at $t = 35$). In this case, since five layers don't span the domain in the $z$-direction, growth in that direction competes with coalescence, as the total mass of the smectic phase is approximately conserved for large $\zeta$. In order to help visualize this process, Fig.~\ref{fig:sep-flow} shows both the velocity $\mathbf{v}$ and the order parameter in the $x-z$ plane at $L_y/2$. At early times, $t = 2.5$, the two cylinders are still closely in contact (at $x = 100$), but a flow forms that induces mass to move from the lateral of the cylinder towards the top and bottom. At a later time, $t = 25$, the flow continues to point opposite to the bridge (which is very thin at this point, as seen in Fig.~\ref{fig:sep}), leading the cylinders to further separate from each other. As we have seen in Fig.~\ref{fig:bridge}, coalescence is slower at a higher $\epsilon$, so at $\epsilon_c$ the coalescence bridge collapses as mass continuously moves away from it. At the top and bottom of the cylinders, we observe the formation of target structures~\cite{knobloch2015spatial}, which are interfacial rings  (the bottom and top droplets shown in the cross-section) induced by the circular geometry of the order parameter. Therefore, while we argue that the capillary induced diffusion leads to a slow coalescence process, in this case the flow dominates the motion of the smectic and inhibits coalescence.

\begin{figure}[ht]
	\centering
    \begin{subfigure}[b]{0.3\textwidth}
        \includegraphics[width=\textwidth]{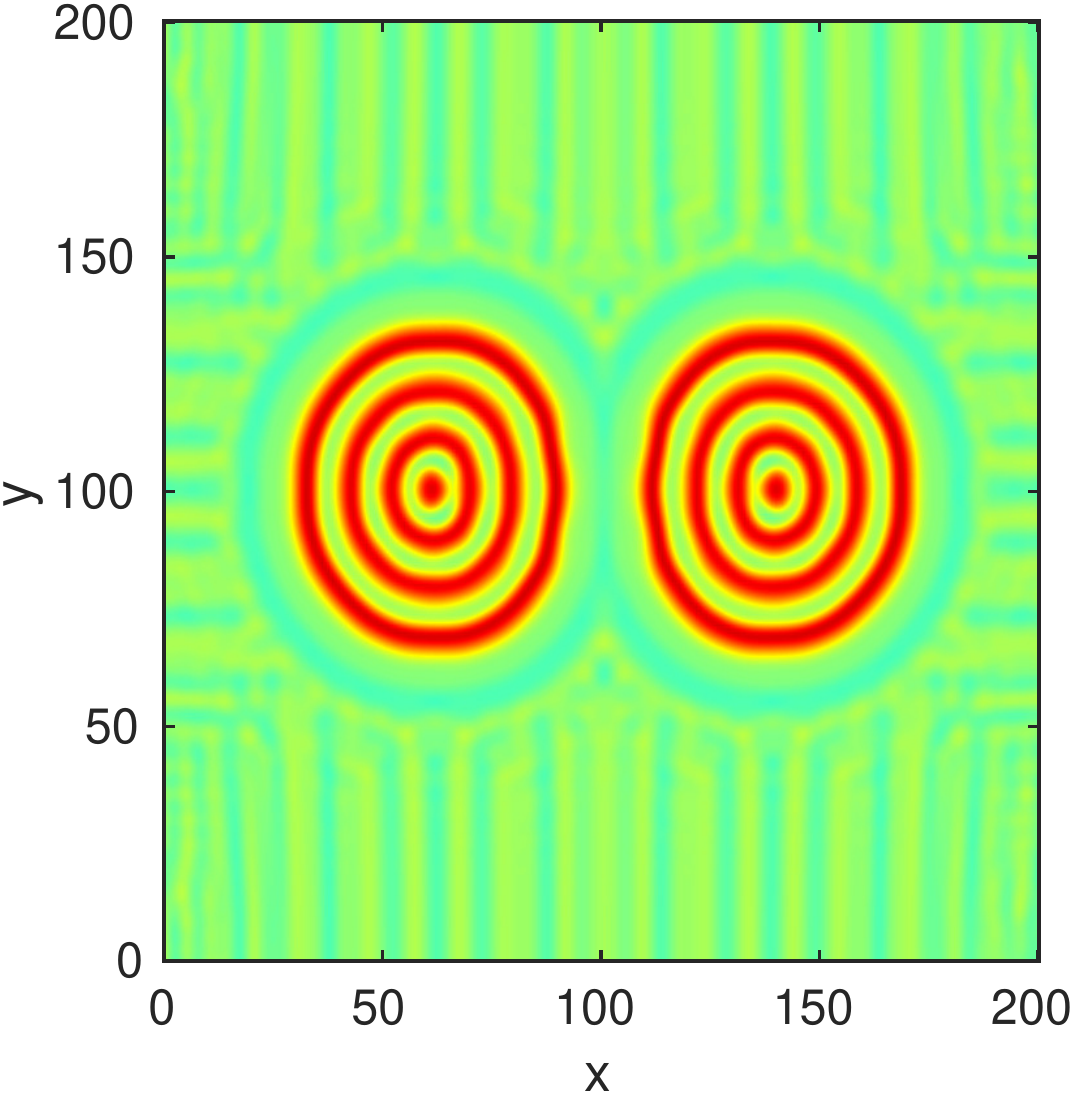}
        \caption{$t = 7.5$}
    \end{subfigure}
    \begin{subfigure}[b]{0.3\textwidth}
        \includegraphics[width=\textwidth]{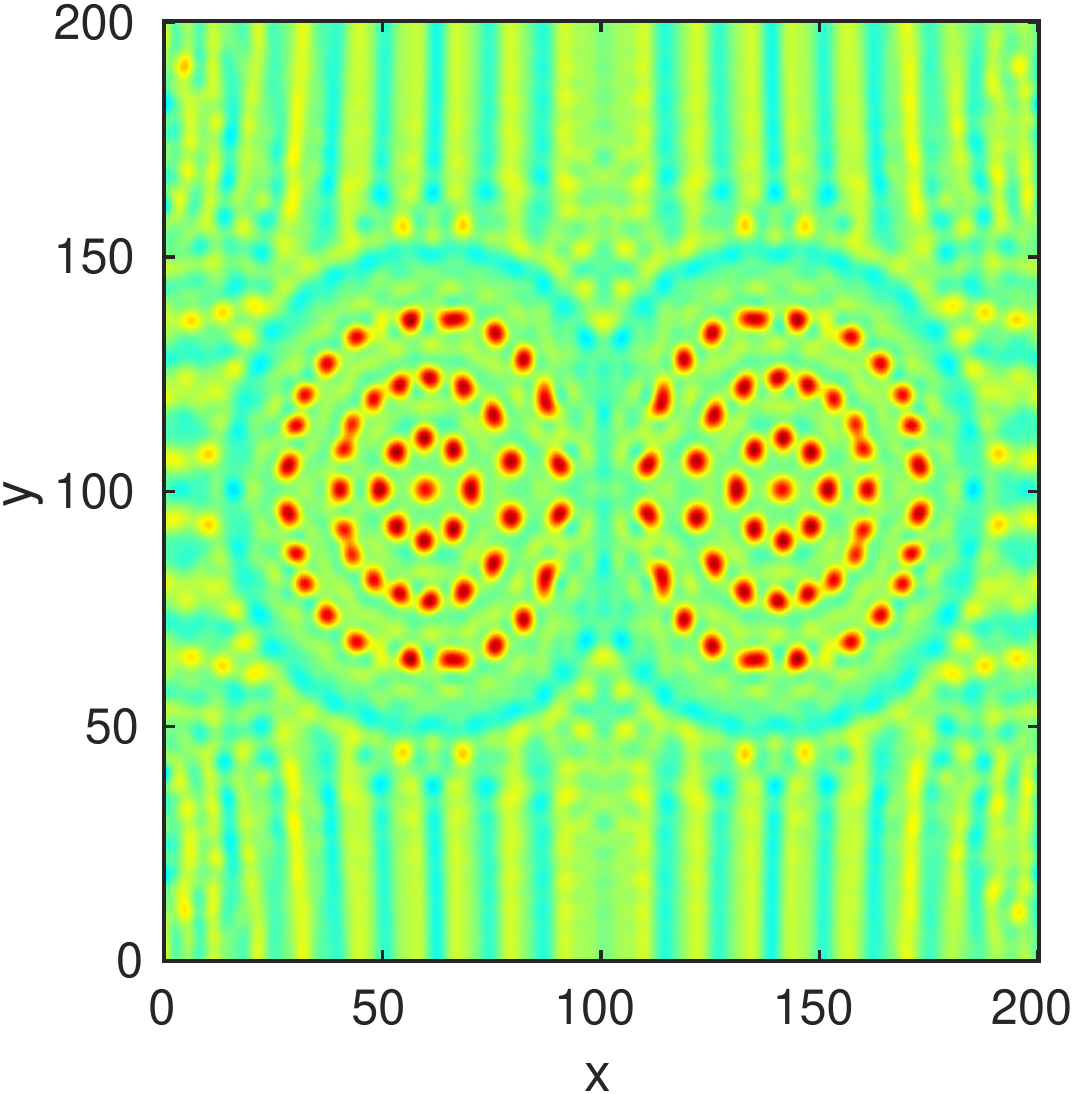}
        \caption{$t = 20$}
    \end{subfigure}
    \begin{subfigure}[b]{0.3\textwidth}
        \includegraphics[width=\textwidth]{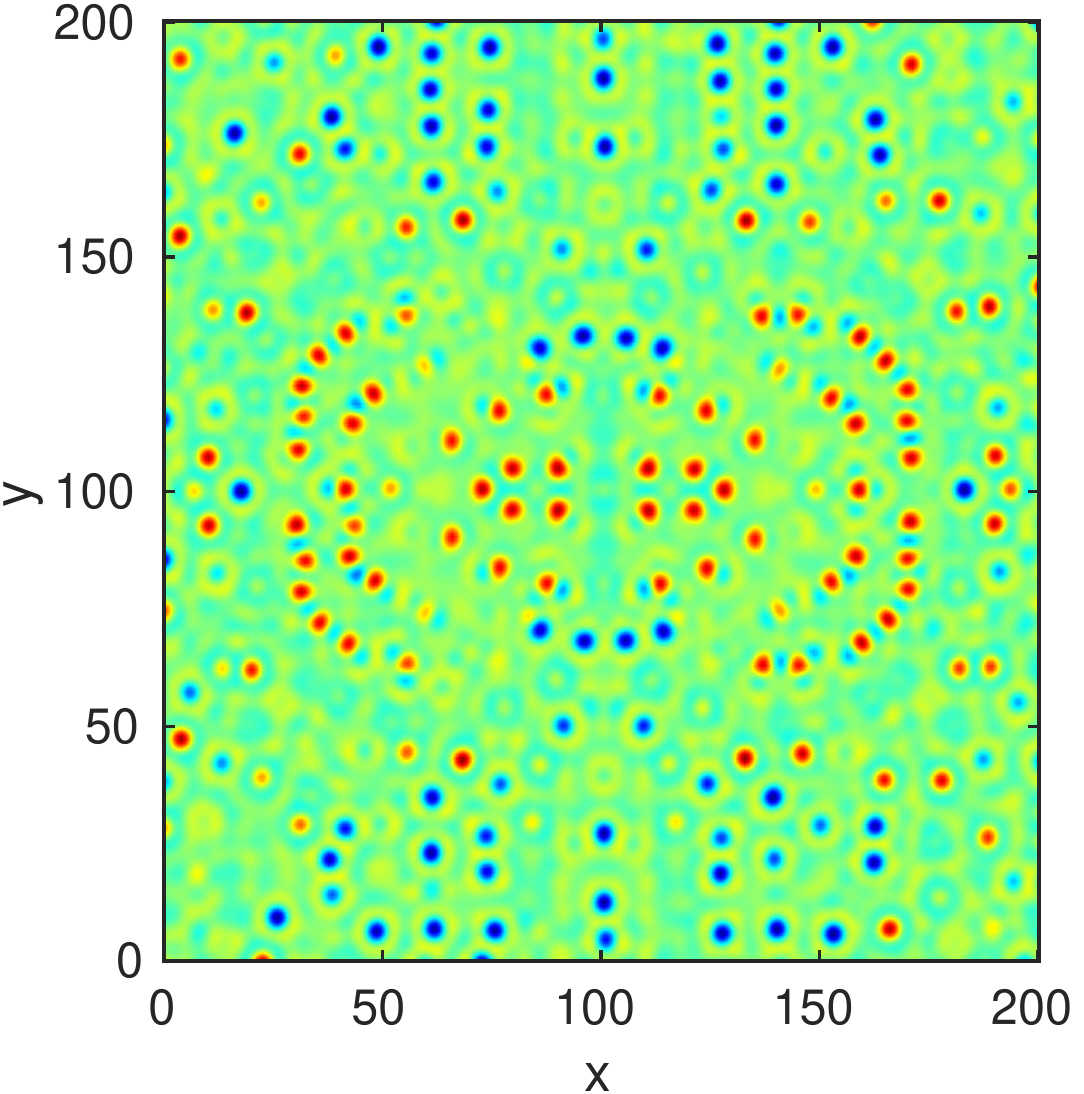}
        \caption{$t = 200$}
    \end{subfigure}
    \caption{Order parameter at the mid height $xy$ plane of a box with $L_x = L_y = 200$ and $L_z  = 25$ showing disintegration of smectic cylinders, at times $t = 7.5$, $t = 20$ and $t = 200$ (steady state). Initial conditions present two smectic tangential cylinders with layers filling the height $L_z$. Parameters are $\epsilon = 0.9$ (isotropic region), $\zeta = 100$, $\beta = 2$, $\gamma = 1$, $\nu = 1$, $\lambda = 1$, $\rho_s = 1$, and $\rho_0 = 0.5$ ($\kappa = 0.3727$).}
    \label{fig:e0d9}    
\end{figure}

We note finally that coalescence does not occur in the isotropic region $\epsilon > \epsilon_c$ for any of the geometries shown in this section, even for cylindrical domains spanning the computational cell. Numerical results are shown in Fig.~\ref{fig:e0d9}, using the same parameters and initial configuration as in Fig.~\ref{fig:cls2d}, with $\epsilon = 0.9$. The smectic region shrinks since $\epsilon > \epsilon_c$. No bridge is initially formed, and the cross-section of the cylinders starts to melt into a target shape, as seen at $t = 7.5$. At time $t = 20$ the target cross-section further breaks into droplets, and at time $t = 200$ the order parameter field shows droplets spread all over the domain. The melting of smectic order when heating suspended films and formation of isotropic or nematic droplets have also been observed experimentally~\cite{schuring2002isotropic,klopp2019structure}, so that the present model may provide further insights on the interaction, arrangement and dynamics of these droplets. In terms of differential equations, these results are strikingly different from classical Swift-Hohenberg dynamics (pure diffusional dynamics of $\psi$), for which the order parameter in the isotropic region would simply disappear in time. We conclude that coalescence requires synergy between flow and diffusive dynamics of the order parameter and density, and it may not occur if one of them becomes antagonistic to the coalescence process.

\subsection{Interactions between focal conic domains}

While topological defects are known to interact in various soft matter systems, little is known about interactions between focal conics, or even about flows on their surface. The formation of a focal conic domain of the type investigated in this paper depends on the balance between the splay energy to form the conic (proportional to splay elastic constant $K_{1}$), and the difference between the surface tensions for parallel and perpendicular molecular anchoring at the the smectic-air interface ($\Delta \sigma = \sigma_{\parallel}-\sigma_{\perp}$). The size of the conic is determined by the balance of the two energies, and it is of the order of $K_{1}/|\Delta \sigma|$~\cite{lavrentovich1994nucleation}. Once an equilibrium array of focal conics is formed, Kim et al.~\cite{kim2016controlling} have reported experiments on sintering of FCDs that showed morphologies in which neighboring focal conic domains interact, for example via thin tunnel like structures that remains from the original film. In order to begin to understand these interactions, we have investigated numerically the evolution of initial configurations that consist of two focal conics, where we have biased the system away from equilibrium. We show three different examples here: (1) we differentially compress the smectic layers of two neighboring focal conic domains, (2) we impose a density gradient in the isotropic phase from one defect towards the other, and (3) we start from an initial condition in which two focal conics overlap. In all cases we use a computational domain with $N = 512 \times 256^2$ grid points,  8 points per $\psi$ wavelength and $q_0 = 1$, so that $L_x = 400$, $L_y = 200$ and $L_z = 200$, and we set the density ratio between phases $\rho_s:\rho_0$ as $2:1$ (with $\rho_s = 1$ and $\rho_0 = 0.5$). Other parameters used are $\beta = 2$, $\gamma = 1$, $\nu = \lambda = 1$, and $\zeta = 1$, which allows for a moderate coupling between the order parameter and density fields.

\subsubsection{Layer compression}

\begin{figure}[ht]
	\centering
    \begin{subfigure}[b]{0.72\textwidth}
        \includegraphics[width=\textwidth]{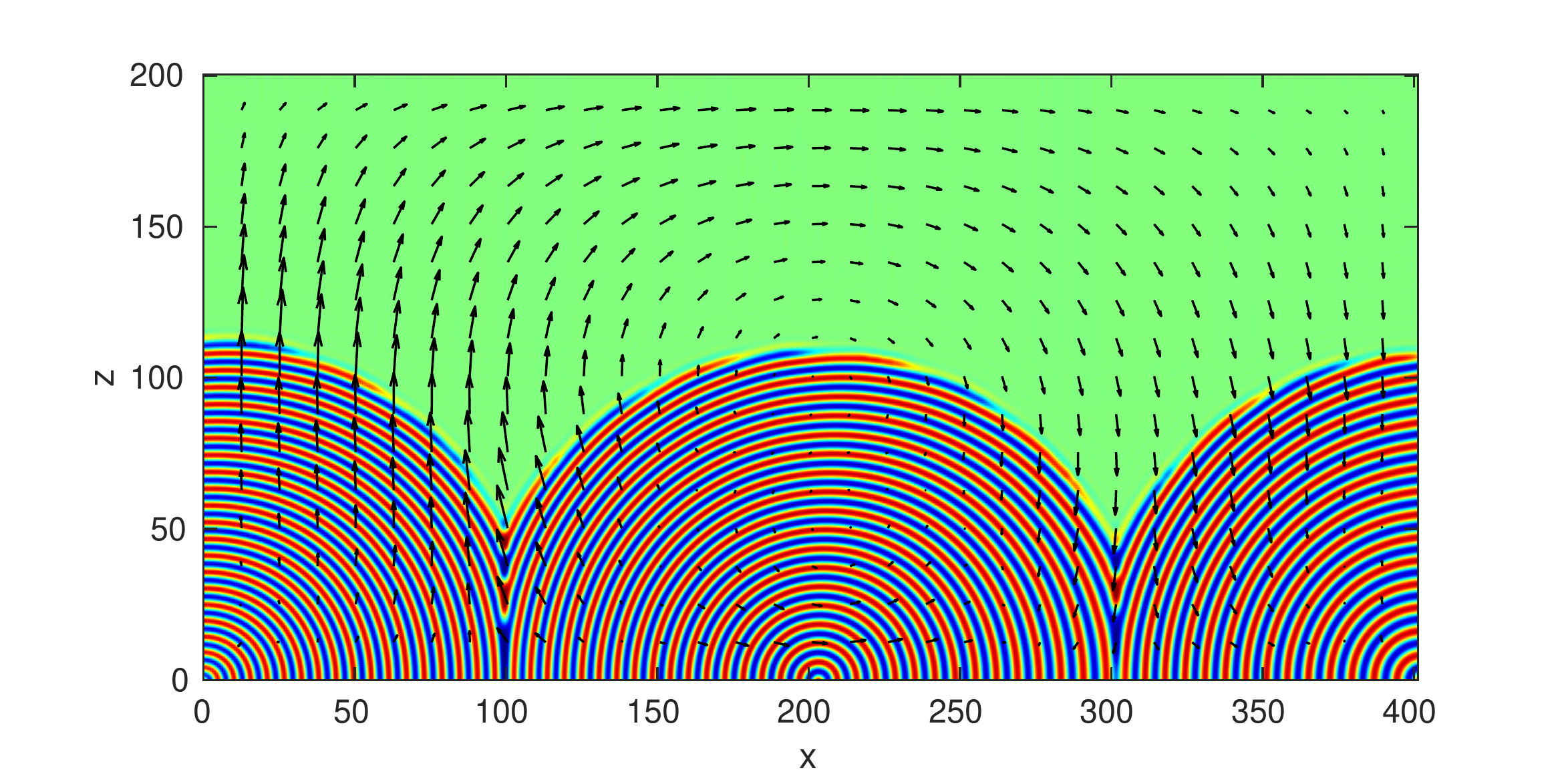}
        \caption{$t = 0.5$}
\end{subfigure}
    \begin{subfigure}[b]{0.72\textwidth}
        \includegraphics[width=\textwidth]{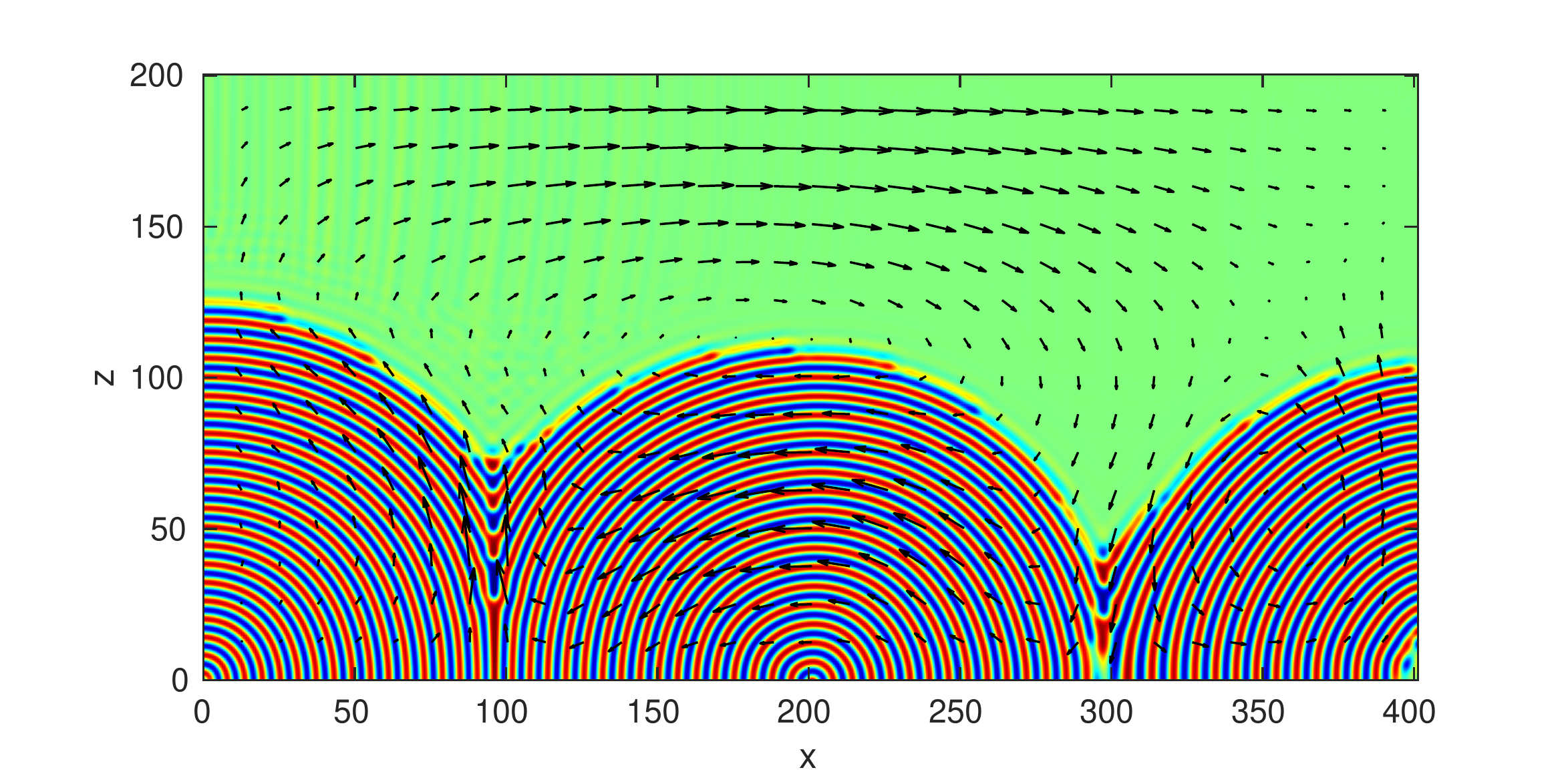}
        \caption{$t = 15$}
    \end{subfigure}
    \caption{Two neighboring focal conics with a varying strain field, showing both the transient velocity field $\mathbf{v}$ and the order parameter $\psi$. The layers of the left focal conic are initially compressed, and those of the right focal conic stretched.}
    \label{fig:2fc-stress}    
\end{figure}

The equilibrium layer spacing, given by $\lambda_0/2$, may be changed by straining the liquid crystal~\cite{clark1973strain}. For arrays of FCDs, due to boundary conditions or proximity to defects, some regions may have layers deviating from $\lambda_0/2$. We impose a variable strain so that the local layer wavenumber increases linearly from $1.2q_0$ at $x = 0$ to $0.8q_0$ at $x = L_x$. Hence, the layers in the left focal conic are initially compressed, and the right ones stretched. Figure~\ref{fig:2fc-stress} shows the velocity $\mathbf{v}$ and order parameter $\psi$ at the middle cross section of the domain ($y = L_y/2$) for $\epsilon = \epsilon_c$. At early times, $t = 0.5$, there is induced flow from the conic in compression to the conic in extension. At $t = 15$ there is a strong flux in the bulk smectic so that the compressed conic grows toward the isotropic phase while increasing the interlayer spacing. On the other side, the conic with stretched layers contracts by decreasing the interlayer spacing. Simultaneous evaporation/condensation and mass transport through the smectic allows the initially compressed (expanded) focal conic to expand (contract) at a later time, so that its layers achieve the required equilibrium layer spacing.

\subsubsection{Imposed density gradient in the isotropic phase}

\begin{figure}[htb!]
	\centering
    \begin{subfigure}[b]{0.72\textwidth}
        \includegraphics[width=\textwidth]{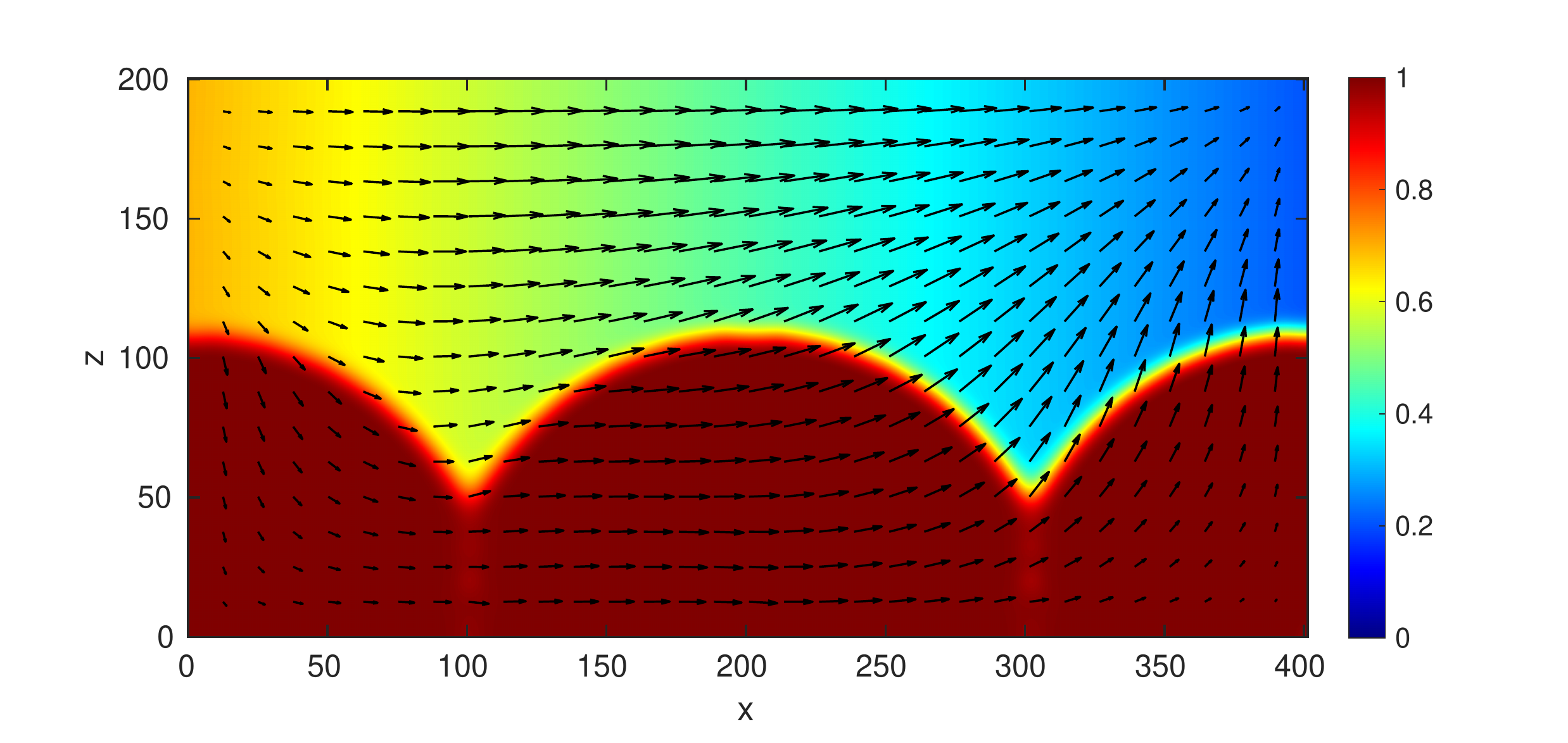}
        \caption{$\rho$, $\;t = 0.5$}
    \end{subfigure}
    \begin{subfigure}[b]{0.72\textwidth}
        \includegraphics[width=\textwidth]{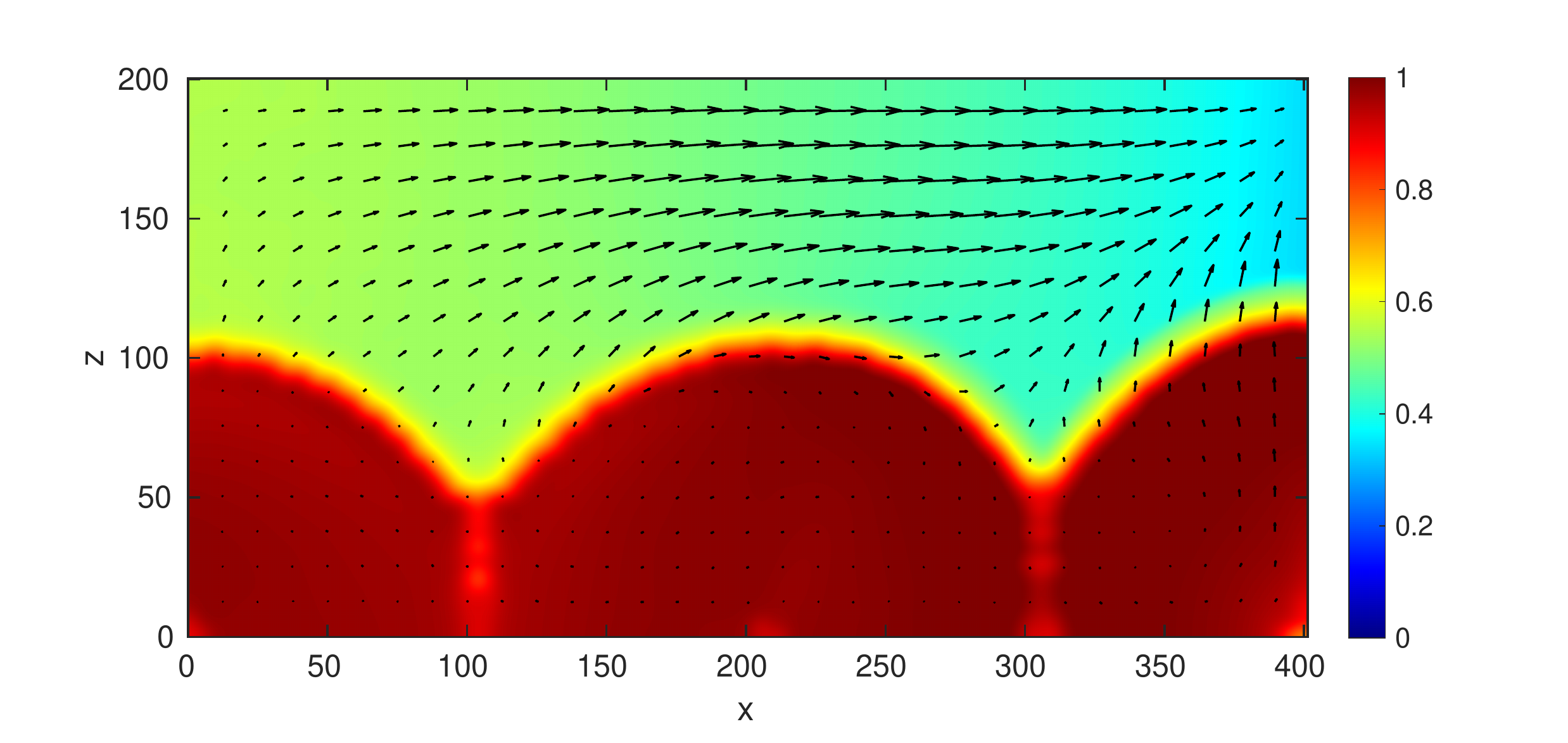}
        \caption{$\rho$, $\;t = 25$}
    \end{subfigure}
    \begin{subfigure}[b]{0.72\textwidth}
        \includegraphics[width=\textwidth]{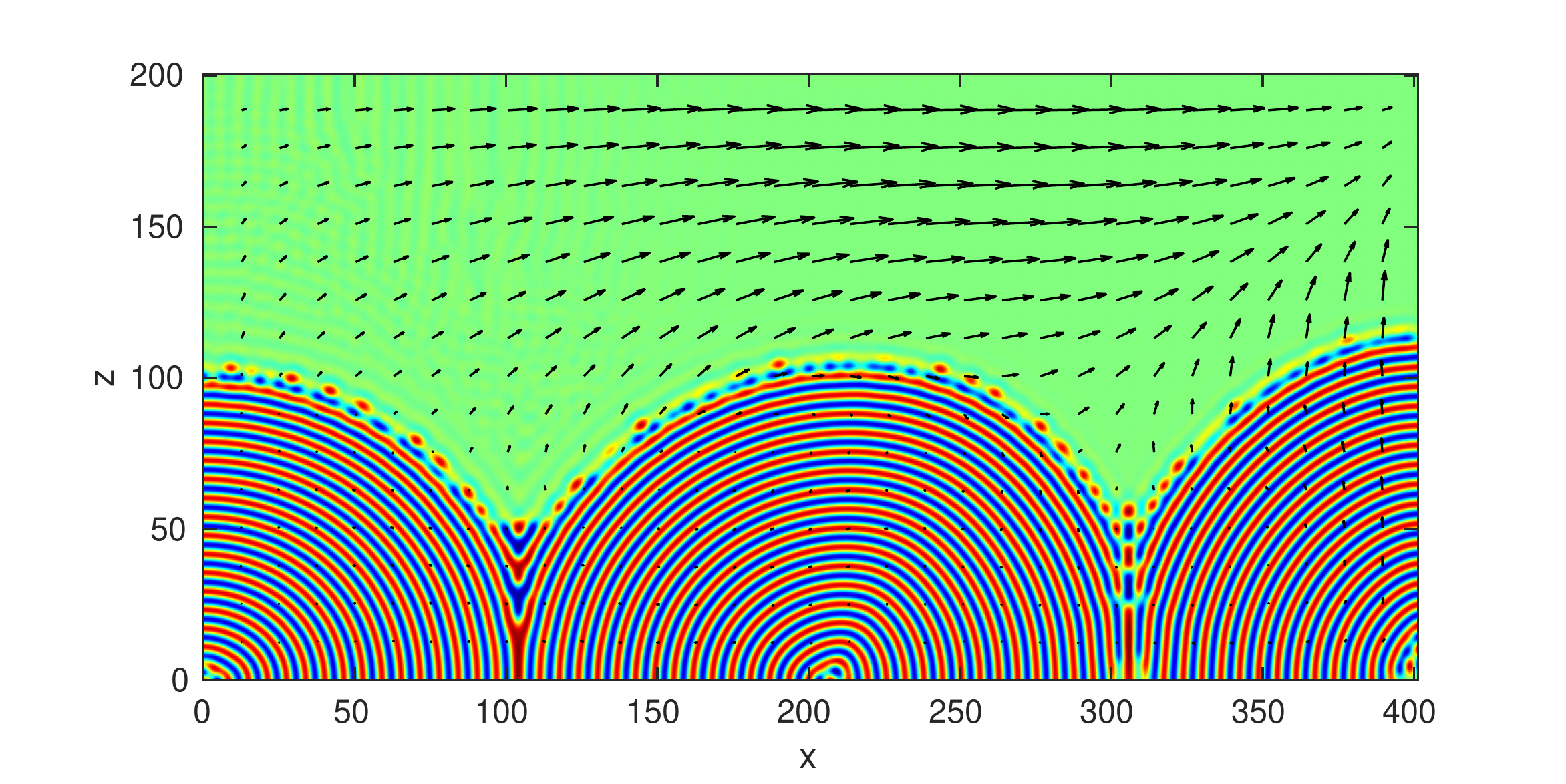}
        \caption{$\psi$, $\;t = 25$}
    \end{subfigure}
    \caption{Two neighboring focal conics in contact with an isotropic phase, with an initial density gradient present in the isotropic phase. Figures show the transient velocity field $\mathbf{v}$ alongside the density (a,b), and with the order parameter (c).}
    \label{fig:2fc-rhog}    
\end{figure}

The second situation we consider is an initial condition consisting of two identical focal conics, with an imposed density gradient in the isotropic phase. The initial density in the isotropic phase is a function of $x$, starting at $\rho = 0.7$ at $x = 0$, and decreasing linearly up to $\rho = 0.2$ at $x = L_x$ (hence deviating from $\rho_0 = 0.5$).  The density of the smectic is constant at $\rho_s = 1$. Figures~\ref{fig:2fc-rhog} (a) and (b) show the velocity field with the density in the background for times $t = 0.5$ and $t = 25$ with $\epsilon = 0.8$. The velocity field goes to the right (so from high density to low density) in the isotropic phase, and also points to the right in the smectic, despite the smectic having uniform density $\rho_s = 1$.  At time $t = 25$, the velocity field is small in the smectic, but flow in the isotropic region continues since the density is still not homogeneous. Figure~\ref{fig:2fc-rhog}(c) shows the order parameter field $\psi$ at time $t = 25$. Note that the radius of the region containing the left focal conic has increased in time due to the imposed density gradient. The center of the left focal conic has also moved from $x = 100$ to the right, and this region now spans up to approximately $x=215$, thereby compressing the right focal conic.

\subsubsection{Overlapping focal conics}

\begin{figure}[htb]
	\centering
    \begin{subfigure}[b]{0.72\textwidth}
        \includegraphics[width=\textwidth]{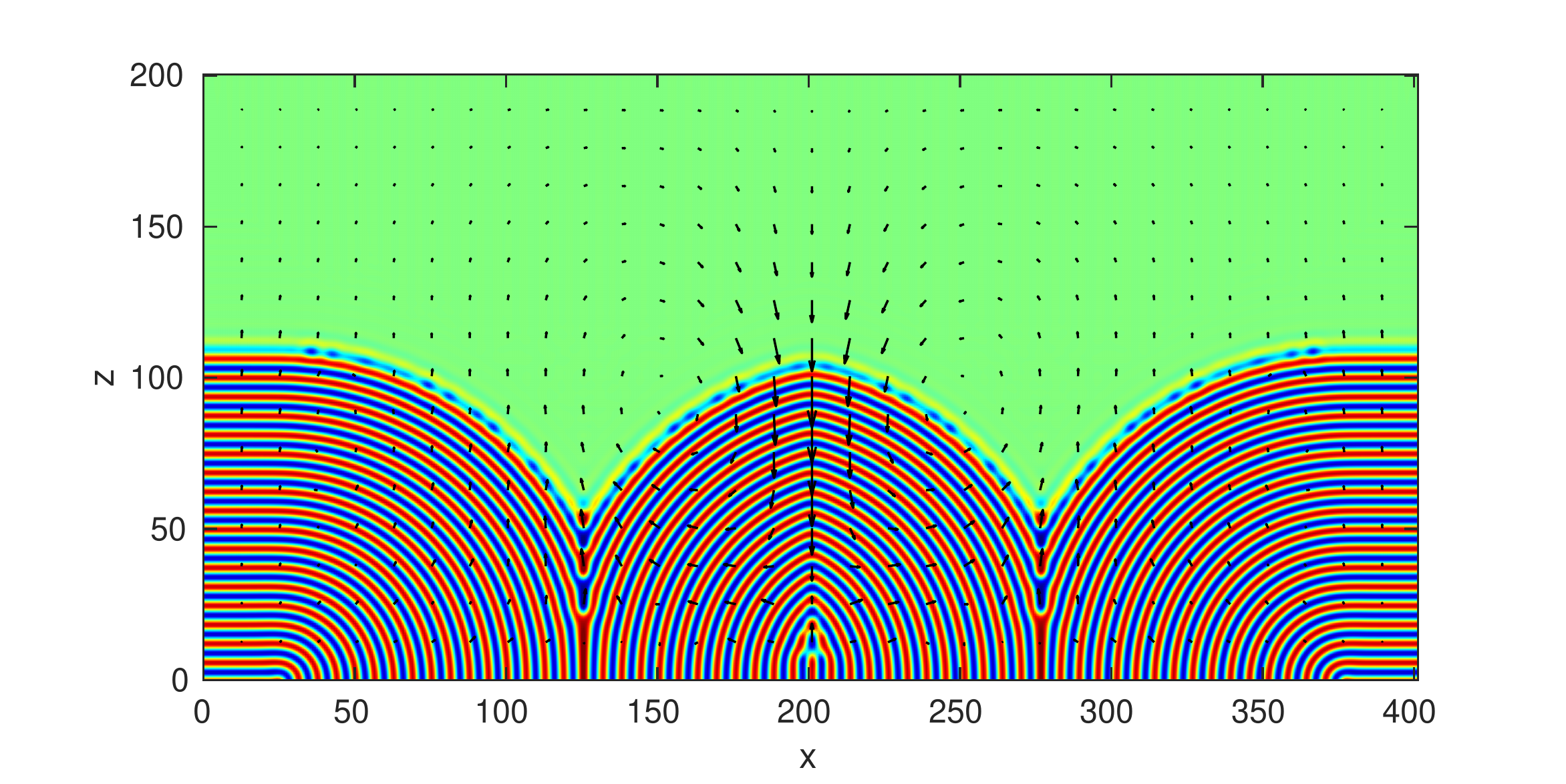}
        \caption{$t = 5$}
    \end{subfigure}
    \begin{subfigure}[b]{0.72\textwidth}
        \includegraphics[width=\textwidth]{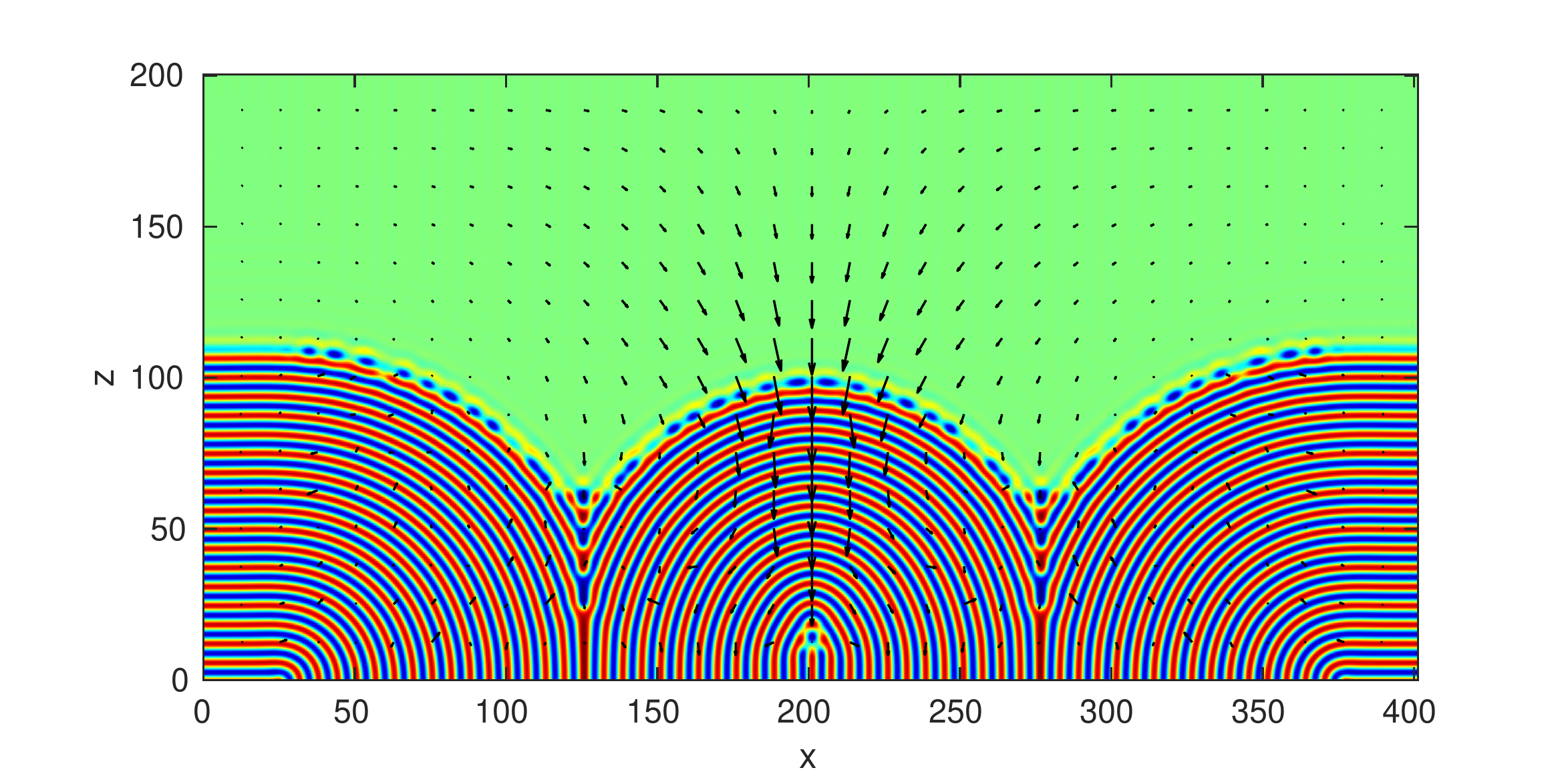}
        \caption{$t = 38$}
    \end{subfigure}
    \caption{Two neighboring focal conics that overlap at $x = 200$, showing both the transient velocity field $\mathbf{v}$ and the order parameter $\psi$. The minimum non-overlapping distance between their cores is $L_e = 200$, while the distance employed as initial condition is $L = 150$.}
    \label{fig:2fc-over}    
\end{figure}

The third example is that of two focal conics with a reduced distance between their centers. Figure \ref{fig:2fc-over} shows the velocity and order parameter fields at the cross section $y = L_y/2$ with $\epsilon = \epsilon_c$. The initial distance between the cores is $L = 150$, whereas the minimum distance under these conditions to isolate two focal conics is $L_e = 200$. This compression creates a region of high positive mean curvature between the two focal conics, inducing strong flow from the isotropic phase towards the smectic in that region. This flow persists for long times as the focal conic domains relax to equilibrium.

\section{Conclusions}

A coupled phase-field and hydrodynamics model has been introduced to describe a weakly compressible two phase system consisting of a smectic (soft modulated phase) in contact with an isotropic fluid of different density (e.g. water, air or the own liquid crystal isotropic state). A non-conserved smectic order parameter is coupled to a conserved mass density so as to accommodate non-solenoidal flows near the smectic-isotropic boundary arising from a density contrast between the two phases. The model energy is a functional of the order parameter and its derivatives, and also includes coupling to a conserved density that has different values in bulk smectic and isotropic regions. For large values of the coupling coefficient, the order parameter becomes approximately conserved. This is the quasi-incompressible limit in which the density is constitutively related to the order parameter. However for smaller values of the coupling coefficient the variations of density and order parameter become independent. In real physical systems, this suggests a temperature dependence of the coupling coefficient: for instance, it would become lower at elevated temperatures in order to allow the smectic to melt/evaporate independently of the density. The model fully incorporates surface driven flows due to local stresses that depend on boundary curvatures, and non-solenoidal flows that arise from density gradients.


The weakly compressible model has been used to describe morphological instabilities in smectic thin films, away from an equilibrium configuration comprising an array of focal conic domains. Experiments show that, upon sintering, focal conics decay into conical pyramids and concentric rings. For a single focal conic domain, our model leads to a transition between focal conics and conical pyramids upon increasing the temperature, mediated, as in the experiments, by localized evaporation of smectic layers. Flows are induced by boundary stresses due to curvature -including Gaussian curvature- through an extended form of equilibrium thermodynamics conditions at a curved surface (the Gibbs-Thomson equation). Furthermore, different boundary conditions apply when smectic layers become exposed in pyramidal domains, which do not have a counterpart in classic thermodynamics. This configuration is associated with tangential flows at the boundary of conical pyramids, which helps explain why this structures persist while focal conics evaporate. Irrotational flows are also induced by boundary motion due to phase density changes. 

As further applications, we consider configurations that mimic focal conic interactions. These include setting up a density gradient in the isotropic phase above two focal conics, adding a nonuniform strain in the smectic layers, and taking the inter-center distance of two conics small enough so the conics overlap. Finally, we discuss the coalescence of cylindrical stacks of smectic layers, motivated by Hopper's theory of coarsening of isotropic fluid cylinders, and experiments on freely-suspended smectic films. Coalescence depends on the competition between capillarity induced diffusion of the order parameter, local curvatures, and mass transport. Instead, if we simulate a heating of the smectic cylinders, numerical results show the melting of smectic order and formation of droplets, a phenomenon that has also been observed in experiments. Future work will address the complex role of permeation in the coalescence of islands and holes, as discussed in recent experiments, and further explore the physics behind the transition into arranged droplets.


\section*{Acknowledgments}

This research has been supported by the Minnesota Supercomputing Institute, and by the Extreme Science and Engineering Discovery Environment (XSEDE) \cite{towns2014xsede}, which is supported by the National Science Foundation under Grant No. ACI-1548562. EV thanks the support from the Doctoral Dissertation Fellowship and from the Aerospace Engineering and Mechanics department, University of Minnesota. This research is also supported by the National Science Foundation under Grant No. DMR-1838977. Part of the work of JV was done during the ACTIVE20 program at the Kavli Institute for Theoretical Physics which is funded by the National Science Foundation under Grant No. NSF PHY-1748958.


\bibliographystyle{vancouver}
\bibliography{wcomp}

\end{document}